\documentclass[final]{elsart}
\usepackage{side}
\usepackage{rotate}
\usepackage{graphicx}
\usepackage{amsmath}
\usepackage{amssymb}
\usepackage{subfigure}
\usepackage{epsfig}
\usepackage{lscape}
\usepackage{lineno}

\journal{Nuclear Physics B}

\begin{document}

\begin{frontmatter}

\title{
A Precision Measurement of Charm Dimuon Production in Neutrino 
Interactions from the NOMAD Experiment 
}

\author[6]{O.~Samoylov}
\author[19]{R.~Petti}
\author[25]{S.~Alekhin}
\author[14]{P.~Astier}
\author[8]{D.~Autiero}
\author[18]{A.~Baldisseri}
\author[13]{M.~Baldo-Ceolin \thanksref{Deceased}}
\author[14]{M.~Banner}
\author[1]{G.~Bassompierre}
\author[9]{K.~Benslama}
\author[18]{N.~Besson}
\author[8,9]{I.~Bird}
\author[2]{B.~Blumenfeld}
\author[13]{F.~Bobisut}
\author[18]{J.~Bouchez \thanksref{Deceased}}
\author[20]{S.~Boyd\thanksref{Now1}}
\thanks[Now1]{Now at University of Warwick, UK}
\author[3,24]{A.~Bueno}
\author[6]{S.~Bunyatov}
\author[8]{L.~Camilleri}
\author[10]{A.~Cardini}
\author[15]{P.W.~Cattaneo}
\author[16]{V.~Cavasinni}
\author[8,22]{A.~Cervera-Villanueva}
\author[11]{R.~Challis}
\author[6]{A.~Chukanov}
\author[13]{G.~Collazuol}
\author[8,21]{G.~Conforto \thanksref{Deceased}}
\thanks[Deceased]{Deceased}
\author[15]{C.~Conta}
\author[13]{M.~Contalbrigo}
\author[10]{R.~Cousins}
\author[9]{H.~Degaudenzi}
\author[8,16]{A.~De~Santo}
\author[16]{T.~Del~Prete}
\author[8]{L.~Di~Lella \thanksref{Now2}}
\thanks[Now2]{Now at Univ. of Pisa and INFN, Pisa, Italy}
\author[8]{E.~do~Couto~e~Silva}
\author[14]{J.~Dumarchez}
\author[19]{H.~Duyang}
\author[20]{M.~Ellis}
\author[3]{G.J.~Feldman}
\author[15]{R.~Ferrari}
\author[8]{D.~Ferr\`ere}
\author[16]{V.~Flaminio}
\author[15]{M.~Fraternali}
\author[1]{J.-M.~Gaillard}
\author[8,14]{E.~Gangler}
\author[5,8]{A.~Geiser}
\author[5]{D.~Geppert}
\author[13]{D.~Gibin}
\author[8,12]{S.~Gninenko}
\author[19]{A.~Godley}
\author[8,22]{J.-J.~Gomez-Cadenas}
\author[18]{J.~Gosset}
\author[5]{C.~G\"o\ss ling}
\author[1]{M.~Gouan\`ere}
\author[8]{A.~Grant \thanksref{Deceased}}
\author[7]{G.~Graziani}
\author[13]{A.~Guglielmi}
\author[18]{C.~Hagner}
\author[22]{J.~Hernando}
\author[3]{P.~Hurst}
\author[11]{N.~Hyett}
\author[7]{E.~Iacopini}
\author[9]{C.~Joseph}
\author[9]{F.~Juget}
\author[11]{N.~Kent}
\author[6]{O.~Klimov}
\author[8]{J.~Kokkonen}
\author[12,15]{A.~Kovzelev}
\author[1,6]{A. Krasnoperov}
\author[19]{J.J.~Kim}
\author[12]{M.~Kirsanov}
\author[12]{S.~Kulagin}
\author[19]{C.T.~Kullenberg}
\author[13]{S.~Lacaprara}
\author[14]{C.~Lachaud}
\author[23]{B.~Laki\'{c}}
\author[15]{A.~Lanza}
\author[4]{L.~La Rotonda}
\author[13]{M.~Laveder}
\author[14]{A.~Letessier-Selvon}
\author[14]{J.-M.~Levy}
\author[19]{J.~Libo} 
\author[8]{L.~Linssen}
\author[23]{A.~Ljubi\v{c}i\'{c}}
\author[2]{J.~Long}
\author[7]{A.~Lupi}
\author[6]{V.~Lyubushkin}
\author[7]{A.~Marchionni}
\author[21]{F.~Martelli}\
\author[18]{X.~M\'echain}
\author[1]{J.-P.~Mendiburu}
\author[18]{J.-P.~Meyer}
\author[13]{M.~Mezzetto}
\author[19]{S.R.~Mishra}
\author[11]{G.F.~Moorhead}
\author[6]{D.~Naumov}
\author[1]{P.~N\'ed\'elec}
\author[6]{Yu.~Nefedov}
\author[9]{C.~Nguyen-Mau}
\author[17]{D.~Orestano}
\author[17]{F.~Pastore \thanksref{Deceased}}
\author[20]{L.S.~Peak}
\author[21]{E.~Pennacchio}
\author[1]{H.~Pessard}
\author[8]{A.~Placci}
\author[15]{G.~Polesello}
\author[5]{D.~Pollmann}
\author[12]{A.~Polyarush}
\author[11]{C.~Poulsen}
\author[6,14]{B.~Popov}
\author[13]{L.~Rebuffi}
\author[24]{J.~Rico}
\author[5]{P.~Riemann}
\author[8,16]{C.~Roda}
\author[8,24]{A.~Rubbia}
\author[15]{F.~Salvatore}
\author[14]{K.~Schahmaneche}
\author[5,8]{B.~Schmidt}
\author[5]{T.~Schmidt}
\author[13]{A.~Sconza}
\author[19]{A.M.~Scott}
\author[11]{M.~Sevior}
\author[1]{D.~Sillou}
\author[8,20]{F.J.P.~Soler}
\author[9]{G.~Sozzi}
\author[2,9]{D.~Steele}
\author[8]{U.~Stiegler}
\author[23]{M.~Stip\v{c}evi\'{c}}
\author[18]{Th.~Stolarczyk}
\author[9]{M.~Tareb-Reyes}
\author[11]{G.N.~Taylor}
\author[6]{V.~Tereshchenko}
\author[19]{X.C.~Tian}
\author[12]{A.~Toropin}
\author[14]{A.-M.~Touchard}
\author[8,11]{S.N.~Tovey \thanksref{Deceased}}
\author[9]{M.-T.~Tran\thanksref{Now6}}
\thanks[Now6]{Now at Ecole Polytechnique Federale de Lausanne, Lausanne, Switzerland}
\author[8]{E.~Tsesmelis}
\author[20]{J.~Ulrichs}
\author[9]{L.~Vacavant}
\author[4]{M.~Valdata-Nappi\thanksref{Now4}}
\thanks[Now4]{Now at Univ. of Perugia and INFN, Perugia, Italy}
\author[6,10]{V.~Valuev}
\author[14]{F.~Vannucci}
\author[20]{K.E.~Varvell}
\author[21]{M.~Veltri}
\author[15]{V.~Vercesi}
\author[8]{G.~Vidal-Sitjes}
\author[9]{J.-M.~Vieira}
\author[10]{T.~Vinogradova}
\author[3,8]{F.V.~Weber}
\author[5]{T.~Weisse}
\author[8]{F.F.~Wilson}
\author[11]{L.J.~Winton}
\author[19]{Q.~Wu\thanksref{Now5}}
\thanks[Now5]{Now at Illinois Institute of Technology, USA}
\author[20]{B.D.~Yabsley}
\author[18]{H.~Zaccone \thanksref{Deceased}}
\author[5]{K.~Zuber}
\author[13]{P.~Zuccon}

\address[1]{LAPP, Annecy, France}
\address[2]{Johns Hopkins Univ., Baltimore, MD, USA}
\address[3]{Harvard Univ., Cambridge, MA, USA}
\address[4]{Univ. of Calabria and INFN, Cosenza, Italy}
\address[5]{Dortmund Univ., Dortmund, Germany}
\address[6]{JINR, Dubna, Russia}
\address[7]{Univ. of Florence and INFN,  Florence, Italy}
\address[8]{CERN, Geneva, Switzerland}
\address[9]{University of Lausanne, Lausanne, Switzerland}
\address[10]{UCLA, Los Angeles, CA, USA}
\address[11]{University of Melbourne, Melbourne, Australia}
\address[12]{Inst. for Nuclear Research, INR Moscow, Russia}
\address[13]{Univ. of Padova and INFN, Padova, Italy}
\address[14]{LPNHE, Univ. of Paris VI and VII, Paris, France}
\address[15]{Univ. of Pavia and INFN, Pavia, Italy}
\address[16]{Univ. of Pisa and INFN, Pisa, Italy}
\address[17]{Roma Tre University and INFN, Rome, Italy}
\address[18]{DAPNIA, CEA Saclay, France}
\address[19]{Univ. of South Carolina, Columbia, SC, USA}
\address[20]{Univ. of Sydney, Sydney, Australia}
\address[21]{Univ. of Urbino, Urbino, and INFN Florence, Italy}
\address[22]{IFIC, Valencia, Spain}
\address[23]{Rudjer Bo\v{s}kovi\'{c} Institute, Zagreb, Croatia}
\address[24]{ETH Z\"urich, Z\"urich, Switzerland}
\address[25]{Inst. for High Energy Physics, 142281, Protvino, Moscow, Russia}

\begin{abstract}

We present our new measurement of the cross-section for charm dimuon production in neutrino-iron interactions
based upon the full statistics collected by the NOMAD experiment. After background subtraction
we observe 15,344 charm dimuon events, providing the largest sample currently available.
The analysis exploits the large inclusive charged current sample - about $9\times 10^6$ events after
all analysis cuts - and the high resolution NOMAD detector to constrain the total systematic uncertainty 
on the ratio of charm dimuon to inclusive Charged Current (CC) cross-sections to $\sim 2\%$.  
We also perform a fit to the NOMAD data to extract the charm production parameters and the strange quark 
sea content of the nucleon within the NLO QCD approximation. 
We obtain a value of $m_c(m_c)=1.159\pm0.075$~GeV/c$^2$ for the running mass of 
the charm quark in the $\overline{\rm MS}$ 
scheme and a strange quark sea suppression factor of $\kappa_s = 0.591 \pm 0.019$ at $Q^2=20$~GeV$^2$/c$^2$.  

\end{abstract}


\end{frontmatter}

\section{Introduction}
\label{sec:intro}

\begin{figure}[htb]
 \begin{center}
  \mbox{\epsfig{file=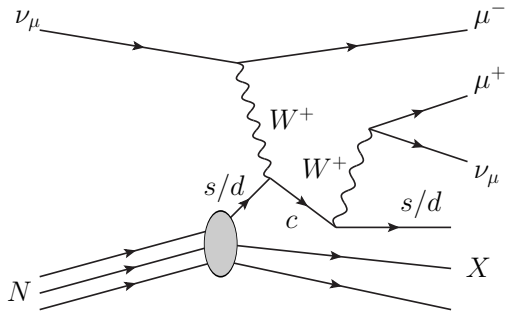,width=0.5\linewidth}}
  \caption {\it Feynman diagram of a $\nu_\mu$ induced Charged Current charm dimuon event.}
  \label{fig:dimuonevt}
 \end{center}
\end{figure}

The process of charm dimuon production stems from the $\nu_\mu$ charged-current (CC) production of a charm quark, 
which semileptonically decays into a final state secondary muon with its electric charge opposite to 
that of the muon from the leptonic CC vertex. Figure~\ref{fig:dimuonevt} shows a schematic Feynman diagram of 
the process.  
The Deep Inelastic Scattering (DIS) production of charm quarks involves the scattering off both 
the strange and the non-strange quark content of the nucleon. 
However, the contributions from $u$- and $d$-quarks are suppressed 
by the small quark-mixing Cabibbo-Kobayashi-Maskawa (CKM) matrix elements. 

A measurement of the cross-section for charm dimuon production in neutrino DIS off nucleons 
provides the most direct and clean probe of $s$, the strange quark sea content of the nucleon. 
Inclusive cross sections are indeed not very sensitive to the strange quark sea content, 
since, in this case, the complementary 
contributions from strange and non-strange distributions are strongly anti-correlated.   
In addition, the flavor selection offered by neutrino and anti-neutrino interactions allows a determination of 
both $s$ and $\bar{s}$ separately. 
The strange quark contribution is particularly important at small values of the parton momentum fractions $x$, 
where the quark distributions are dominated by the sea. This kinematic region is crucial in high-energy 
hadron collisions for the study of many processes and therefore an accurate determination of the strange 
sea is required for the interpretation of experimental data.
The knowledge of the strange quark sea content of the nucleon and of charm production is also the dominant 
contribution to the systematic uncertainties in the electroweak 
measurements from (anti)neutrino DIS interactions~\cite{nutev-sin2w}.  

Traditionally, charm dimuon production in (anti)neutrino interactions has always been measured 
in massive calorimeters, mainly composed of passive heavy materials (iron, marble, lead etc.) in order 
to obtain a sizeable number of events~\cite{Abramowicz:1982zr,Vilain:1998uw,Astier:2000us,Bazarko:1994tt,Goncharov:2001qe,nutev:ssbar07,:2008xp}. 
This detection technique, while relatively simple and efficient, is ultimately limited by the coarse 
resolution of the detectors and by the resulting systematic uncertainties.    

The high resolution NOMAD data allow a new level of precision in the measurement of charm dimuon 
production in neutrino interactions.  
The use of a low density target located inside a magnetic spectrometer measures the momentum and 
emission angle of all secondary particles produced in neutrino interactions. 
Additionally, the detector offers the capability of making redundant {\em in situ} measurements of 
all relevant backgrounds, minimizing the reliance on Monte Carlo simulation  (MC) 
of the hadronization process. 
For the first time a high resolution reconstruction of neutrino interactions can be achieved from  
high statistics samples, typical of the most massive calorimeters but without the corresponding limitations. 
Furthermore, the neutrino spectrum in NOMAD is well suited to study charm production close to the charm threshold, 
providing enhanced sensitivity to the charm production parameters.   

This paper is organized as follows. Section~\ref{sec:charmprod} gives a short overview of the 
charm dimuon production in neutrino charged current interactions. In Section~\ref{sec-nomad} we describe the 
neutrino beam and the NOMAD detector used for our measurement. Section~\ref{sec:selection} explains our 
event selection and the corresponding cuts. In Section~\ref{sec:analysis} and Section~\ref{sec:unfolding} we 
describe the analysis scheme used for our precision measurement and the unfolding procedure needed to   
extract the final cross-sections, respectively. We present our final NOMAD results in Section~\ref{sec:results} 
and provide a detailed discussion of systematic uncertainties in Section~\ref{sec:syst}. 
Section~\ref{sec:charmfit} discusses the extraction of charm production parameters and of the 
strange quark sea content of the nucleon from NOMAD data.  
Finally, Section~\ref{sec:summary} gives a summary of the main results achieved.

\section{Charm dimuon production in neutrino interactions}
\label{sec:charmprod}

The differential cross section for charm quark production in CC neutrino DIS 
off nucleon or nuclear target can be written as: 
\begin{eqnarray}
\frac{d\sigma_c^\nu}  
{dxdy}
=\frac{G_F^2ME}
{\pi(1+Q^2/M_W^2)^2}\left[\left(1-y-\frac{Mxy}{2E}\right)
F^\nu_{2,c} (x,Q^2)
+\right. \nonumber\\
\left. +\frac{y^2}{2}
F^\nu_{T,c} (x,Q^2) +  
y\left(1-\frac{y}{2}\right)
xF^\nu_{3,c}(x,Q^2)
\right],
\label{eqn:cs}
\end{eqnarray}
where $x$, $y$, and $Q^2$ are common DIS variables, $E$ 
is the neutrino energy, $G_F$ is the Fermi constant,  
$M$ and $M_W$ are the nucleon and $W$-boson masses, respectively,  
and $F^\nu_{2,T,3}$ are the corresponding structure functions (SFs).
For an isoscalar target, 
assuming the usual isospin relations between the proton and
neutron quark distributions, we have in the LO QCD approximation~\footnote{\it We give the 
Leading Order (LO) approximation for illustration purpose only. The entire analysis described in this paper is 
performed in the Next to Leading Order (NLO) or Next to Next to Leading Order (NNLO) approximation, 
including both the cross-section calculation and the acceptance corrections.}:  
\begin{eqnarray}
F_{\rm 2,c}^{\nu N} (x,Q^2)
=2\xi \left[\left \vert V_{cs} \right \vert ^2
s(\xi,Q^2) 
+ \left \vert V_{cd} \right \vert ^2
\frac{u(\xi,Q^2) + d(\xi,Q^2)}{2} \right],
\nonumber \\
F_{\rm T,c}^{\nu N} = xF_{\rm 3,c}^{\nu N} =\frac{x}{\xi}F_{\rm 2,c}^{\nu N}, 
~~~~~~~~~~~~~~~~~~~~~~~~
\label{eqn:sf}
\end{eqnarray}
where $u,d,s$ are the light quark distributions in the proton,
$\xi=x(1+m_{\rm c}^2/Q^2)$ is the slow-rescaling variable appearing  
in the kinematics of $2\to2$ parton scattering 
with one massive particle in the final state~\cite{Barnett:1976ak},  
and $m_{\rm c}$ is the charm quark mass. The values of the CKM 
matrix elements
$V_{cs}= 0.97334$ and $V_{cd}=0.2256$~\cite{PDG12} suggest that
the strange quark contribution dominates 
the cross section of Eq.~(\ref{eqn:cs}) at small $x$. 
In the NLO QCD approximation 
the structure functions of Eq.~(\ref{eqn:sf}) get an additional $O(\alpha_{\rm s})$ 
contribution from the gluon-radiation and gluon-initiated 
processes~\cite{Gottschalk:1980rv}.

Due to their larger mass, the sea strange quarks are suppressed with respect to the 
$u$ and $d$ sea quarks. The strange sea suppression factor: 
\begin{equation} 
\label{eq:kappas} 
\kappa_s(Q^2)=\frac{\int_0^1 x\left[ s(x,Q^2)+\overline{s}(x,Q^2)\right]dx}
{\int_0^1 x\left[ \overline{u}(x,Q^2)+\overline{d}(x,Q^2)\right] dx},
\end{equation} 
increases with the momentum transfer $Q^2$.  

Experimental data on neutrino induced DIS charm production is typically collected 
on heavy targets (e.g. iron), rather than on free nucleons. Nuclear corrections 
must then be applied to the structure functions entering the $\sigma_c$ cross-section 
in order to take into account the effect of the nuclear medium. 
In general, nuclear corrections in neutrino interactions are different from those for 
charged-lepton interactions~\cite{Kulagin:2007ju} and can be described to a good accuracy 
by the model of Ref.\cite{Kulagin:2007ju,Kulagin:2004ie,Kulagin:2010gd}.

In the LO the dimuon cross section 
is related to the corresponding cross section for charmed-quark production as:  
\begin{equation}
\frac{d\sigma_{\mu\mu}}{dxdydz} =   
\frac{d\sigma_{\rm c}}{dxdy} 
\sum_{\rm h} f_{\rm h} D_{\rm c}^h(z) Br(h\to \mu X),   
\label{eqn:frag}
\end{equation}
where $f_{\rm h}$ is the production fraction of the charmed hadron $h$,   
$D_{\rm c}^h(z)$ is the fragmentation function of the charm quark 
into a given charmed hadron $h=D^0,D^+,D_s^+,\Lambda_{\rm c}^+$
carrying a fraction $z$ of the charm quark momentum, 
and $Br(h\to \mu X)$ is the corresponding inclusive branching ratio 
for the muon decays~\footnote{\it The normalization is defined such that $\sum f_{\rm h} =1$.}. 
Here $\Lambda_{\rm c}$ refers to a generic charmed baryon.  
At NLO the coefficient functions entering the SFs 
calculation depend, in general, on $z$ as well.    
The charm fragmentation function $D_{\rm c}(z)$ defines the energy of the 
outgoing charmed hadron and, in turn, of the secondary muon    
produced in the semileptonic decay. 
Assuming a universal $D_{\rm c}(z)$ for all charmed hadrons and integrating over $z$,    
Eq.~(\ref{eqn:frag}) becomes $d\sigma_{\mu\mu}/dxdy = D_c(z) B_\mu d\sigma_{\rm c} / dxdy$,   
where $B_\mu=\sum f_{\rm h} Br(h\to\mu X)$ is the effective 
semileptonic branching ratio, and the universal fragmentation function can be described 
by the Collins-Spiller parameterization~\cite{Collins:1984ms}: 
\begin{equation} 
\label{eq:frag}
D_{\rm c} (z) = \left[ \frac{1-z}{z} - \varepsilon_{\rm c} \frac{2-z}{1-z} \right] (1+z)^2 \left[ 1 - \frac{1}{z} - \frac{\varepsilon_{\rm c}}{1-z} \right]^{-2}
\end{equation}  
with $\varepsilon_{\rm c}$ as a free parameter. This parameterization has a more accurate asymptotic behavior in the
limit of $z\to 1$ than the Peterson form~\cite{Peterson:1982ak}.
The charmed fractions $f_{\rm h}$ depend on the incoming neutrino energy.
This fact can be explained by the contributions from quasi-elastic $\Lambda_c$ 
and diffractive $D_s$ production.
Measurements of $f_{\rm h}$ and $B_\mu$ in neutrino interactions were 
performed by the E531~\cite{Ushida:1988rt,Bolton:1997pq}
and CHORUS~\cite{KayisTopaksu:2005je,KayisTopaksu:2011} experiments using the emulsion
detection technique, which allowed them to reconstruct exclusive final states.

\section{Beam and Detector}
\label{sec-nomad}

The Neutrino Oscillation MAgnetic 
Detector (NOMAD) experiment at CERN used a 
neutrino beam  produced by 
the 450~GeV protons from the 
Super Proton Synchrotron (SPS) incident on
 a beryllium target and producing  
secondary $\pi^{\pm}$, $K^{\pm}$, and $K^0$ mesons. 
The positively charged mesons were focussed by 
two magnetic horns into a 290~m long evacuated decay pipe. Decays of  
$\pi^{\pm}$,  $K^{\pm}$, and $K^0_L$  
produced the SPS neutrino beam. 
The average neutrino flight path to  NOMAD was 628~m, the detector being 
836~m downstream of the Be-target.  
The SPS beamline  and the neutrino flux incident 
at NOMAD are described in~\cite{Astier:2003rj}.  
The $\nu$-flux in NOMAD can be estimated from the 
$\pi^{\pm}$ and $K^{\pm}$ production measurements in 
proton-Be collision by the SPY experiment 
~\cite{SPY1,SPY2,SPY3} and by an  earlier measurement conducted by 
Atherton {\it et al.}~\cite{ATHERTON}. The  $E_\nu$-integrated relative composition of 
$\nu_\mu : \bar{\nu}_{\mu} : \nu_e :\bar{\nu}_e$ CC events, constrained {\it in situ} by the 
measurement of CC-interactions of each of the  neutrino species,  is 
$1.00: 0.025: 0.015:0.0015$. Thus,  97.5\% of the events    
are induced by neutrinos with a small  anti-neutrino contamination. 

The NOMAD apparatus, described in~\cite{NOMAD-NIM},  
consisted of several sub-detectors. The active 
target comprised 132 planes of $3 \times 3$~m$^2$ drift chambers~\cite{DCH} (DCH)    
with an average density similar to that of  liquid 
hydrogen (0.1~gm/cm$^3$). 
On average, the equivalent material in the DCH 
encountered by 
particles produced in a $\nu$-interaction  
was about half of a radiation  length 
and a quarter of a hadronic interaction length ($\lambda$).  
The fiducial mass of the NOMAD DCH-target,  2.7 tons, was  
composed  primarily of carbon (64\%), oxygen (22\%), nitrogen (6\%), 
and hydrogen (5\%) yielding an effective atomic number, 
A=12.8, similar to carbon.
Downstream of the DCH, there were nine modules of transition radiation 
detectors~\cite{TRD} (TRD), followed by a preshower (PRS) and a lead-glass 
electromagnetic calorimeter~\cite{ECAL} (ECAL). 
The ensemble of DCH, TRD, and PRS/ECAL was placed within 
a dipole magnet providing a 0.4~T magnetic field orthogonal 
to the neutrino beam line. 
Downstream  of the magnet was a hadron calorimeter, 
followed by two muon stations each comprising large area 
drift chambers (MCH) and separated by an iron filter. 
The two stations, placed at 8- and 13-$\lambda$'s downstream of 
the ECAL, provided a clean identification of the muons. 
The charged tracks in the DCH were measured with an 
approximate  momentum ($p$)  resolution of  
${\rm {\sigma(p)/p = 0.05/\sqrt{L} \oplus 0.008p/\sqrt{L^5} }}$  
($p$ in GeV/$c$ and $L$ in meters) 
with unambiguous charge separation in the energy range of interest. 
The energy deposition from $e$ and $\gamma$ were measured in ECAL with an 
energy resolution of ${\rm {\sigma(E)/E = 1.04\% + 3.22\%/\sqrt{E} }}$ 
($E$ in GeV). 
The experiment recorded over 1.7 million 
neutrino interactions in the active drift-chamber target in 
the range ${\mathcal {O}}(1) \leq E_\nu \leq 300$~GeV.

The detector was suspended from iron pillars (the ‘I’s) at the two ends of the magnet. 
The front pillar was instrumented with scintillators to provide an additional massive active 
target for neutrino interactions, the front calorimeter (FCAL). 
The FCAL consisted of 23 iron plates 4.9 cm thick and separated by 1.8 cm gaps.
Twenty out of the 22 gaps were instrumented with long scintillators, which were read out on both
ends by 3 in photomultipliers. The dimensions of the scintillators were
$175 \times 18.5 \times 0.6$ cm$^3$. To achieve optimal light collection and a reasonable
number of electronic channels five consecutive scintillators along the beam axis were 
ganged together by means of twisted light guides to form a module. Ten such modules were 
placed above each other to form a stack. Four stacks were placed along the beam axis.
The area of the FCAL "seen" by the neutrino beam was $175 \times 190$ cm$^2$. The
detector had a depth of about five nuclear interaction lengths and a total fiducial 
mass of about 17.7 tons, composed mainly of iron. The energy deposition in FCAL 
was measured with a resolution ${\rm {\sigma(E)/E = 104\%/\sqrt{E} }}$ ($E$ in GeV). 
The experiment recorded over 18 million neutrino interactions in the FCAL  
in the range ${\mathcal {O}}(1) \leq E_\nu \leq 300$~GeV.  

In front of the FCAL there was a plane of scintillators, $V_8$, used to veto charged particles 
entering the detector. Before and after the TRD there were two planes of scintillators, 
$T_1$ and $T_2$, whose coincidence provided the main trigger for charged tracks in the 
drift chamber volume. 

\begin{figure}
\begin{center}
\includegraphics[width=0.9\textwidth,angle=0]
{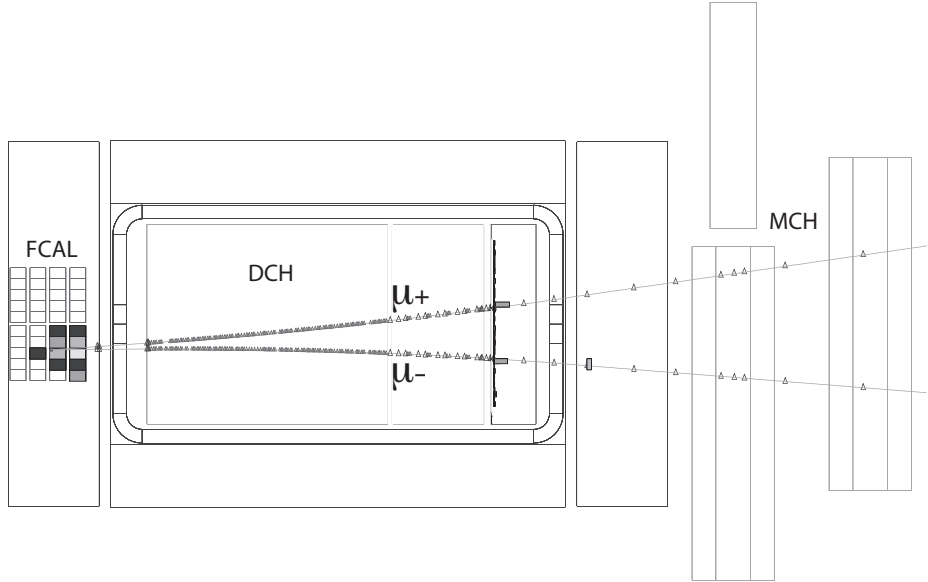}
\caption{\it   
Candidate of opposite signed dimuon event occurring in Stack 2 of the FCAL. 
The two tracks are the oppositely charged muons in the event and the detector is shown from the side. 
The grey shading in each FCAL module indicates different levels of energy deposition mostly from the hadronic shower produced in the interaction.}
\label{fig:evtdimuon}
\end{center}
\end{figure}

The schematic of the FCAL and the DCH-tracker with a charm dimuon event candidate is 
shown in Figure~\ref{fig:evtdimuon}.

\section{Event Selection}
\label{sec:selection}

\subsection{Trigger and Calibration} 

The main trigger selection used for the analysis of FCAL data requires an energy deposition in the FCAL of at least
3.15 m.i.p. - m.i.p. is the signal of a minimum ionizing particle crossing one FCAL stack - and no signal in the $V_8$ veto scintillator plane: $\bar{V}_8 \times FCAL$  -  (FCAL trigger). 
This trigger vetoes through-going muons and has a live time of $90\pm3\%$.

A second independent trigger selection with lower threshold is used to measure the
FCAL trigger efficiency from data (FCAL$^\prime$ trigger). For this trigger an energy deposition
of at least 1.2 m.i.p. in the FCAL is requested, in addition to the coincidence with
a signal from the scintillator plane in the TRD region: $\bar{V}_8 \times T_1 \times T_2
\times FCAL^\prime$. The live time is $90\pm3\%$.

We measure the efficiency of the FCAL trigger directly from data in the following way:
\begin{equation}
 \epsilon_{FCAL} = \frac{N^{DATA}_{FCAL \& FCAL^\prime}}{N^{DATA}_{FCAL^\prime}}
\end{equation}
We determine the FCAL trigger efficiency separately for each of the 4 stacks and each of the
4 years of data taking, for a total of 16 configurations. The final trigger efficiency is obtained 
by averaging over the different stacks and years of data taking, weighted by the corresponding numbers 
of $\nu_\mu$ CC events identified in the data. 

In order to take into account the saturation of the readout electronics, 
which resulted in a reduced value of the ADC counts observed in the data at high energies, 
we apply a correction function to the ADC response of each 
individual FCAL module in the Monte Carlo. This correction function  
is obtained by minimizing the $\chi^2$ between the total energy distributions in the 
FCAL for $\nu_\mu$ CC events in data and Monte Carlo. 

The energy deposited in each stack, $F^s$, is calculated as the sum of the energy depositions
(in m.i.p.) of the ten individual FCAL modules (in m.i.p.)in the stack:
\begin{equation}
 F^s = \sum\limits_{i \leq 10} F^s_i, \qquad s=1,2,3,4.
\end{equation}
The relative calibration of individual modules in m.i.p. (ADC to m.i.p) is performed
by using the energy
deposition of high energy muons from SPS secondary beams crossing the FCAL in between neutrino spills.
The muon tracks are reconstructed in the drift chambers and extrapolated back to the FCAL.
The absolute energy deposition in GeV is then obtained by dividing the values of
$F^s$ by the appropriate mip/GeV conversion factor $P_0$:
\begin{equation}
\label{eq:lincalib}
 E_{\rm FCAL} = \frac{1}{P_0} \sum\limits_{s=1}^4 F^s, \mbox{ where }  P_0 = 2.388 \pm 0.006 \mbox{~mip}/\mbox{GeV}
\end{equation}
The constant $P_0$ is obtained from the default MC simulation by comparing the reconstructed
to the simulated energy and is validated by the analysis of single muon events in data.

After determining the response function between the simulated energy and the energy reconstructed in FCAL 
$E_{\rm FCAL}^{\rm sim} = {\mathcal{F}}(E_{\rm FCAL})$, we extract the inverse function 
${\mathcal{F}}^{-1}$ from the MC simulation by comparing separately the simulated and reconstructed 
shower energy for events originated in each stack. The reconstructed shower energy is obtained by 
subtracting the energy loss by the muons from the total energy measured in FCAL.

\subsection{Analysis cuts} 
\label{sec:cuts} 

The analysis of FCAL data proceeds with the selection of two independent samples: a) inclusive 
$\nu_\mu$ CC events with a single negatively charged muon; b) dimuon events. 
All the applied cuts are relatively loose in order to avoid potential biases of the samples and to 
retain a large signal efficiency. In addition, we try to use similar selection cuts (if applicable)  
for both the dimuon and the $\nu_\mu$ CC samples. These criteria minimize systematic uncertainties 
from the event selection procedure. We identify two different dimuon samples: the Opposite Sign Dimuons (OSDM) 
with two muons of opposite charge, and the Like Sign Dimuons (LSDM) with two muons of the same charge 
(background sample). 

The momenta of the identified muons, $p_{\mu_{cc}}$ for the primary muon and $p_{\mu_c}$ for the secondary one, 
are measured with high precision with the DCH located in the low density spectrometer following FCAL. 
We define the primary muon as the one with the largest transverse momentum relative to the beam direction. 
The total hadronic energy of the event, 
$E_{\rm Had}$, is calculated from the sum of the energy of the hadronic shower measured in FCAL and the 
energy of the secondary muon $E_{\mu_c}$ and of any other charged track measured in the drift chambers, 
minus the energy loss of the primary and secondary muons in FCAL.   
The reconstructed neutrino energy is obtained as the sum of the energy of the primary muon and the total 
hadronic energy: $E_\nu=E_{\mu_{cc}}+E_{\rm Had}$. The DIS kinematic variables are reconstructed 
in the following way: 
\begin{eqnarray} \nonumber  
Q^2 & = & 4 E_\nu E_{\mu_{cc}} \sin^2 \theta_{\mu_{cc}} \\ \nonumber  
\nu & = & E_\nu - E_{\mu_{cc}} \\ \nonumber 
x_{Bj} & = & Q^2/(2 M \nu ) \\ \nonumber 
y_{Bj} & = & \nu / E_\nu  \\ \nonumber 
z & = & E_{\mu_{cc}} / \nu  \nonumber 
\end{eqnarray}  
where $\theta_{\mu_{cc}}$ is the angle of the primary muon with respect to the neutrino beam direction 
in the lab frame and $M$ is the mass of the proton. 

The applied cuts are:  
\begin{itemize}
 \item[1] FCAL trigger and good run.  
 \item[2] One negatively charged muon, $\mu_{cc}$, identified by the MCH.
 \item[3*] A second muon, $\mu_c$, identified by the Muon System: either a $\mu^+$ from $c$-quark production/background 
or a $\mu^-$ from background.
 \item[4] Fiducial volume inside one of the 4 FCAL stacks. The coordinates of the 
primary vertex, $x^{PV}_{ext}$ and $y^{PV}_{ext}$, are determined from the extrapolation of the muon(s) to the middle $z$ point of the most 
upstream stack containing energy and are constrained to: $|x^{PV}_{ext}|<80$~cm and $|y^{PV}_{ext}|<90$~cm.

 \item[5*] Time difference between the two muons, as measured from the first hit in the drift chambers, 
less than 5ns to reject accidental backgrounds. 
 \item[6*] Require a leading negative muon as the one with the largest transverse momentum relative to the beam direction, $p^T_{\mu_{cc}} > p^T_{\mu_{c}}$. This cut rejects the background from charm dimuon events originated by  
the small anti-neutrino contamination of the beam. We require the presence of only one negative muon 
in the event for the $\nu_\mu$ CC selection.  
 \item[7*] Energy of the hadron shower without the energy of the muon from charm decay ($E_{\rm Had} - E_{\mu_c}$) less 
than $100$~GeV and $E_{\nu} < 300$~GeV. 
 \item[8] $x_{b_j} < 1$. 
 \item[9] Energy of the primary muon $\mu_{cc}$ more than $3$~GeV. 
We note that $3$~GeV is practically the minimal energy 
for muons to reach the MCH and be identified as muons in the NOMAD detector.
 \item[10] Energy of the secondary muon more than $3$~GeV or $E_{\rm Had} > 3$~GeV for CC.
 \item[11] Four-momentum transfer $Q^2 > 1$~GeV$^2/$c$^2$.
\end{itemize}
where the cuts marked with * refer mainly to the dimuon sample.
The main goal of the selection is to ensure that the events are well measured in FCAL.
We limit our analysis to the region $Q^2 > 1$~GeV$^2/$c$^2$ in which we can realiably
calculate the cross-sections within the parton model. It must be also noted that the
impact of the $Q^2$ cut on the charm sample is negligible, due to the intrinsic production threshold.
Tables~\ref{tab:evtnumucc} summarizes the effect of the $\nu_\mu$ CC selection. 
A subsample ($\sim 17\%$) of the data analyzed for this 
paper was the subject of an earlier NOMAD publication~\cite{Astier:2000us}.   

We normalize the number of $\nu_\mu$ CC events in the MC to the ones observed in the data after the 
fiducial volume and leading muon cuts (cut 6 in Table~\ref{tab:evtnumucc}). 
The number of charm dimuon events in MC is normalized to the number of inclusive CC events 
multiplied by the cross-section ratio in our model. 

\begin{table}[htb]
\begin{center}
\small
\begin{tabular}{||c|c|c|c|c||}
\hline
\hline
 & \multicolumn{2}{c}{MC} & \multicolumn{1}{|c|}{DATA} & DATA/MC \\
\hline
 Cut              &  Rec.    & Eff.   & Rec.     & \\
\hline\hline  
 $1$  (Trig.)                 & 12143746 & ---    & 12401729 & 1.021 \\
 $2$  ($\mu_{cc}^-$)               & 12126348 & ---    & 12298205 & 1.014 \\
 $4$  (FV)                    & 10639388 & 76.1\% & 10757864 & 1.011 \\
 $6$  (Lead. $\mu^-_{cc}$)         & 10636157 & 76.1\% & 10636157 & 1.000 \\
 $7$  ($E_{had}^{up}$)        & 10582711 & 75.7\% & 10576596 & 0.999 \\
 $8$  ($x_{B_j}$)             & 10359121 & 74.1\% & 10381255 & 1.002 \\
 $9$ ($E_{\mu_{cc}}$)        & 10354170 & 74.1\% & 10376815 & 1.002 \\
 $10$ ($E_{\mu_c,had}^{low}$) &  9730058 & 69.6\% &  9615738 & 0.988 \\
 $11$ ($Q^2$)                 &  9175383 & 65.8\% &  8759065 & 0.954 \\
\hline
\hline
\end{tabular}
\normalsize
\caption {\it 
Event selection for $\nu_{\mu}$ CC events in data and MC. The top row
shows the raw number of MC events generated in the fiducial volume and used for the normalization
of the efficiency. All the other MC numbers have been normalized to data after the fiducial
volume and leading muon cuts (cut 6). The ratio of data and normalized MC is also given in the
last column.}
\label{tab:evtnumucc}
\end{center}
\end{table}

\section{Analysis scheme} 
\label{sec:analysis} 

The analysis measures the ratio of the charm dimuon cross-section to the inclusive CC cross-section,
as a function of the kinematic variables:
\begin{equation}
{\mathcal{R}}_{\mu \mu} (x) \equiv \frac{\sigma_{\mu \mu}}{\sigma_{cc}} \simeq \frac{N_{\mu \mu}(x)}{N_{cc} (x)}, 
\end{equation}
where $x = E_{\nu}, x_{B_j}, \sqrt{\hat{s}}$ and the partonic center of mass energy squared 
is defined as $\hat{s}=Q^2(1/x_{Bj}-1)$. 
The ratio ${\mathcal{R}}_{\mu \mu}$ provides a large cancellation of all systematic uncertainties
affecting both the numerator and the denominator.

\begin{figure}[htb]
   \hspace*{-0.90cm}\mbox{\epsfig{file=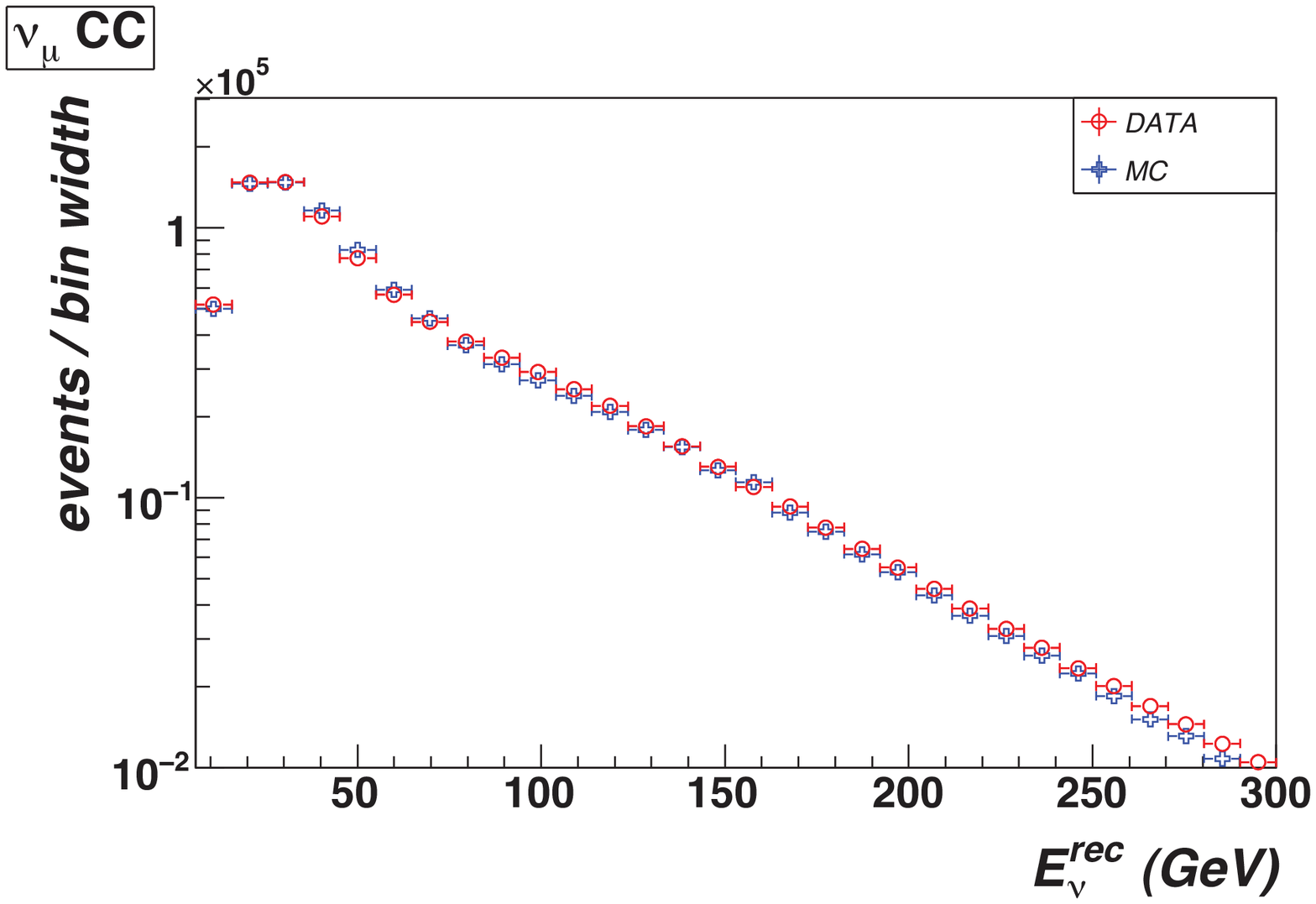,width=0.55\linewidth}\hspace*{-0.20cm}\epsfig{file=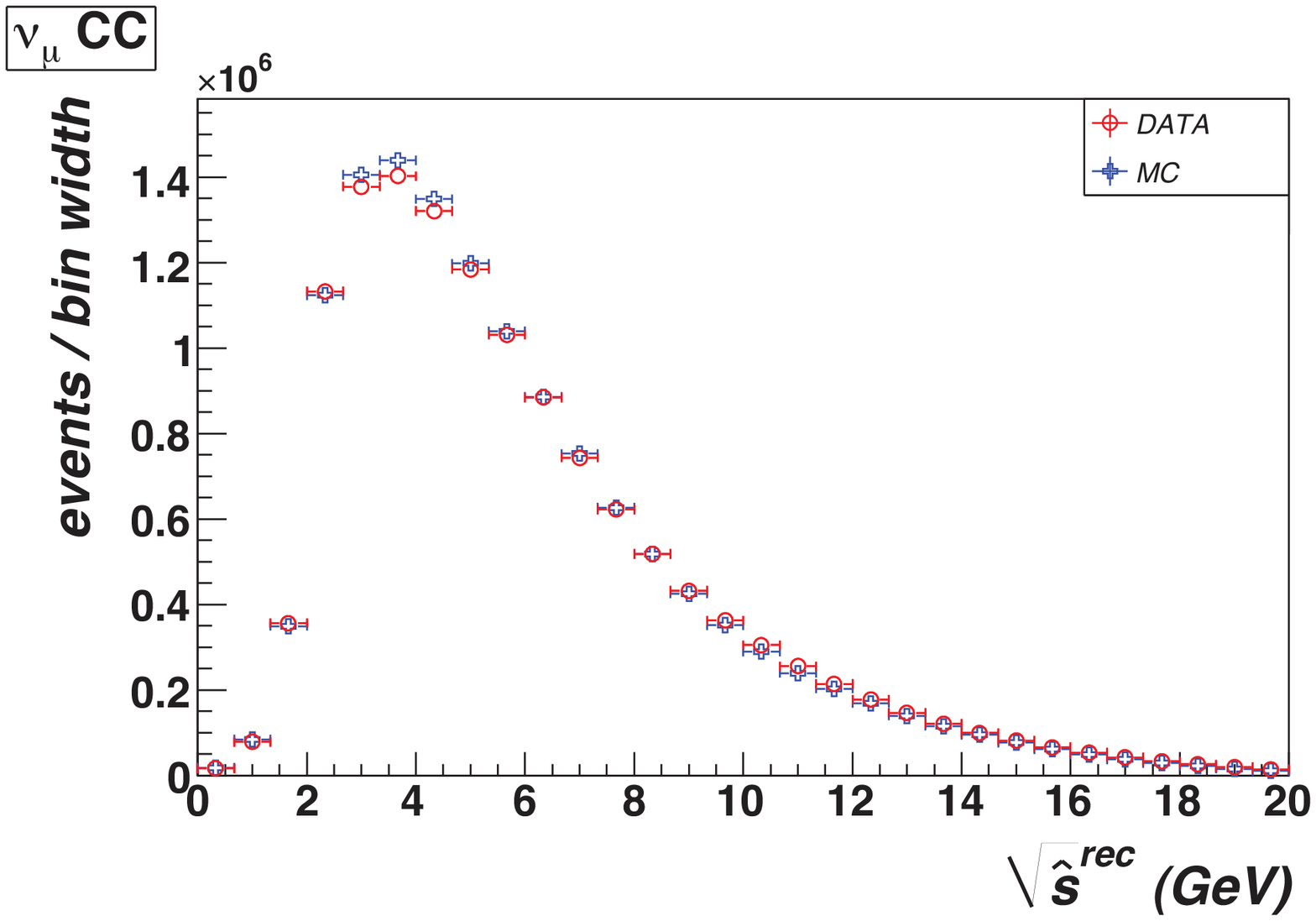,width=0.55\linewidth}}   
   \hspace*{-0.90cm}\mbox{\epsfig{file=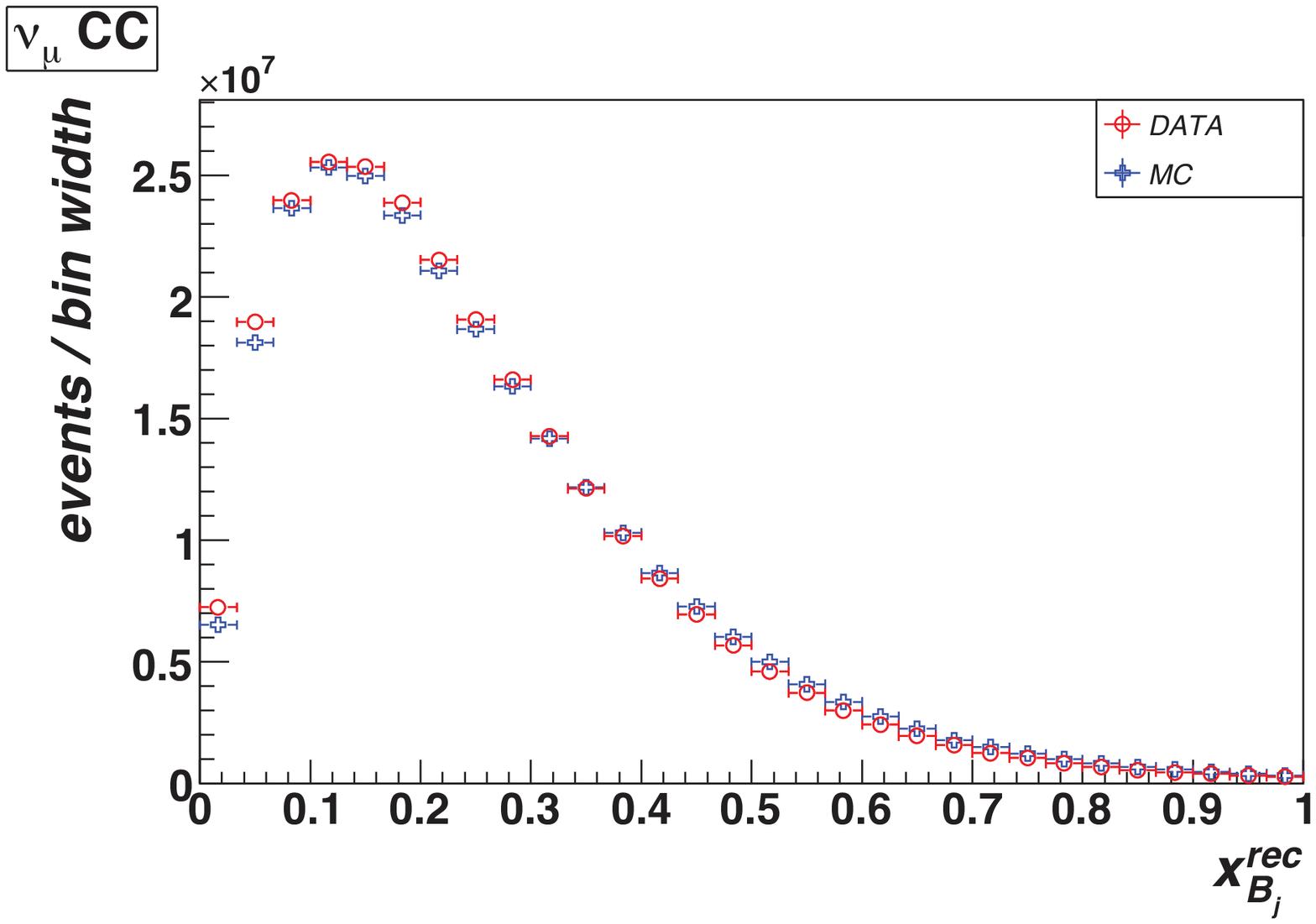,width=0.55\linewidth}\hspace*{-0.20cm}\epsfig{file=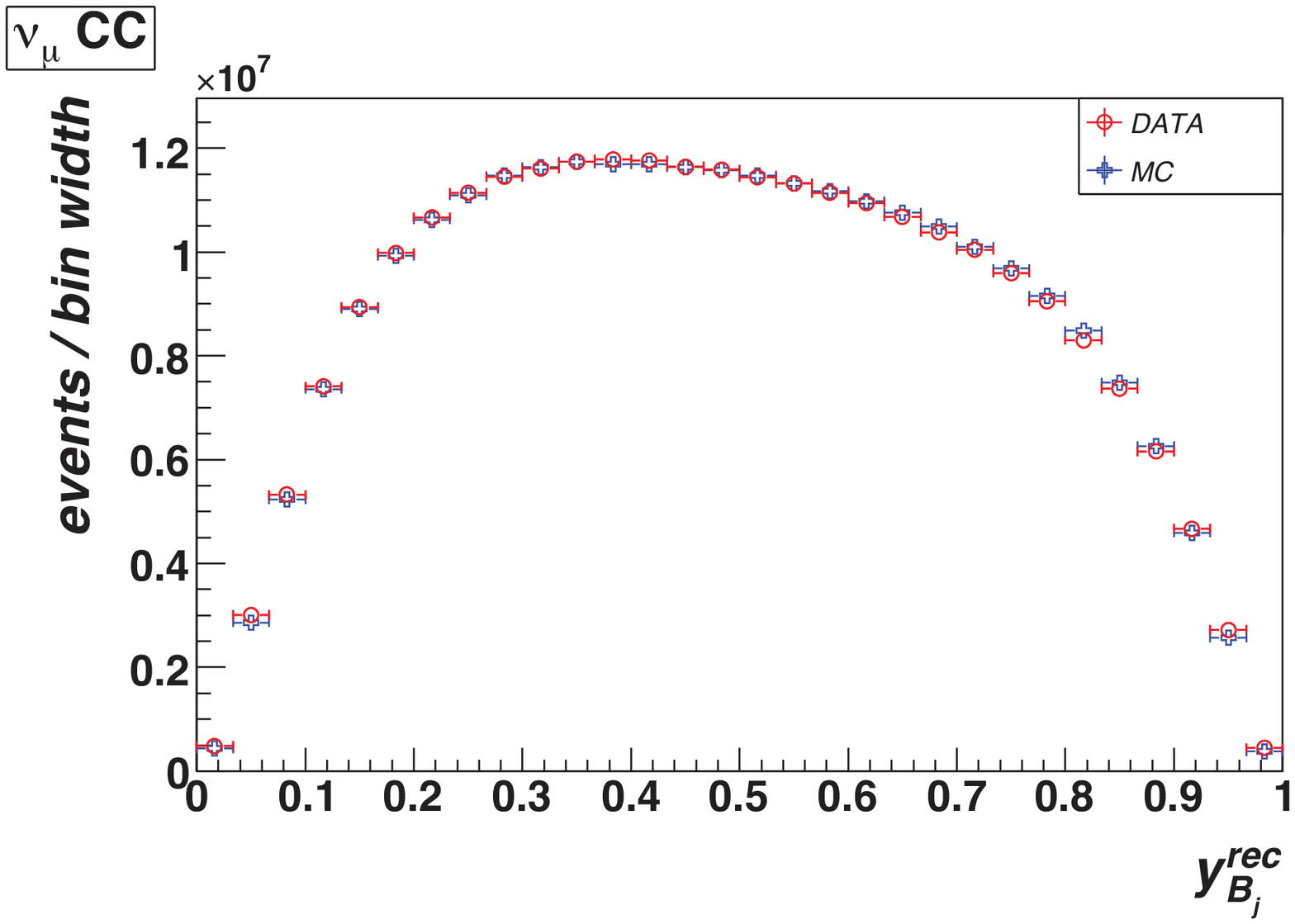,width=0.55\linewidth}}
 \begin{center}
  \caption {\it Distributions of reconstructed kinematic variables in $\nu_{\mu}$ CC events:
visible neutrino energy (top left), center of mass energy $\sqrt{\hat{s}}$ (top right), $x$-Bjorken (bottom left), $y$-B
jorken (bottom right).
Data are shown as circles while MC points are shown as crosses.}
  \label{fig:numucc_rec}
 \end{center}
\end{figure}

The $\nu_{\mu}$ CC events in the data are well reconstructed and have a negligible background.
Figure~\ref{fig:numucc_rec} shows good agreement between data and Monte Carlo for the reconstructed 
kinematic variables in $\nu_\mu$ CC events.   

The charm dimuon events are determined from the opposite sign dimuons measured in
the data after subtracting the background: 
\begin{equation}
 N^{^{\mbox{\tiny DATA}}}_{\mbox{\tiny $\mu\mu_c$}} = N^{^{\mbox{\tiny DATA}}}_{\mbox{\tiny $\mu\mu^+$}} - N^{^{\mbox{\tiny DATA}}}_{\mbox{\tiny $\mu\mu^+_{bg}$}}
\end{equation}
The background to the opposite sign dimuon process arises from muonic decays of hadrons - 
primarily $\pi^+$ and $K^+$ mesons - produced in the hadronic shower or by hadrons which punch through 
to the MCH, thereby being misidentified as muons. The background from charm dimuon events originating  
by the interaction of the small anti-neutrino contamination of the beam is suppressed by the 
requirement of a leading muon with negative charge (cut 7). This selection cut correctly 
identifies the leading muon in about 95\% of the events (Table~\ref{tab:evtdimuon}). 
Since $\bar{\nu}_\mu$ represent only about 5\% of the $\nu_\mu$ flux, the 
final background from anti-neutrino charm dimuons will be at the level of $10^{-3}$.  
Similarly, other background sources like trident production, the overlap of a neutrino and 
an anti-neutrino event, or the production and subsequent muonic decay of $J/\psi$ in 
Neutral Current interactions are found to be negligible. 

\begin{figure}[htb] 
 \begin{center}
   \mbox{\epsfig{file=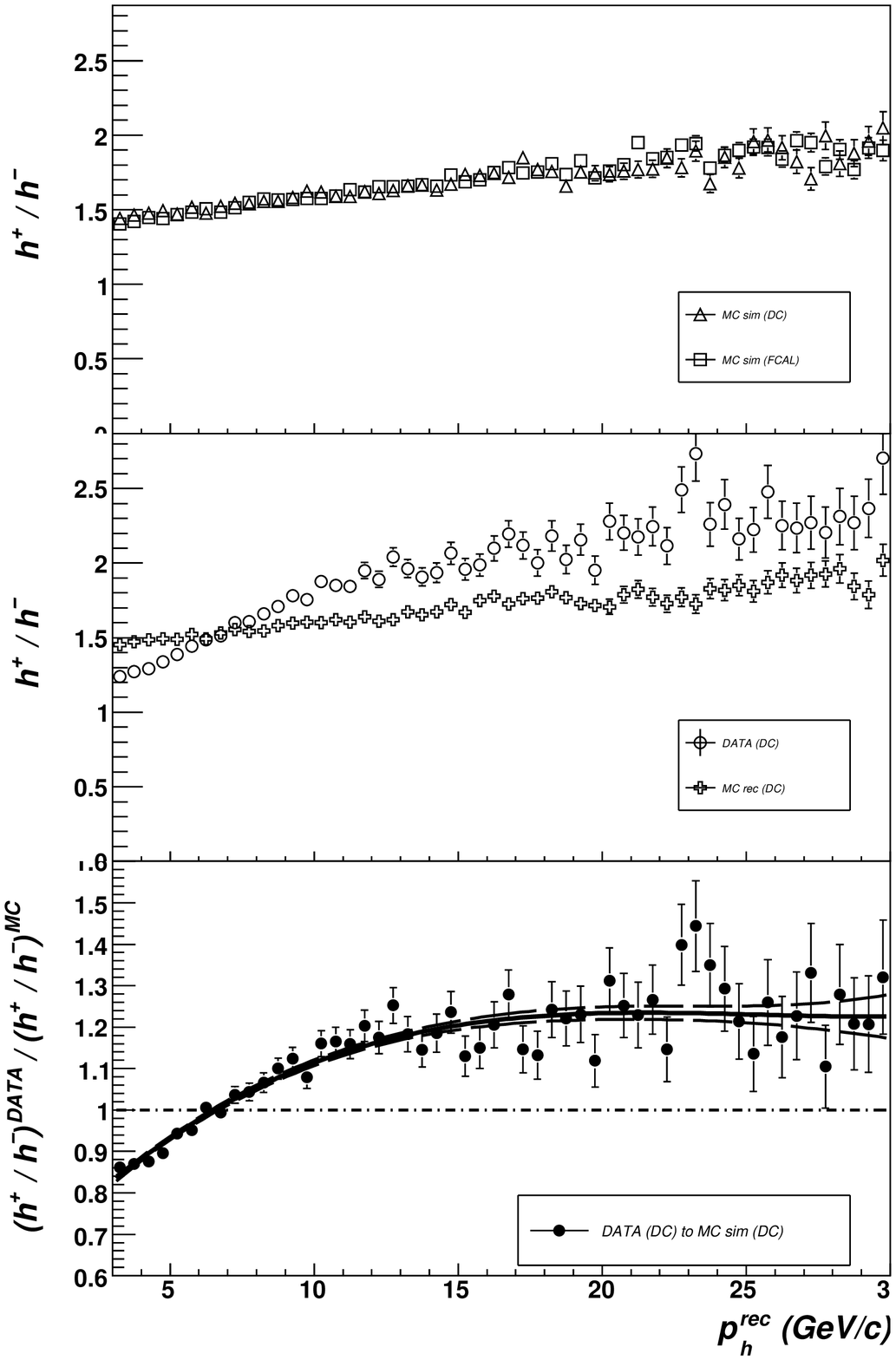,width=0.9\linewidth}} 
  \caption {\it Top plot: ratio between postively charged and negatively charged mesons $h^+/h^-$ 
as a function of momentum in DCH (Carbon target) and FCAL (Iron target) Monte Carlo. 
Middle plot: ratio $h^+/h^-$ in DCH data and DCH reconstructed Monte Carlo. 
The bottom plot shows the measured double ratio $\omega_{h^+}$ between DCH data and the corresponding MC, 
with our fit (solid line: central value; dashed lines: $\pm 1 \sigma$ band) used for the re-weighting of the MC.}
  \label{fig:had_dch}
 \end{center}
\end{figure}

\begin{figure}[htb] 
   \hspace*{-0.90cm}\mbox{\epsfig{file=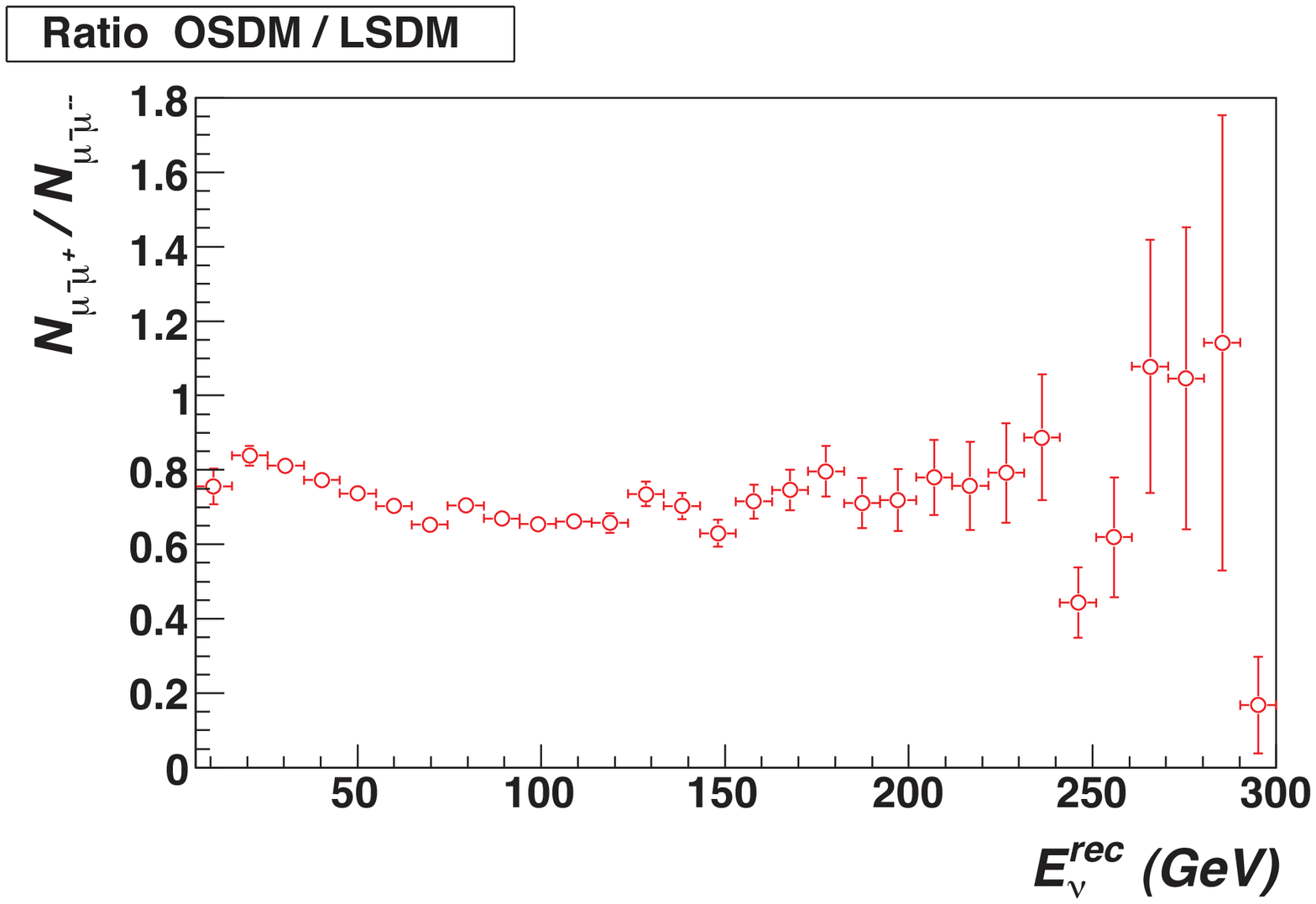,width=0.55\linewidth}\hspace*{-0.20cm}\epsfig{file=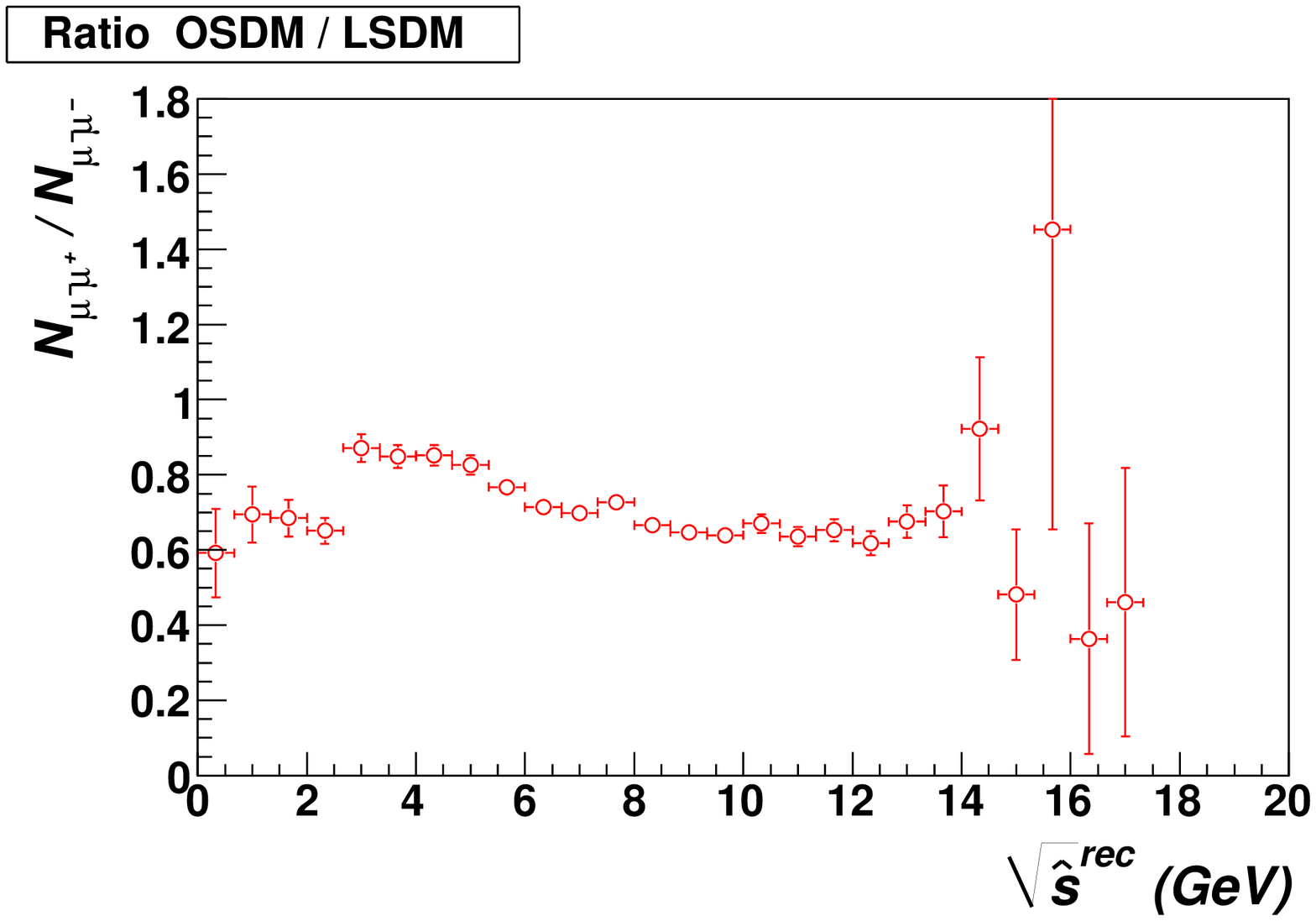,width=0.55\linewidth}}   
   \hspace*{-0.90cm}\mbox{\epsfig{file=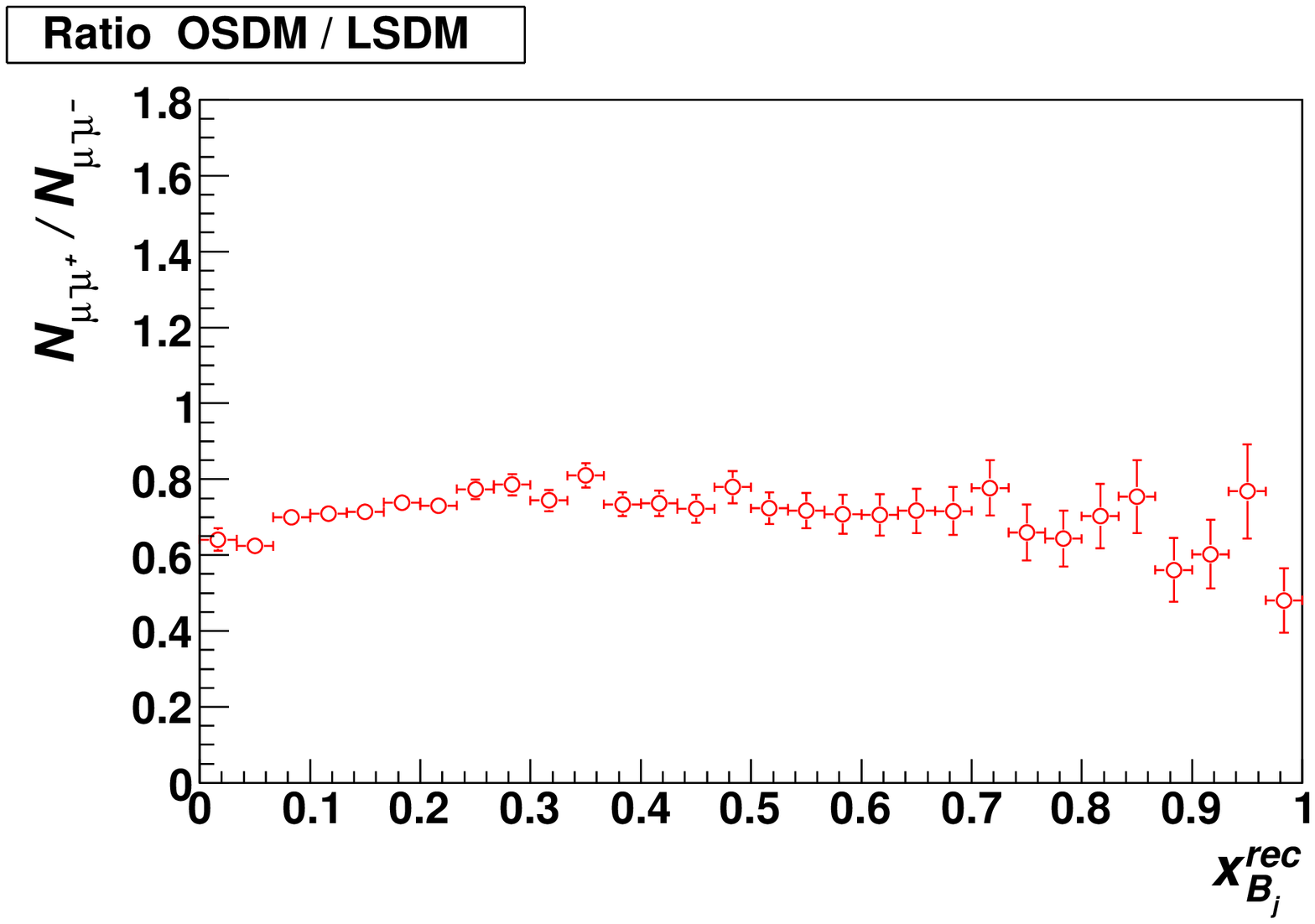,width=0.55\linewidth}\hspace*{-0.20cm}\epsfig{file=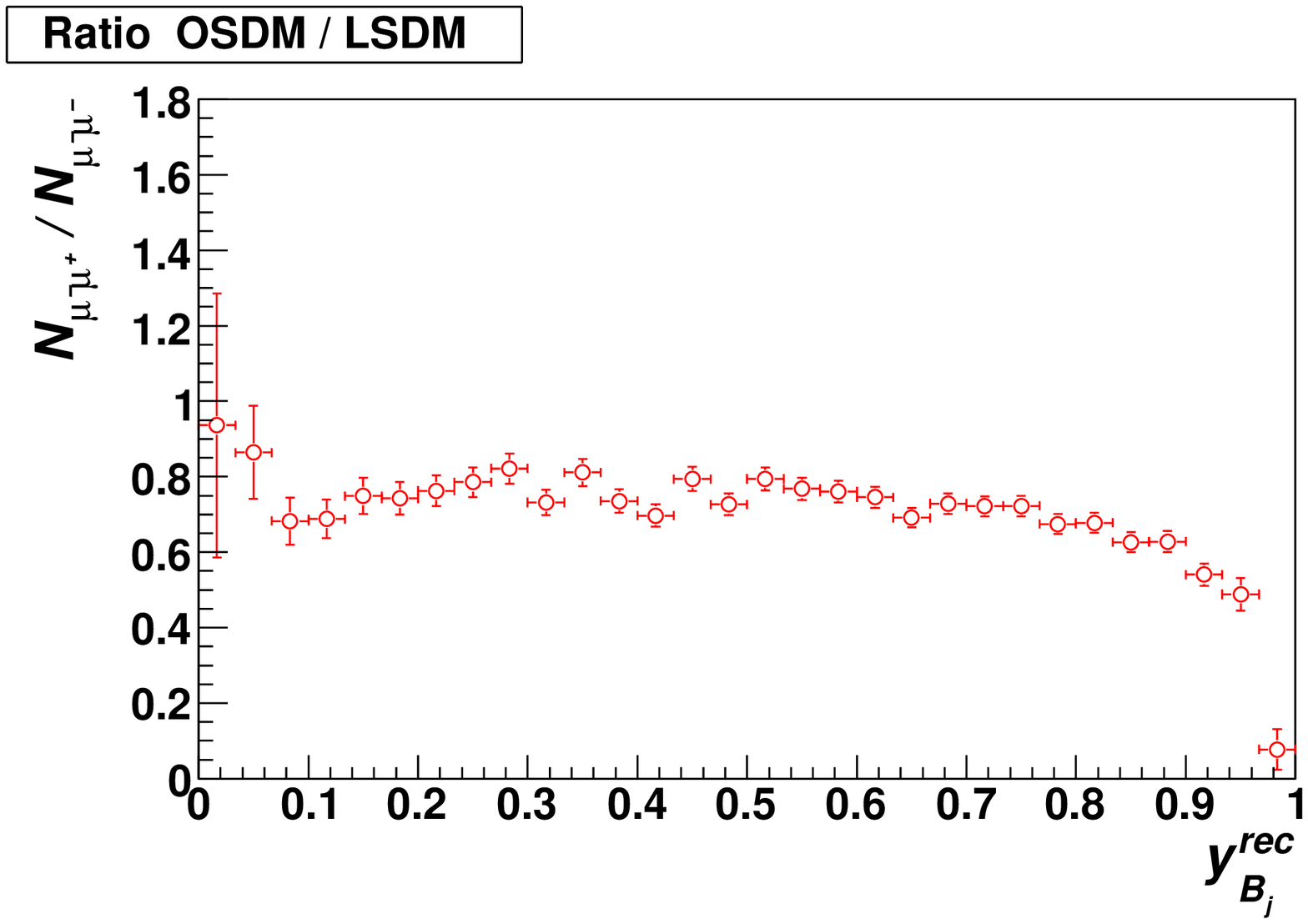,width=0.55\linewidth}}   
   \hspace*{-0.90cm}\mbox{\epsfig{file=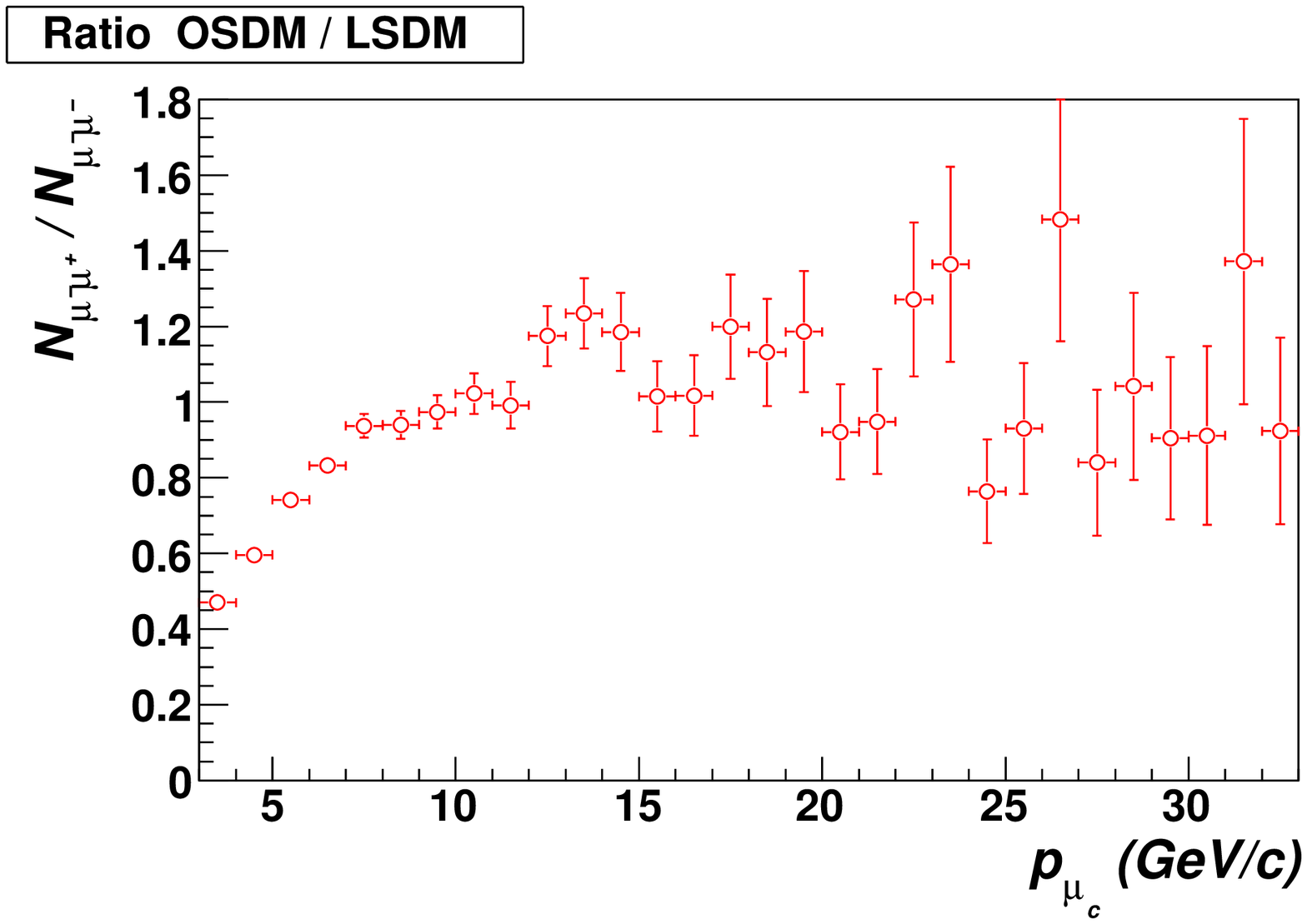,width=0.55\linewidth}\hspace*{-0.20cm}\epsfig{file=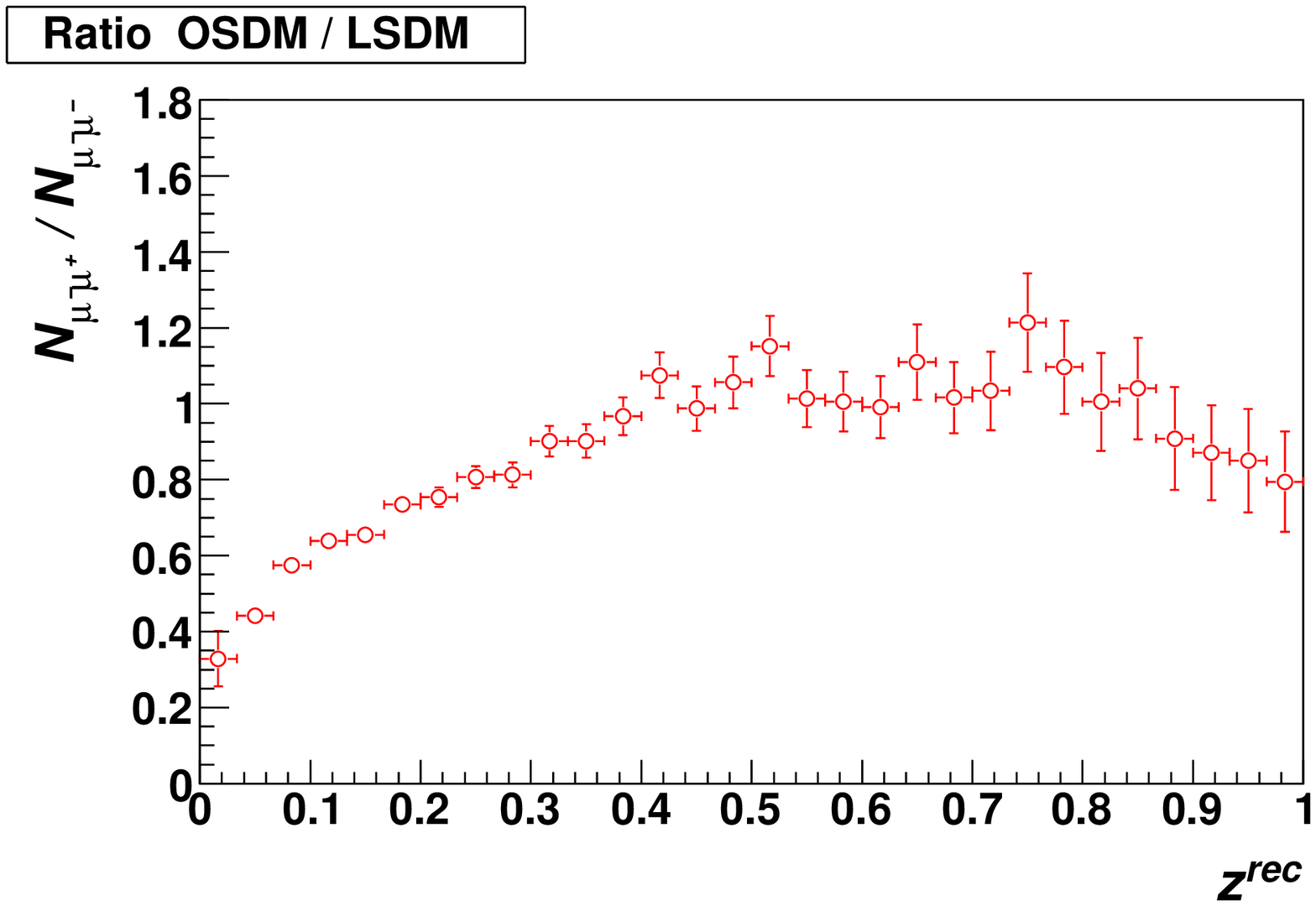,width=0.55\linewidth}}   
  \begin{center}
  \caption {\it Values of the ratio between OSDM background and LSDM as a function of kinematic variables: 
visible neutrino energy (top left), center of mass energy $\sqrt{\hat{s}}$ (top right), $x$-Bjorken (middle left), $y$-B
jorken (middle right), $\mu^+$ momentum (bottom left), fraction of the energy of the hadron shower carried by the $\mu^+
$ (bottom right).
The circles give the fully calibrated ratio after re-weighting with the $h^+/h^-$ ratio measured in DCH data.}
  \label{fig:bkgscale}
 \end{center}
\end{figure}

\begin{figure}[htb]
   \hspace*{-0.90cm}\mbox{\epsfig{file=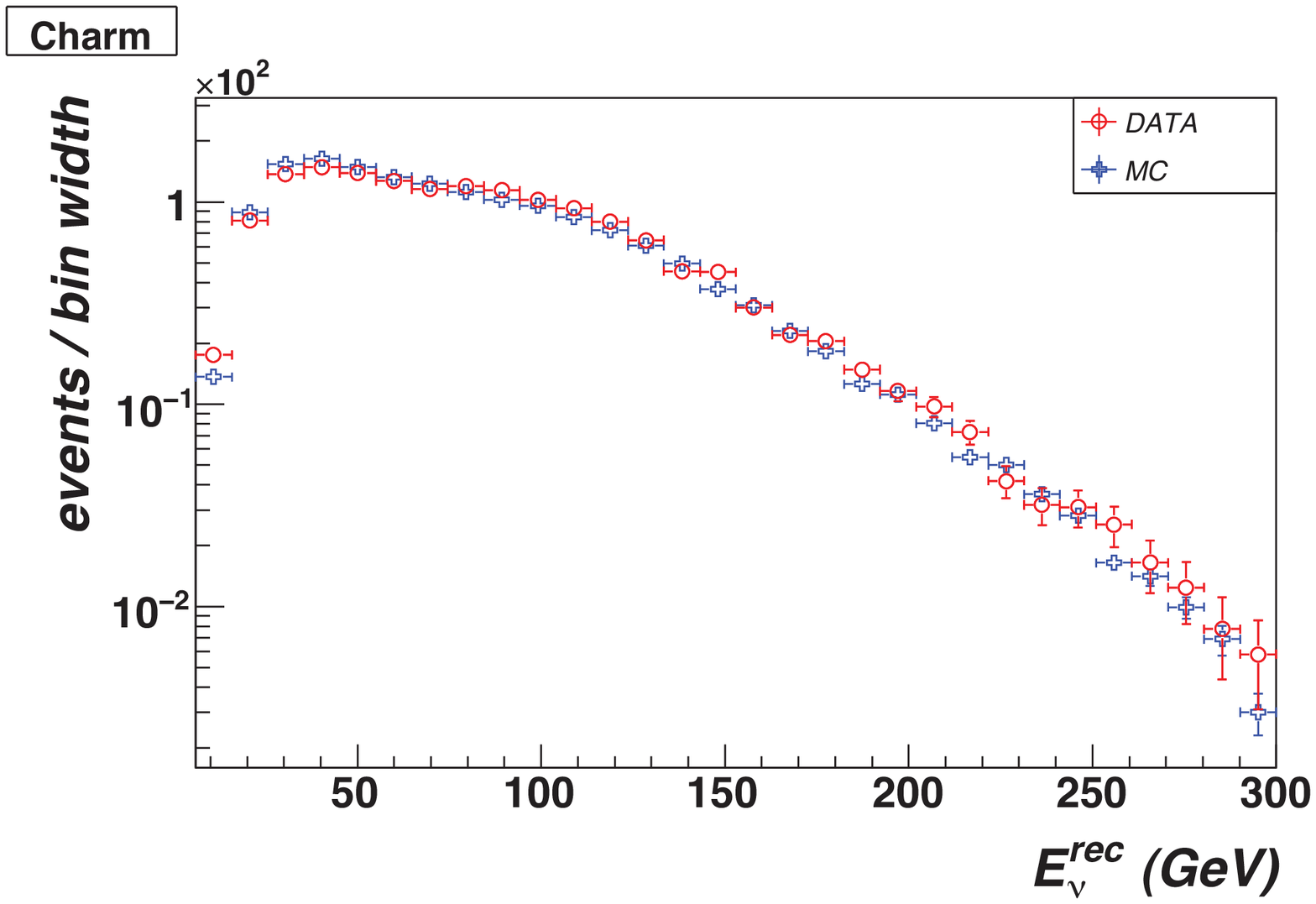,width=0.55\linewidth}\hspace*{-0.20cm}\epsfig{file=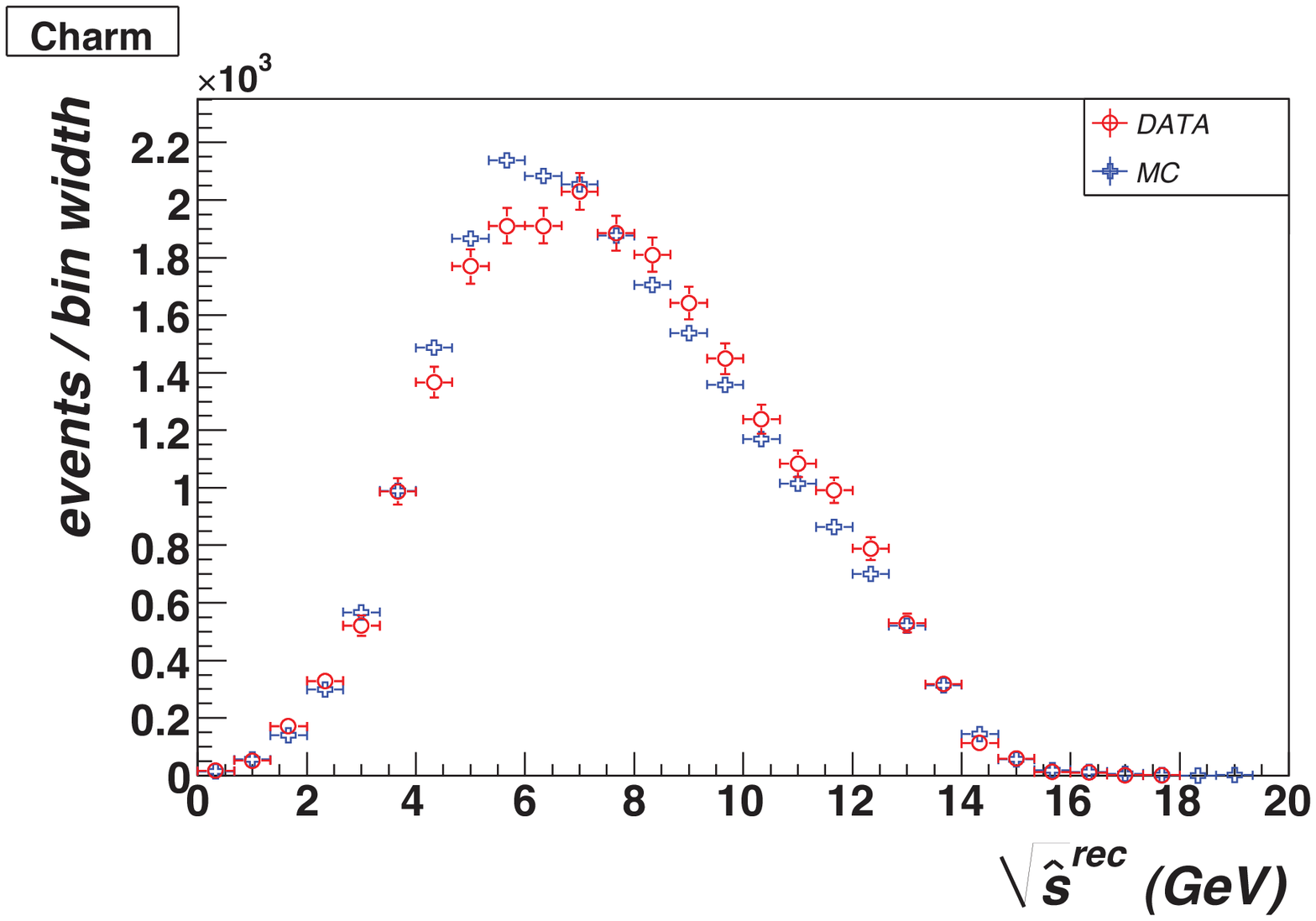,width=0.55\linewidth}}   
   \hspace*{-0.90cm}\mbox{\epsfig{file=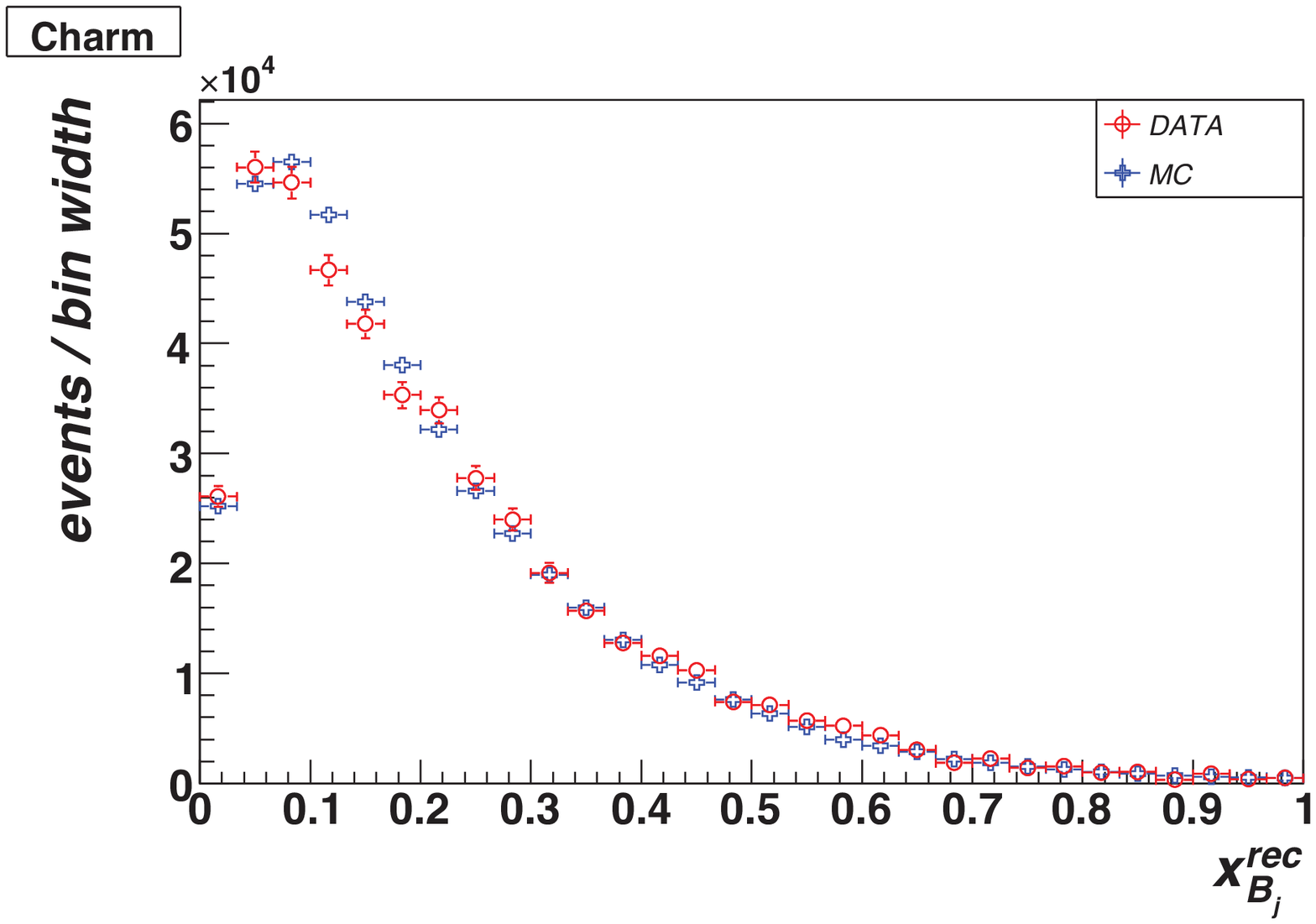,width=0.55\linewidth}\hspace*{-0.20cm}\epsfig{file=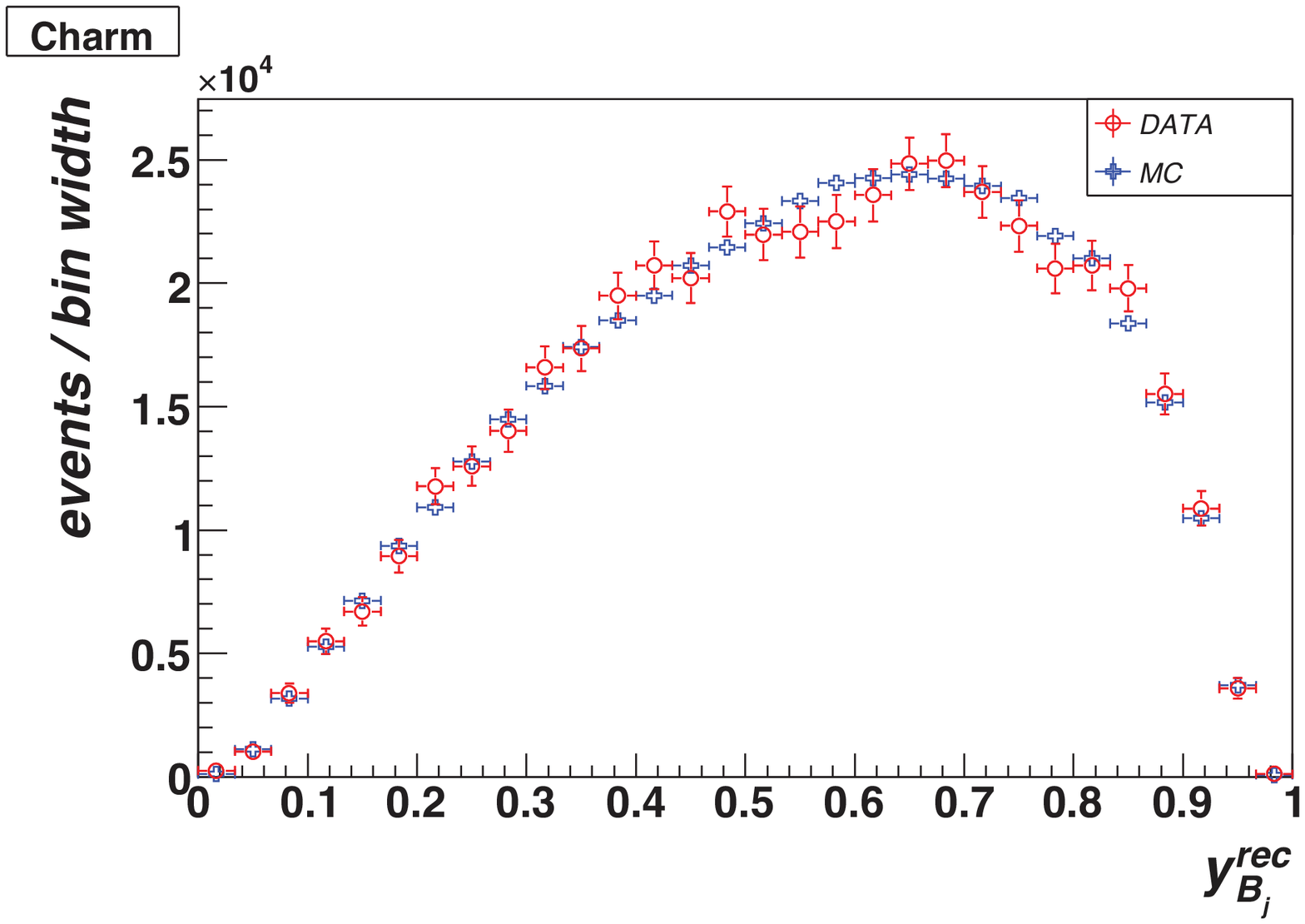,width=0.55\linewidth}}   
   \hspace*{-0.90cm}\mbox{\epsfig{file=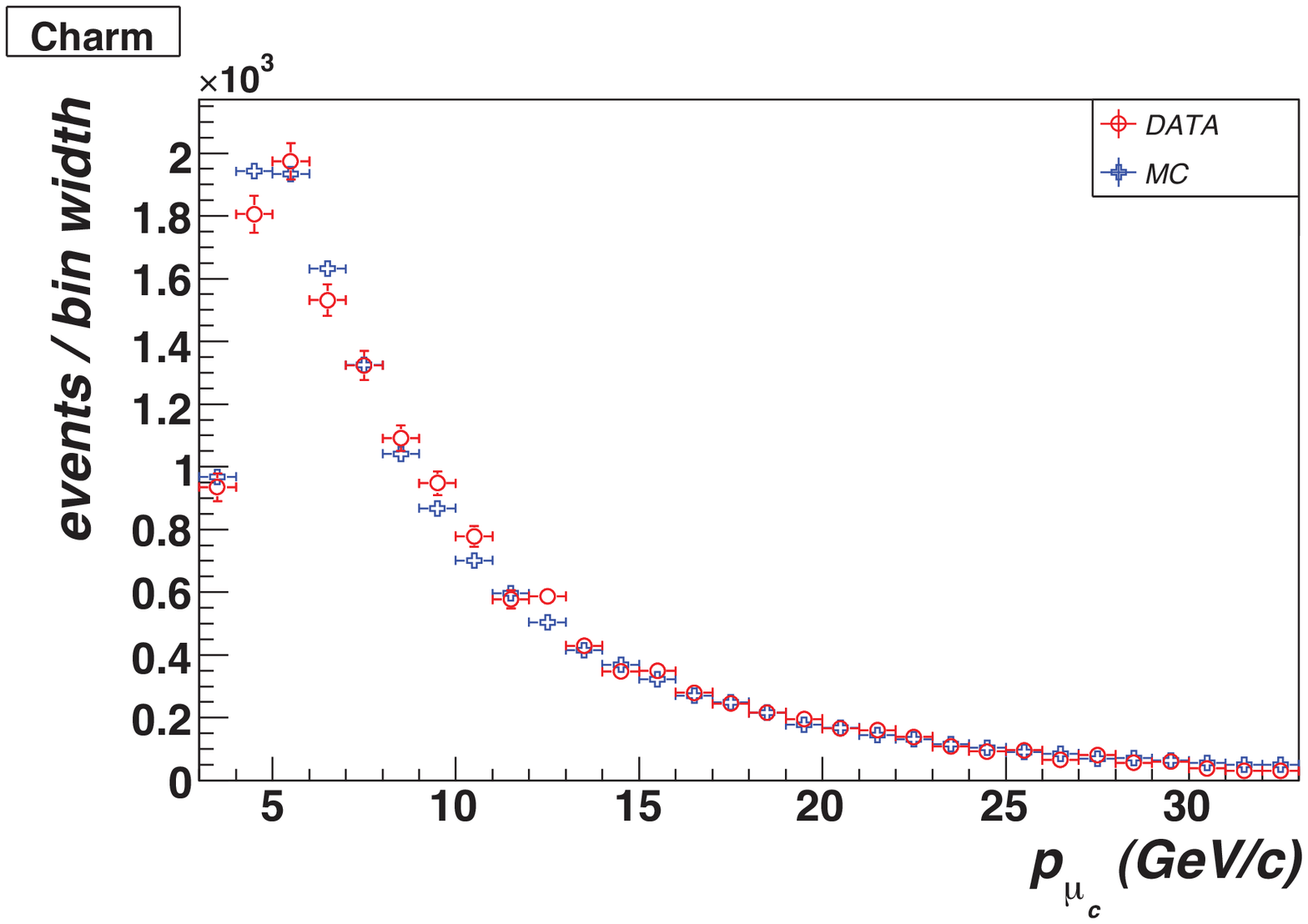,width=0.55\linewidth}\hspace*{-0.20cm}\epsfig{file=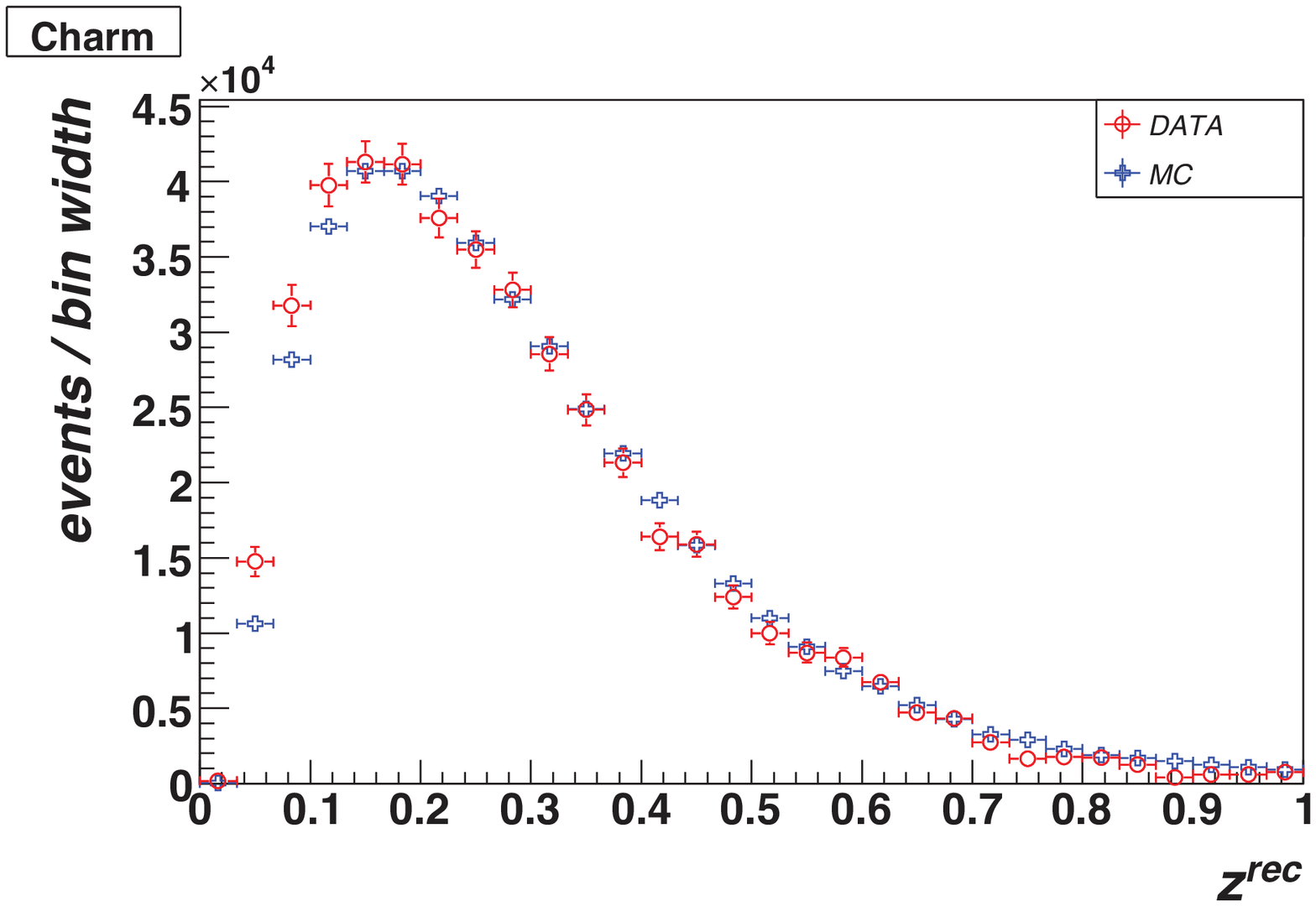,width=0.55\linewidth}}   
 \begin{center}
  \caption {\it Distributions of reconstructed kinematic variables in charm dimuon events:
visible neutrino energy (top left), center of mass energy $\sqrt{\hat{s}}$ (top right), $x$-Bjorken (middle left), $y$-B
jorken (middle right), $\mu^+$ momentum (bottom left), fraction of the energy of the hadron shower carried by the $\mu^+
$ (bottom right).
Data, after background subtraction, are shown as circles while MC points are shown as crosses.}
  \label{fig:charm_rec}
 \end{center}
\end{figure}

The background events $N^{^{\mbox{\tiny DATA}}}_{\mbox{\tiny $\mu\mu^+_{bg}$}}$ are estimated
from the LSDM events, measured in the data, in which both muons have the same negative charge, 
 ($\mu^- \mu^-$), multiplied by a scale factor extracted from the MC:
\begin{equation}
N^{^{\mbox{\tiny DATA}}}_{\mbox{\tiny $\mu\mu^+_{bg}$}} = N^{^{\mbox{\tiny DATA}}}_{\mbox{\tiny $\mu\mu^-$}} \cdot \left( N^{^{\mbox{\tiny MC}}}_{\mbox{\tiny $\mu\mu^+_{bg}$}} / N^{^{\mbox{\tiny MC}}}_{\mbox{\tiny $\mu\mu^-$}} \right)
\end{equation}
where the scale factor is given by the ratio of opposite sign to like sign dimuon events
originated from background events.
\begin{figure}[htb] 
   \hspace*{-0.90cm}\mbox{\epsfig{file=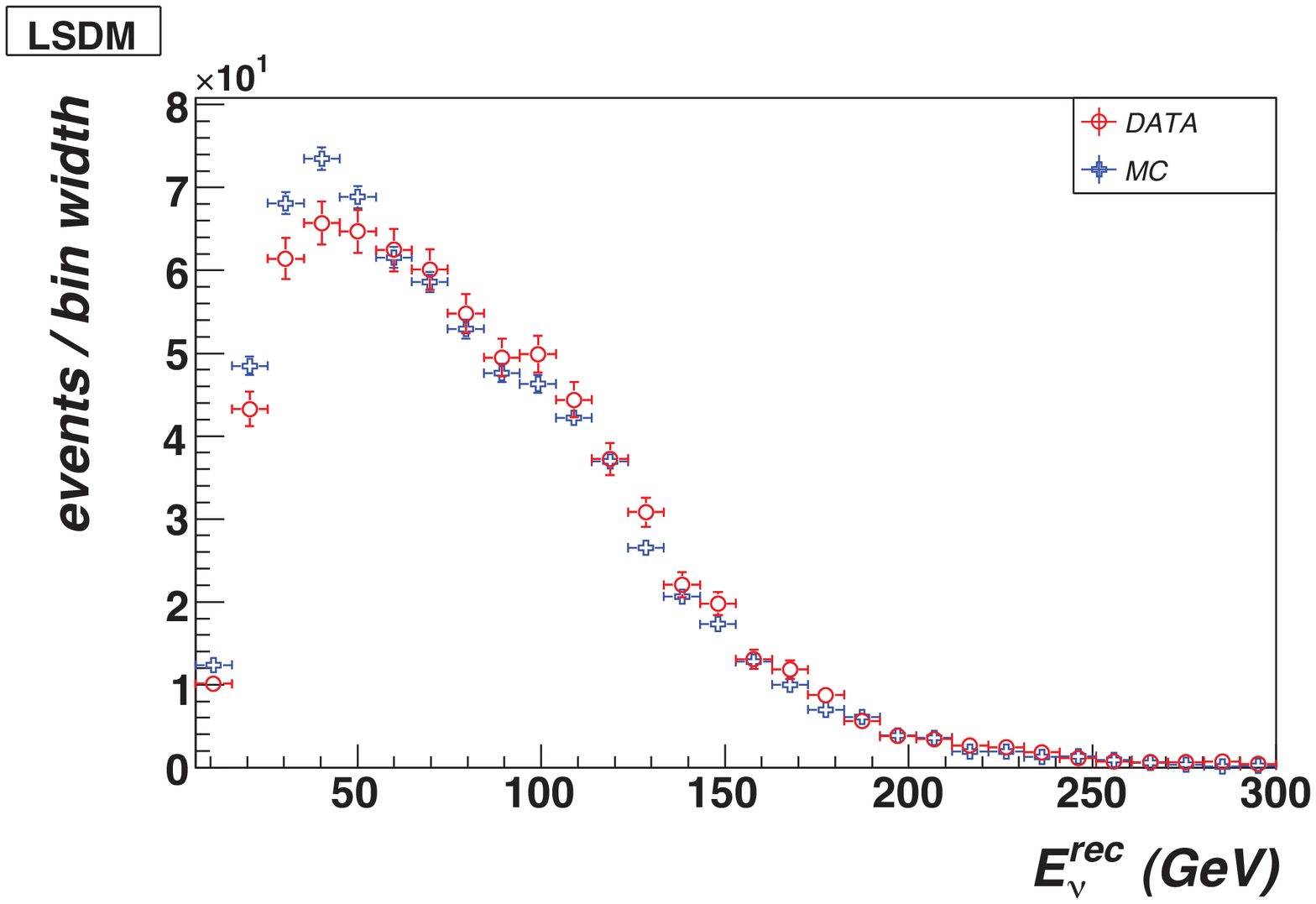,width=0.55\linewidth}\hspace*{-0.20cm}\epsfig{file=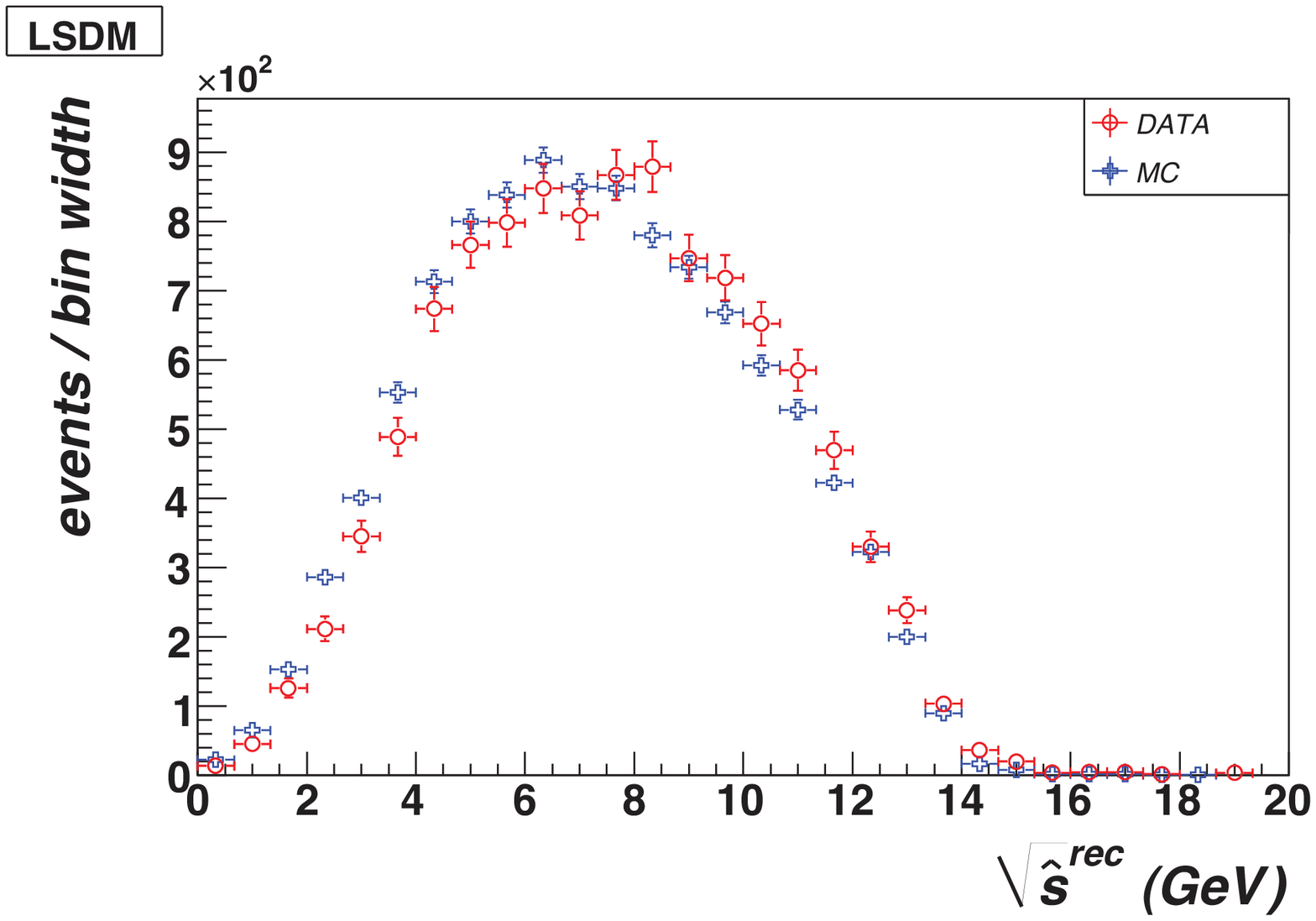,width=0.55\linewidth}}   
   \hspace*{-0.90cm}\mbox{\epsfig{file=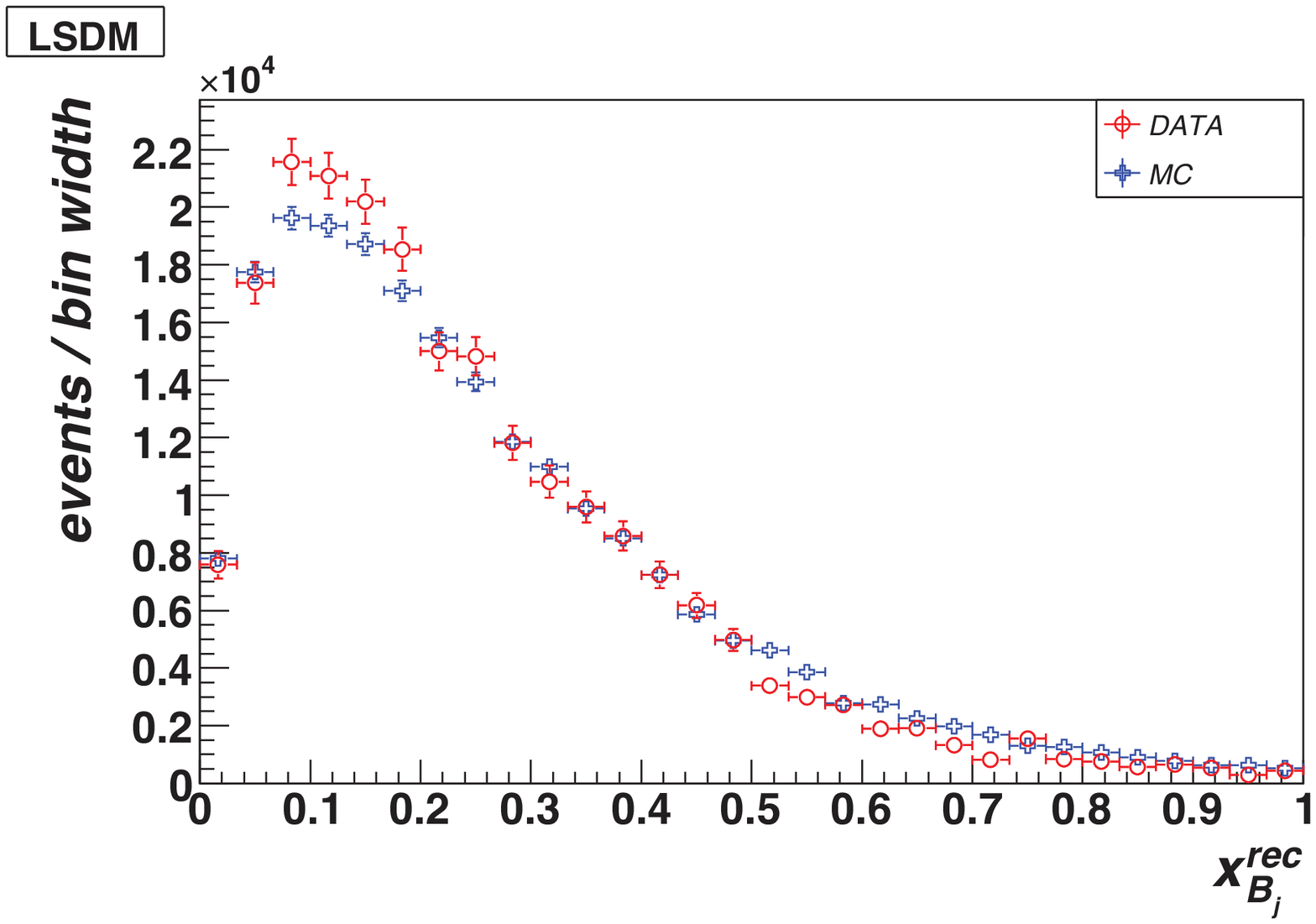,width=0.55\linewidth}\hspace*{-0.20cm}\epsfig{file=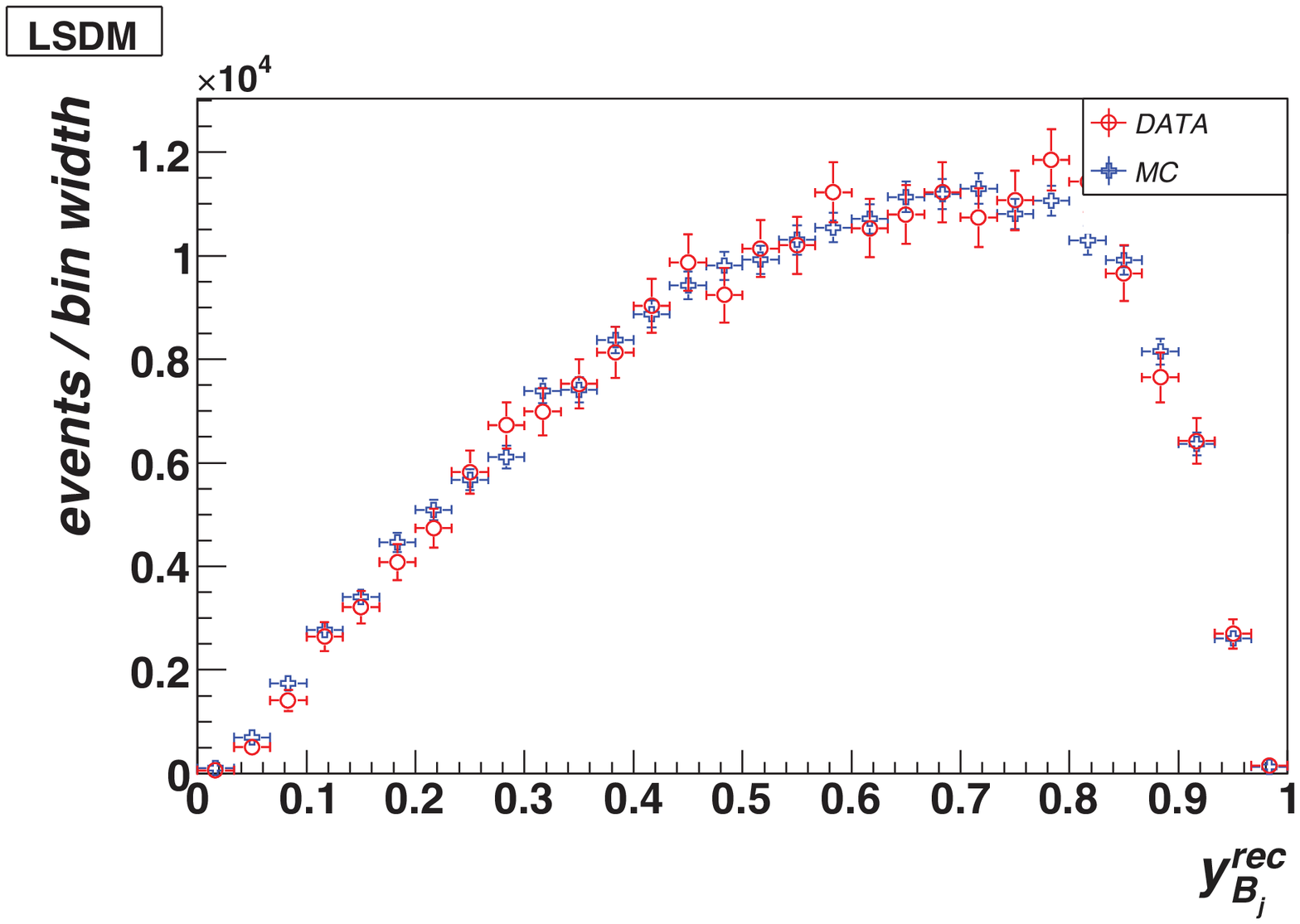,width=0.55\linewidth}}   
   \hspace*{-0.90cm}\mbox{\epsfig{file=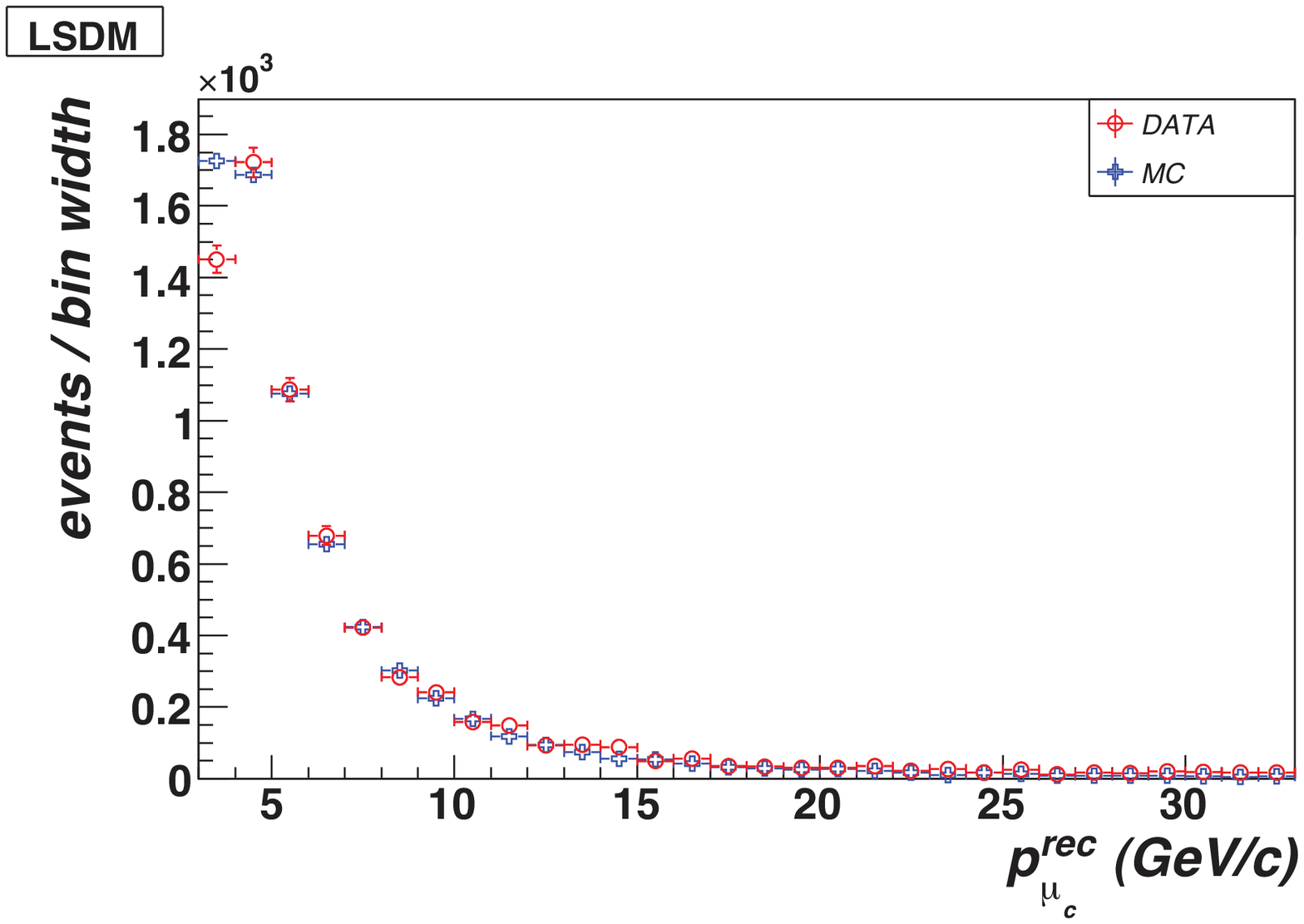,width=0.55\linewidth}\hspace*{-0.20cm}\epsfig{file=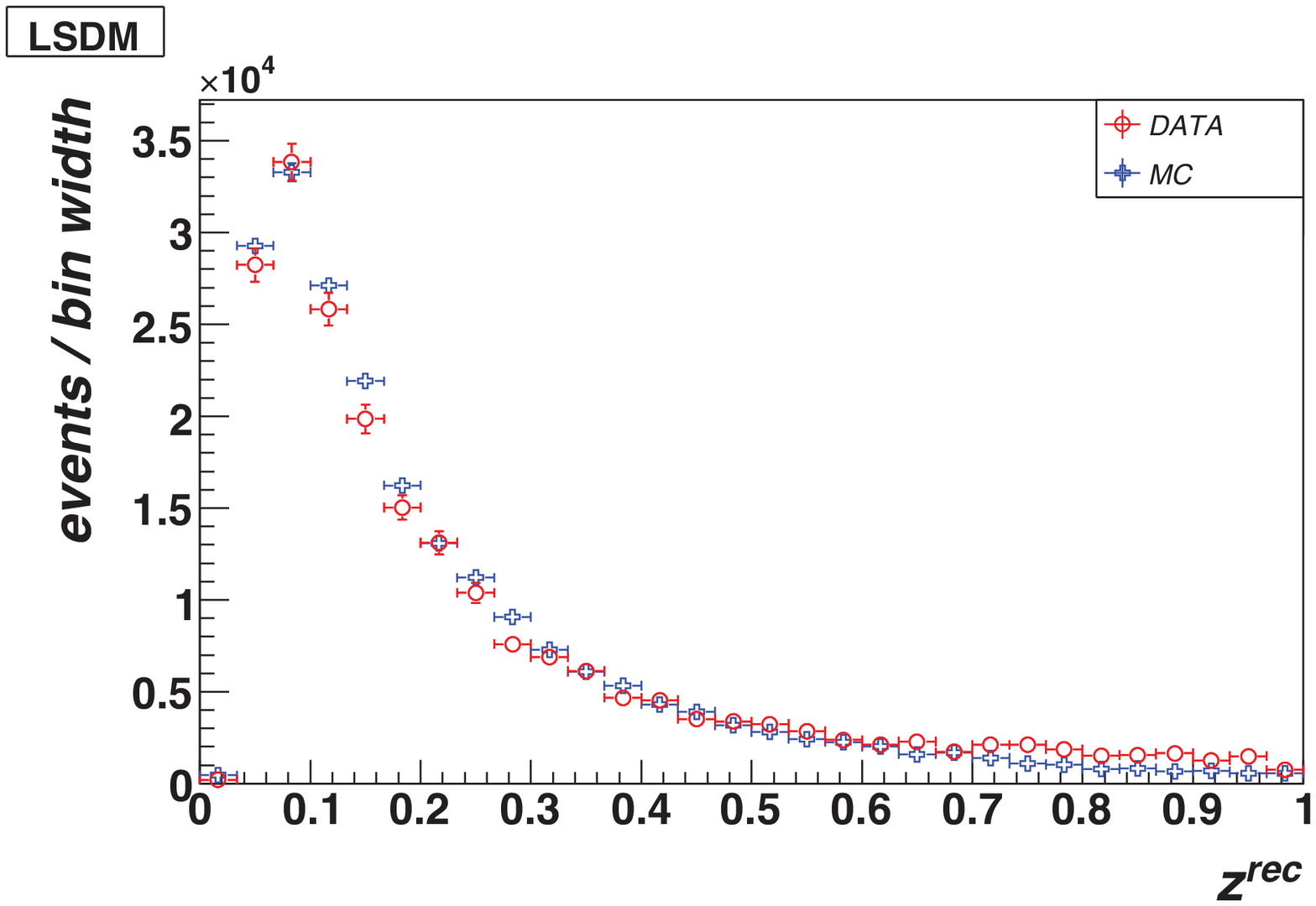,width=0.55\linewidth}}   
  \begin{center}  
  \caption {\it Distributions of reconstructed kinematic variables in like-sign dimuon events:
visible neutrino energy (top left), center of mass energy $\sqrt{\hat{s}}$ (top right), $x$-Bjorken (middle left), $y$-B
jorken (middle right), $\mu^+$ momentum (bottom left), fraction of the energy of the hadron shower carried by the $\mu^+
$ (bottom right).
Data are shown as circles while MC points are shown as crosses.}
  \label{fig:bkg_lsdm}
 \end{center}
\end{figure}
In order to reduce the MC statistical uncertainties we generated a total of about
$80 \times 10^6$ $\nu_\mu$ CC events fully reconstructed in FCAL.
Figure~\ref{fig:bkg_lsdm} shows the distributions of the LSDM events in FCAL data and MC.
The general agreement is adequate since LSDM events from MC are never
directly used in our analysis. Rather, we only use the {\it ratio} of OSDM to LSDM
background events in MC. This ratio is very sensitive to the details of the
fragmentation of the hadronic system, in particular at low momenta. For this reason we cannot
rely on the Monte Carlo simulation at the level of precision of a few percent.
Instead, since the background scale is determined by
the ratio of positively charged to negatively charged mesons inside the hadronic system
produced by the fragmentation of partons in DIS events, we {\it measure} this
ratio as a function of the meson momentum from the NOMAD data originated in the
light Drift Chamber target (DCH). The differences of the inclusive 
fragmentation variable distributions between target nuclei (carbon in DCH vs. iron in FCAL) are found to be negligible 
based on a direct comparison between the corresponding MC samples. This comparison is
shown in the top plot of Figure~\ref{fig:had_dch}.
We exclude charged tracks which are either identified as electrons/positrons by the
TRD or have a range consistent with protons. We then subtract the small residual
proton/electron/positron contamination by using the MC. These contaminations are small even before rejecting identified tracks.
The average $K/\pi$ ratio in neutrino interactions is only about 7\% and therefore
the uncertainty introduced in the measurement by the different $K$ and $\pi$ lifetimes
is negligible.
After measuring the ratio of positively charged to negatively charged mesons from DCH
data, we re-weight each positive meson originated from the hadronic system in FCAL
events according to the measured ratio $\omega_{h^+}$:
\begin{equation}
\label{eq:bkgwgt}
 W^{\mbox{\tiny{MC}}} = \prod\limits_{h^+} \omega_{h^+}
\end{equation}
As a result, the entire background estimate for the charm dimuon sample is based upon
the data themselves, which are used {\it both} for the LSDM and for the background scale.
We emphasize that only the use of a low density target embedded inside a magnetic 
spectrometer in NOMAD has allowed the use of this procedure. 
Figure~\ref{fig:had_dch} shows the measured ratio $h^+/h^-$ from the DCH data, as well as
a comparison with the corresponding MC simulation and the function $\omega_{h^+}$.
The calibration of the background through the re-weighting procedure is a crucial step
in the analysis and allows a substantial improvement in the description of the charm
dimuon data by the Monte Carlo simulation. Without the use of data from the low density
DCH target it would have not been possible to lower the energy threshold on the
secondary muon down to 3 GeV as well as to reduce the systematic uncertainty on the
background subtraction. Figure~\ref{fig:bkgscale} shows the background scale
$N^{^{\mbox{\tiny MC}}}_{\mbox{\tiny $\mu\mu^+_{bg}$}} / N^{^{\mbox{\tiny MC}}}
_{\mbox{\tiny $\mu\mu^-$}}$ as a function of the main kinematic variables
with the re-weighting procedure according to Eq.~(\ref{eq:bkgwgt}).
Table~\ref{tab:evtdimuon} summarizes signal and background events after each
selection cut. After all cuts we retain 20,479 OSDM events in the data, of which
15,344 are genuine charm signal (75\%) and 5,135 are background (25\%).
The final distributions of various kinematic variables for the charm data after background 
subtraction are given in Figure~\ref{fig:charm_rec}.

\begin{table}[htb]
\begin{center}
\small
\begin{tabular}{||c|c|c|c|c|c|c|c||}
\hline
\hline
 & \multicolumn{2}{c}{MC} & \multicolumn{4}{|c|}{DATA} & DATA/MC  \\
\hline
 Cut              & Rec.   & Eff.  & OSDM  & LSDM  & Bg. & Charm & \\
\hline\hline  
 $4$  (FV)               & 18783 & 27.0\% & 30955 & 33127 & ---  & ---   & ---    \\
 $5$  (Time)                  & 18671 & 26.9\% & 26739 &  9488 & 6565 & 20174 & 1.080  \\
 $6$ (Lead. $\mu_{cc}^-$)          & 18144 & 26.1\% & 24642 &  9488 & 6366 & 18276 & 1.007  \\
 $7$  ($E_{had}^{up}$)        & 16287 & 23.4\% & 21637 &  7763 & 5403 & 16234 & 0.997  \\
 $8$  ($x_{B_j}$)             & 16180 & 23.0\% & 21256 &  7524 & 5308 & 15948 & 0.985  \\
 $9$ ($E_{\mu_{cc}}$)        & 16173 & 23.0\% & 21245 &  7518 & 5307 & 15938 & 0.985  \\
 $10$ ($E_{\mu_c,had}^{low}$) & 16019 & 22.8\% & 20949 &  7324 & 5269 & 15680 & 0.978  \\
 $11$ ($Q^2$)                 & 15684 & 22.5\% & 20479 &  7148 & 5135 & 15344 & 0.978  \\
\hline
\hline
\end{tabular}
\normalsize
\caption {\it 
Event selection for dimuon events in data and MC. The efficiency is normalized to the raw number of 
MC events generated in the fiducial volume, as in Table~\ref{tab:evtnumucc}.  
All the other MC numbers have been normalized to $\nu_\mu$ CC data after
the fiducial volume and leading muon cuts (cut 6), by taking into account the ratio of charm dimuon cross-section to the inclusive CC cross-section, which are calculated analytically.
The number of background events is calculated from the LSDM data multiplied by the scale
factor obtained after re-weighting for the $h^+/h^-$ ratio measured in DCH data.
} 
\label{tab:evtdimuon}
\end{center}
\end{table}


Table~\ref{tab:expstat} shows a comparison of our final charm dimuon sample with the 
previous measurements. The new NOMAD measurement has the highest available statistics of neutrino induced 
charm dimuon events. In particular, the measurement described in this paper has about 3 times the statistics 
of NuTeV and CCFR, which till now have been the only source of information on strange sea quark distributions 
in global Parton Distribution Function (PDF) fits~\cite{Alekhin:2008mb}. 
Furthermore, the NOMAD sample can reach the lowest 
energy threshold, giving additional sensitivity to the charm production parameters.

\begin{table}[htb]
 \begin{center}
  \begin{tabular}{||c|c|c|c|c||}
   \hline
   \hline
    & Exp.    & Publ.    & Stat. ($N_{\mu\mu}$) & $E_{\nu}$ (GeV) \\
   \hline
   \underline{$\nu N$} & & & & \\
   \hline
    & CDHS~\cite{Abramowicz:1982zr} & 1982 & ~8,600   & 30-250 (20) \\
    & CHARM II~\cite{Vilain:1998uw} & 1999 & ~3,100   & 35-290 (24) \\
    & NOMAD~\cite{Astier:2000us}  & 2000 & ~2,714 & 14-300 (27) \\ 
    & CCFR~\cite{Bazarko:1994tt,Goncharov:2001qe} & 2001 & ~5,030  & 30-600 (150) \\
    & NuTeV~\cite{Goncharov:2001qe,nutev:ssbar07} & 2001 & ~5,102   & 20-400 (157.8) \\
    & CHORUS~\cite{:2008xp}         &  2008 & ~8,910   & 15-240 (27) \\
    & NOMAD   &  2013 & {\bf 15,344} & {\bf 6}-300 (27) \\
   \hline
   \underline{$\bar{\nu} N$} & & & & \\
   \hline
    & CDHS~\cite{Abramowicz:1982zr} &  1982 & ~2,000   & 30-250 \\
    & CHARM II~\cite{Vilain:1998uw} &  1999 & ~~~700   & 35-290 \\
    & CCFR~\cite{Bazarko:1994tt,Goncharov:2001qe} &  2001 & ~1,060 & 30-600 \\
    & NuTeV~\cite{Goncharov:2001qe,nutev:ssbar07} &  2001 & ~1,458   & 20-400 \\
    & CHORUS~\cite{:2008xp} &  2008 & ~~~430           & 10-240 \\
   \hline
   \hline
  \end{tabular}
  \caption {\it Summary of the observed numbers of charm dimuon events from existing measurements in neutrino 
and anti-neutrino interactions. The average neutrino energy of each experiment is given in parenthesis in the 
last column. The NOMAD analysis described in this paper has the largest statistics and the 
lowest energy threshold. The latter is particularly useful for the determination 
of the charm quark mass.}
  \label{tab:expstat}
 \end{center}
\end{table}

\section{Unfolding procedure}
\label{sec:unfolding} 

The detector smearing and acceptance corrections require an input model for the
cross-sections and fragmentation functions. For the inclusive $\nu_\mu$ CC sample
the model is largely independent from the NOMAD data. However, for the charm sample
the NOMAD dimuon data are eventually used to determine the charm production parameters,
which, themselves, are inputs to the experimental acceptance corrections.
In our analysis we follow an iterative procedure. First, we use an input model which
is fully independent from NOMAD data and we verify its consistency with the
NOMAD data. After this step we add NOMAD data to the global PDF fits to improve the
precision on the charm production parameters. We then feed back the modified model
into the acceptance calculation and iterate until convergence. In the following sections
we describe in details the model used and the corrections applied.

\subsection{Cross-section weights}
\label{sec:MCwei} 

Neutrino interactions are simulated using a modified version of LEPTO 6.1~\cite{lepto} and JETSET 7.4~\cite{jetset} 
with $Q^2$ and $W^2$ cutoff parameters removed. 
Final state interactions within the target nucleus are described by DPMJET~\cite{dpmjet}. 
A full detector simulation based on GEANT 3.21~\cite{geant} is then performed. 
The default DIS cross-sections are calculated in the LO approximation with the parton density functions  
from the GRV-HO parameterizations~\cite{grv94} available in PDFLIB~\cite{pdflib}. 
This simulation does not
adequately describe the charm production process since it does not include any rescaling
mechanism to take into account the large mass of the charm quark. Furthermore, no
electroweak, nuclear and High Twist corrections are included.

In order to achieve an accurate description of data, we implement a re-weighting procedure
for the charm cross-section:
\begin{equation}
\omega_{\mu \mu} (E_\nu, x_{Bj}, y_{Bj}) = \frac{\sigma^{^{\mbox{\tiny AKP}}}_{\mu \mu} (E_\nu, x_{Bj}, y_{Bj})}{\sigma^{^{\mbox{\tiny LEPTO}}}_{\mu \mu} (E_\nu, x_{Bj}, y_{Bj})}
\end{equation}
where $\sigma^{^{\mbox{\tiny LEPTO}}}_{\mu \mu}$ is the original LEPTO cross-section used
to generate the MC events and $\sigma^{^{\mbox{\tiny AKP}}}_{\mu \mu}$ is the new cross-section
obtained from an analytical calculation~\cite{Kulagin:2007ju,Kulagin:2004ie,Kulagin:2010gd,Alekhin:2008mb,Alekhin:2007fh,Arbuzov-Bardin}. The charm cross-section is calculated in the NLO QCD approximation for the
heavy quark structure functions, in a factorization scheme with three light flavors in the initial
state (FFS)~\cite{Alekhin:2008mb}. The Target Mass Corrections (TMC) are implemented following the prescription
by Georgi and Politzer~\cite{Georgi:1976ve}. The impact of the dynamical High Twist corrections to the charm production
is evaluated by applying a simple rescaling for the quark charge to the phenomenological
twist-4 terms extracted from the inclusive lepton-nucleon cross-sections~\cite{Alekhin:2007fh}.
We apply nuclear corrections using the calculations of Refs.~\cite{Kulagin:2007ju,Kulagin:2004ie,Kulagin:2010gd}.
This calculation takes into account a number of different effects including the nucleon Fermi motion
and binding, neutron excess, nuclear shadowing, nuclear pion excess and the off-shell
correction to bound nucleon Structure Functions (SF).
The electroweak corrections, including one-loop terms,
are calculated according to Ref.~\cite{Arbuzov-Bardin} within the framework of the parton
model. The parameters related to charm production like the mass of the charm quark and the
strange quark sea distribution are fixed to the ones extracted from the global PDF fit including
NuTeV and CCFR charm dimuon data~\cite{Alekhin:2008mb} at this stage. This allows a
consistency check with a calculation fully independent from NOMAD data.

We extract the LEPTO cross-section from the NOMAD MC by simulating $10^9$ events with
an input flux which is chosen to be inversely proportional to the neutrino energy $\propto 1/E_\nu$.We then bin the events in the $(E_\nu, x_{Bj}, y_{Bj})$ space and smooth the corresponding
histograms with an interpolation procedure.

Finally, we apply an additional re-weighting to the charm events to take into account the
effect of the charm fragmentation, which is described by the Collins-Spiller 
parameterization ~\cite{Collins:1984ms} (see Eq.~(\ref{eq:frag})). This function describes the probability 
for a charmed hadron to carry a given fraction $z = P_L(h_c) / P_L^{\rm max}$ of the charmed quark 
logitudinal momentum and is defined by one free parameter $\varepsilon_{\rm c}$.
Figure~\ref{fig:charm_rec} shows a comparison between data and weighted MC for
different kinematic variables in charm dimuon events.

We apply a similar re-weighting procedure to the inclusive $\nu_\mu$ CC events (single muons).
The model used for the inclusive CC structure functions on iron is the same described above for 
charm production. The light quark contributions to the SFs are calculated in the NNLO QCD
approximation.
Figure~\ref{fig:numucc_rec} shows a comparison between data and weighted MC for different
kinematic variables in $\nu_\mu$ CC events.

\subsection{Binning and hadronic energy correction}
\label{sec:ehadcal}

The binning of the data is performed in such a way that the size of each bin is comparable to
the experimental resolution within that given bin. This procedure allows a reduction of
correlations among different bins, thus minimizing systematic uncertainties.
Overall we have 19 bins for
$E_\nu$, 14 bins for $x_{Bj}$ and 15 bins for $\sqrt{\hat{s}}$.

\begin{figure}[htb]
 \begin{center}
   \mbox{\hspace*{-0.5cm}\epsfig{file=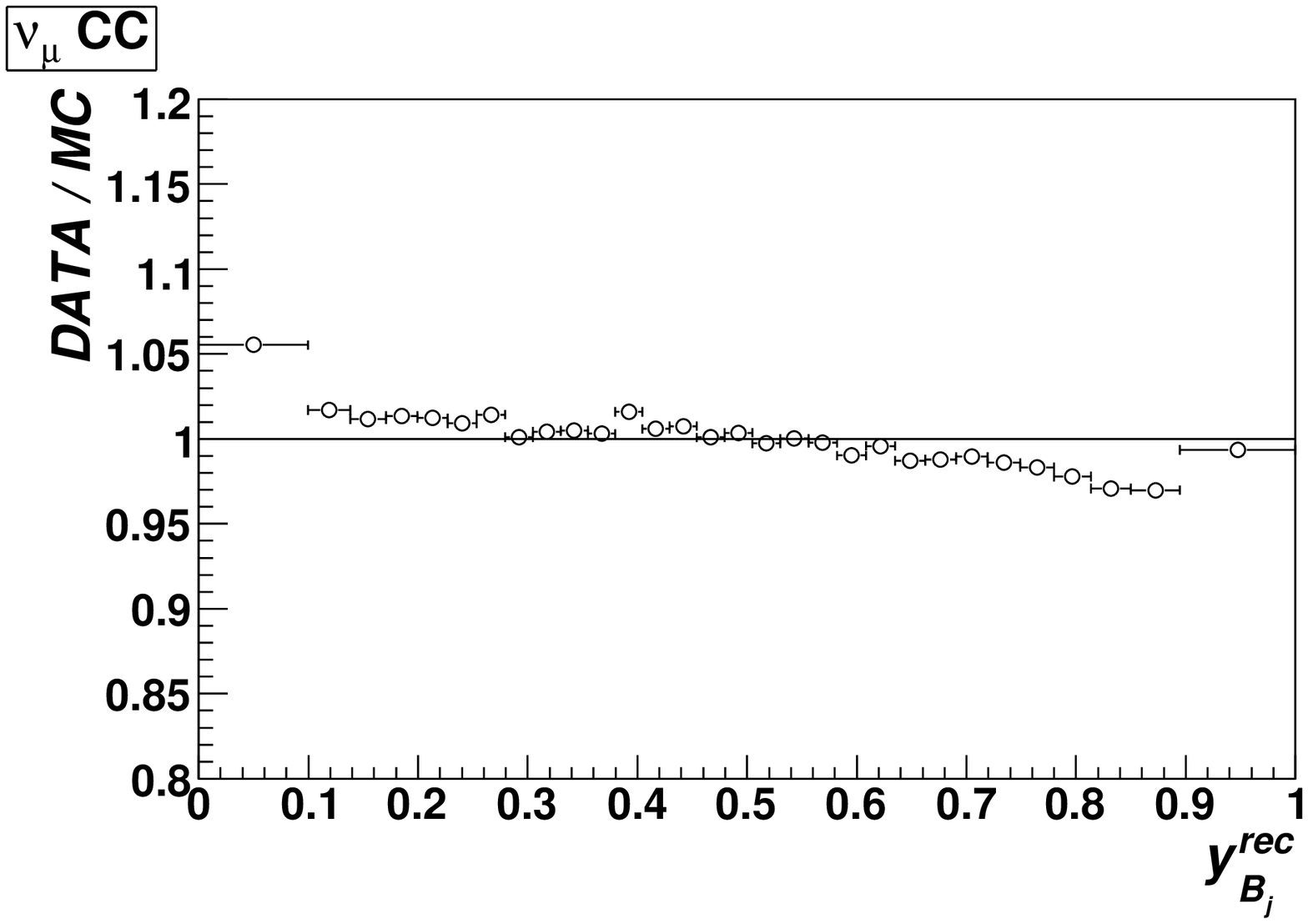,width=0.55\linewidth}\hspace*{-0.3cm}\epsfig{file=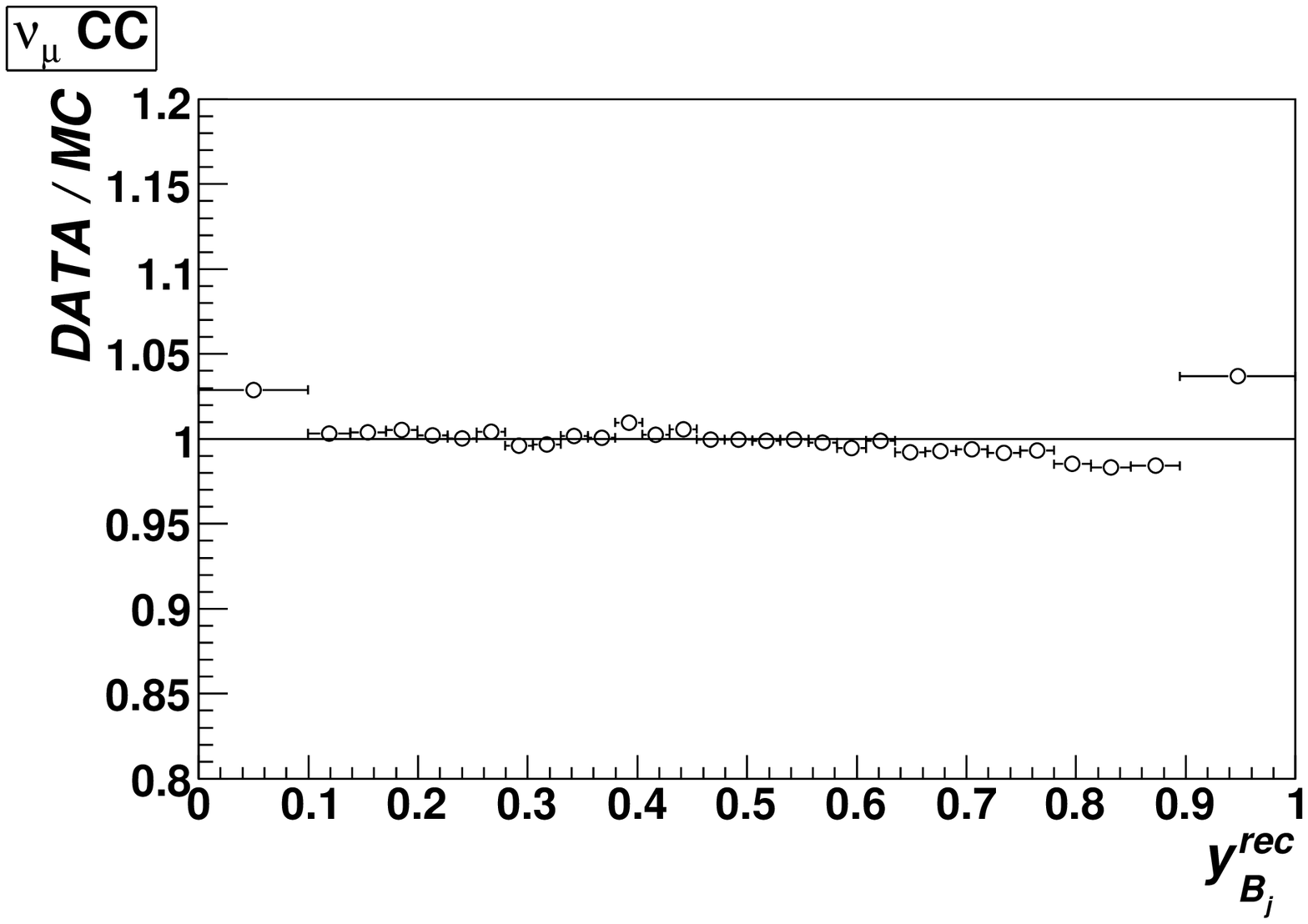,width=0.55\linewidth}}
  \caption {\it Ratio of data to MC for the $y_{Bj}$ distribution before (left plot)
and after (right plot) the calibration of the global hadronic energy scale.}
  \label{fig:ybjcalib}
 \end{center}
\end{figure}

After defining the binning we perform a calibration of the global hadronic energy scale.
This procedure corrects for potential discrepancies between data and MC related to
the simulation of neutral and charged particles in the hadronic jet orginated
by the neutrino-nucleus interactions. We perform the calibration of the hadronic
energy scale by using the $y_{Bj} = E_{\rm Had}/E_\nu$ distribution in
inclusive $\nu_\mu$ CC events (single muon events). For each of the 19 bins in
the {\it reconstructed} visible energy $E_\nu$, we multiply the hadronic energy $E_{\rm Had}$ in MC
events by a free scale factor $k_{\rm H}$, and we determine the optimal value of $k_{\rm H}$
by minimizing the $\chi^2$ calculated from the $y_{\rm Bj}$ distribution in data and MC.
This technique relies upon the precise measurement of $E_\mu$ in the drift chambers 
(see Section~\ref{sec:flux}).
The best fit values for $k_{\rm H}-1$ range from -0.1\% to -3.7\%,
depending upon the bin considered. Finally we interpolate the corrections for each bin
with a spline function in order to have a smooth behavior of the hadronic energy scale
as a function of the visible energy which can be extrapolated to different binning
definitions. 
The use of a separate calibration factor $k_{\rm H}$ for each $E_\nu$ bin effectively
takes into account differences in the development of the hadronic shower as a function
of the neutrino energy (e.g. missing particles, fragmentation etc.). In general, we
observe that the hadronic energy correction increases with the neutrino energy.
As can be seen from Figure~\ref{fig:ybjcalib} the calibration of the global hadronic energy
improves the agreement between data and MC for the $y_{Bj}$ distribution.

In order to estimate the corresponding uncertainties
on $k_{\rm H}$ we inflate the MC errors until the values of $\chi^2$/dof at
the minimum is equal to unity for each bin. We then calculate the $1\sigma$ error
band as the range in $k_{\rm H}$ which is resulting in $\Delta \chi^2 = 1.0$.
The uncertainties obtained for all bins 
range from 0.3\% at low energy to about 1\% at high energy.

\subsection{Smearing matrix and efficiency correction}

After re-weighting the MC events to our cross-section model, we unfold the detector response
from the measured data for both the inclusive $\nu_\mu$ CC and the charm dimuon events.
To this end we first determine the smearing matrix and the efficiency corrections from
the MC simulation:
\begin{equation}
N_i^{rec} (x^{rec}) = \sum_j r_{ij}(x^{rec}, x^{sim}) \times \epsilon_j (x^{sim}) \times
N_j^{sim}( x^{sim})
\end{equation}
where $x^{rec}$ and $x^{sim}$ are the reconstructed and simulated variable $x$
($x=E_\nu, x_{Bj}, \sqrt{\hat{s}}$). The inverse of the above relation provides the
unfolded measurement:
\begin{equation}
N_j^{sim} (x^{sim}) = \sum_i \epsilon_i^{-1} ( x^{sim}) \times r^{-1}_{ji} ( x^{sim}, x^{rec})
\times N_i^{rec} ( x^{rec})
\end{equation}
The impact of non-diagonal terms in the smearing matrix
is reduced because the bin size is comparable to the experimental resolution.

We validate the unfolding procedure by splitting the available MC events in two
independent samples. The first sample (biased) is used to extract the smearing matrix
and the efficiency correction. The second half of the MC sample (unbiased) is used as
fake data to determine the unfolded distributions. These latter are then compared
with the input simulated variables of the unbiased sample. Our results indicate that 
we can reproduce the input distributions
in the unbiased sample with a good accuracy for both $\nu_\mu$ CC and charm dimuon events.
Finally we compare the unfolded distributions obtained from FCAL data and MC with
an analytical calculation performed by convoluting our cross-section model with
the neutrino flux. 

\subsection{Bin centering correction}

The definition of the bin size is based on the procedure described in Section~\ref{sec:ehadcal} for the
{\em reconstructed} variables. The boundaries of the corresponding bins in the {\em simulated}
variables are then slightly adjusted in order to keep the same statistical uncertainty as
in the reconstructed variables.

Since the ${\mathcal R}_{\mu \mu}$ functions vary inside the chosen bins,
we need to apply a bin centering correction to the unfolded data.
To this purpose we use our model calculation convoluted with the NOMAD flux.
For each bin, we find the value of the kinematic variable on the horizontal axis
($E_\nu, x_{Bj}$ or $\sqrt{\hat{s}}$) for which the analytical function ${\mathcal R}_{\mu \mu}$
is equal to the corresponding average value inside the bin. We then assign the
measured value of ${\mathcal R}_{\mu \mu}$ for the bin considered to this calculated
point on the horizontal axis.

\section{Results}
\label{sec:results} 

\subsection{Charm fragmentation}
\label{sec:charmfrag}

The measured kinematic distributions for charm events are sensitive to the charm
fragmentation function, which gives the probability for the charmed hadron to
carry a given fraction $z$ of the longitudinal momentum of the hadronic system.
We model the charm fragmentation function with the Collins-Spiller
parameterization~\cite{Collins:1984ms} shown
in Eq.~(\ref{eq:frag}), which has a more accurate asymptotic behavior in the
limit of $z\to 1$ than the Peterson form~\cite{Peterson:1982ak}.
The charm fragmentation function is folded with the NLO charm cross-section~\cite{Alekhin:2008mb}
through the re-weighting procedure implemented for all our MC events.
This procedure allows a variation of the free parameter $\varepsilon_{\rm c}$ in the
Collins-Spiller fragmentation function, together with the charm production parameters
in the charm cross-section. 

We use two kinematic distributions to detemine
$\varepsilon_{\rm c}$ from our data: the energy of the secondary muon from charm decay,
$E_{\mu_c}$, and the scaling variable $x_{Bj}$. The first one gives the most
sensitive variable to fragmentation and has the advantage that it is largely
independent from the details of the development of the hadronic shower inside FCAL,
since the muon momentum is precisely measured in the drift chambers.
Some additional sensitivity can be obtained from $x_{Bj}$, while the
remaining kinematic variables do not add any substantial contribution.
We perform a simultaneous fit to both $E_{\mu_c}$ and $x_{Bj}$ by varying
$\varepsilon_{\rm c}$ on an event-by-event basis in our MC re-weighting.
The results are shown in Table~\ref{tab:fragfit}.
The correlation with the mass of the charm quark $m_c$ and with the strange
sea parameterization turns out to be small in our fit since we mostly rely
on the energy of the secondary muon $E_{\mu_c}$ to extract $\varepsilon_{\rm c}$.
A two-dimensional fit to $E_{\mu_c}$ and $x_{Bj}$ with both $\varepsilon_{\rm c}$ and
$m_c$ as free parameters results indeed in a $\chi^2$ surface which is flat
as a function of $m_c$.

\begin{table}[htp]
\begin{centering}
\begin{tabular}{l|ccc} \hline
Experiment & NOMAD ($E_{\mu_c}$, $x_{Bj}$)~~~~~ & E531 ($z_{\rm C}$)~~~~~ & NOMAD + E531  \\ \hline\hline
$\varepsilon_{\rm c}$ & $0.165^{+0.035}_{-0.029}$ & $0.169 \pm 0.036$  & $0.165 \pm 0.025$  \\ \hline
\end{tabular}
\caption{\it Best fit values for the Collins-Spiller fragmentation parameter 
obtained from the NOMAD dimuon data, from the E531 charm emulsion data~\cite{Ushida:1988ru}, and from 
the combined data set.} 
\label{tab:fragfit}
\end{centering}
\end{table}

In order to reduce the uncertainty in the determination of $\varepsilon_{\rm c}$, we
also consider the direct measurement of charm production performed in nuclear emulsions 
by the E531~\cite{Ushida:1988ru} experiment, which had an average 
neutrino energy comparable to NOMAD. This latter condition is crucial for our analysis,  
since the distribution of the measured fragmentation variable is expected to become softer 
by increasing the center of mass energy, due to gluon radiation. 
We re-fit the $z_{\rm c}$ distribution of the
charmed mesons~\footnote{\it In nuclear emulsions the fragmentation variable $z_c$ is defined as the ratio 
between the energy of the primary charmed hadron and the energy trasfer $\nu$. This variable is different from 
the $z$ measured in NOMAD, which is the ratio between the energy of the secondary muon originated in the 
semileptonic decay of the charmed hadrons and the energy transfer $\nu$.} 
published by E531 with the Collins-Spiller function.
The value of $\varepsilon_{\rm c}$ that we obtain from E531 data is
shown in Table~\ref{tab:fragfit} and is in good agreement with the value
from the NOMAD analysis. We then use both the NOMAD and E531 data
in a combined fit and obtain:
\begin{equation} 
\label{eqn:epsilonc}
\varepsilon_{\rm c} = 0.165 \pm 0.025 \pm 0.002
\end{equation} 
where the first uncertainty is due to statistics and the second one to systematics 
(see Section~\ref{sec:syst}). This is our final result, which will be used in the following analysis.
It is interesting to note that this result is consistent with the 
expectation $\varepsilon_{\rm c} = \langle k_T^2 \rangle /m_c^2$~\cite{Collins:1984ms} where 
$\langle k_T^2 \rangle$ represents the size of the hadron in momentum space and $m_c$ is the 
charm quark mass (see Section~\ref{sec:charmfit}). 

For the sake of comparison with previous 
measurements, we also performed a fit to the combined NOMAD and E531 data sets with the 
Peterson parameterization~\cite{Peterson:1982ak} of the charm fragmentation function and 
obtained $\varepsilon_{\rm p} = 0.068^{+0.009}_{-0.008} \pm 0.001$.   

\subsection{Charm semileptonic branching ratio $B_\mu$}
\label{sec:bmu} 

The effective charm semileptonic branching ratio $B_\mu$ is expected to increase with increasing 
neutrino energy because the charmed fractions $f_h$ are themselves neutrino energy dependent. 
This in turn is due to the lower branching ratio to muon of the $\Lambda_c^+$ and $D_s^+$ 
coupled with the fact that quasi-elastic $\Lambda_c^+$ and diffractive $D_s^+$ production 
make a more important contribution at low energies. Since NOMAD dimuon data extend down to
$E_{\nu} \sim 6$ GeV we need to take into account the energy dependence of $B_\mu$
in our analysis.

\begin{figure}[htb]
 \begin{center}
  \mbox{\epsfig{file=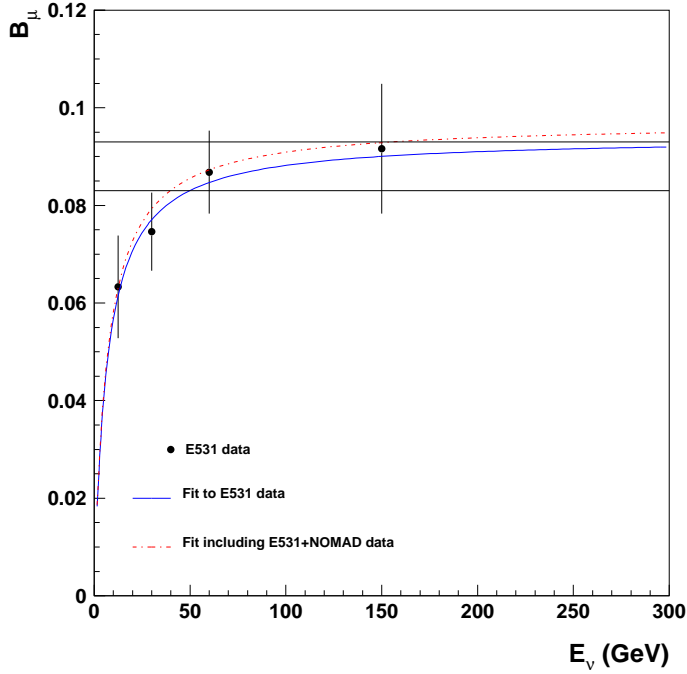,width=0.7\linewidth}}
  \caption {\it The semileptonic branching ratio $B_{\mu}$ as a function of the neutrino energy. 
The solid curve represents a fit to the E531 emulsion data (points with error bars) with the
function in Eq.~(\ref{eqn:bmu}), while the dashed-dotted line is the result of our global fit  
including E531 and NOMAD data (see Section~\ref{sec:charmfit}). The horizontal lines show the $\pm 1 \sigma$ band
obtained from a fit to NuTeV and CCFR charm dimuon data assuming a  value of
$B_\mu$ independent of energy~\cite{Alekhin:2008mb}.}
  \label{fig:bmu}
 \end{center}
\end{figure}

The only existing measurement of the charmed fractions $f_h$ as a function of the
neutrino energy comes from a re-analysis~\cite{Bolton:1997pq} of the data from the
E531 experiment~\cite{Ushida:1988rt,Ushida:1988ru}. A new determination of
$B_\mu$ as a function of the neutrino energy was obtained in Ref.~\cite{Alekhin:2008mb}
from the E531 emulsion data. We fit the data from Ref.~\cite{Alekhin:2008mb}
with the following smooth function:
\begin{equation}
\label{eqn:bmu}
B_{\mu} (E_{\nu}) = \frac{a}{1 + b/E_{\nu}}
\end{equation}
which has two free parameters $a$ and $b$. The values of the parameters obtained from
a fit to E531 data are given in Table~\ref{tab:bmufit}. Figure~\ref{fig:bmu} shows the
results of the fit together with the $1\sigma$
uncertainty band on $B_\mu$ obtained from a global energy independent fit to the NuTeV and CCFR charm dimuon data
for $E_\nu>30$ GeV~\cite{Alekhin:2008mb}. The E531 data are consistent with the
constant value of $B_\mu$ extracted at high energy from NuTeV and CCFR data in the
common energy range.
In the following we use Eq.~(\ref{eqn:bmu}) to parameterize
$B_\mu$ in our analysis. If we add the NOMAD dimuon data to the E531 data
and extract the $a$ and $b$ parameters from the corresponding fit 
(described in Section~\ref{sec:charmfit}) we
substantially reduce the uncertainties on $a$ and $b$, as can be
seen from Table~\ref{tab:bmufit}.

\begin{table}[htp]
\begin{centering}
\begin{tabular}{l|ccc} \hline
Experiment & $a$ & $b$ (GeV)  \\ \hline\hline
E531 & $0.094 \pm 0.010$  & ~~~~~$6.6 \pm 3.9$  \\
E531+NOMAD &  $0.097 \pm 0.003$  & ~~~~~$6.7 \pm 1.8$  \\ \hline 
\end{tabular}
\caption{\it Coefficients of the energy dependent function used to parameterize $B_\mu$
obtained from the E531 charm emulsion data~\cite{Ushida:1988ru,Alekhin:2008mb} and from 
the combined fit including NOMAD dimuon data (see Section~\ref{sec:charmfit}).} 
\label{tab:bmufit}
\end{centering}
\end{table}

\subsection{Ratios ${\mathcal R}_{\mu \mu}$}

\begin{figure}[p]
 \begin{center}
   \mbox{\epsfig{file=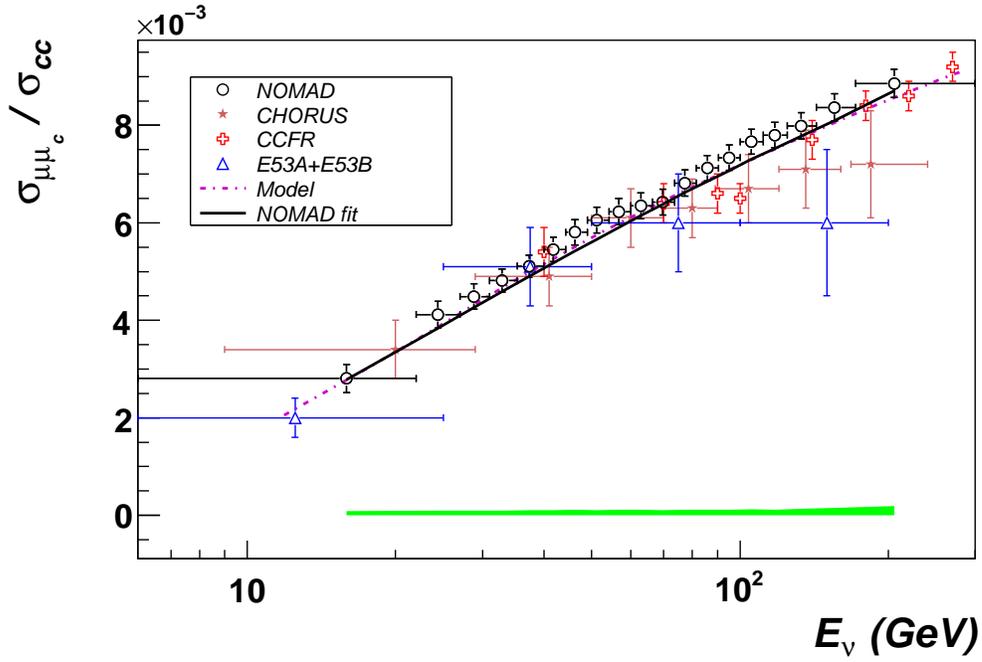,width=1.0\linewidth}} 
   \mbox{\epsfig{file=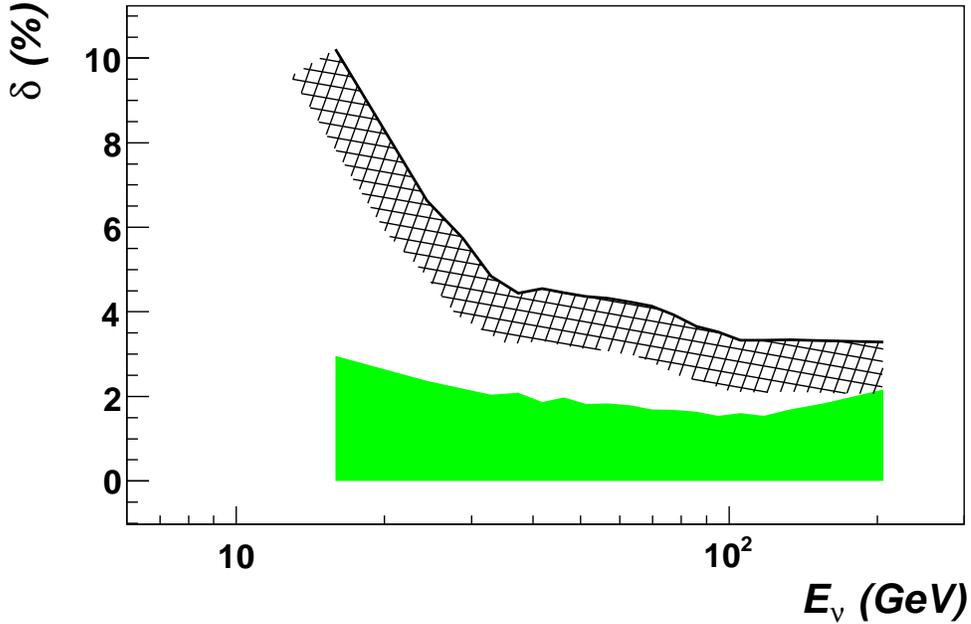,width=1.0\linewidth}}
  \caption {\it Final ratio ${\mathcal{R}}_{\mu\mu}$ between charm dimuon cross-section and inclusive
$\nu_\mu$ CC cross-section as a function of the neutrino energy. Both statistical
and systematic uncertainties are shown and a bin centering correction was applied.
The bottom plot gives the relative statistical (black curve) and systematic (green band)
uncertainties in percentage. 
The solid curve represents the result of our QCD fit to NOMAD and E531 data, while the dashed line 
describes an analytical calculation fully independent from NOMAD data and
based upon the cross-section model of Section~\ref{sec:MCwei}. 
A comparison with previous
measurements~\cite{Bazarko:1994tt,:2008xp,Baker:1985} is also given in the top plot for completeness.}
  \label{fig:prodcharmfull1}
 \end{center}
\end{figure}

\begin{figure}[p]
 \begin{center}
   \mbox{\epsfig{file=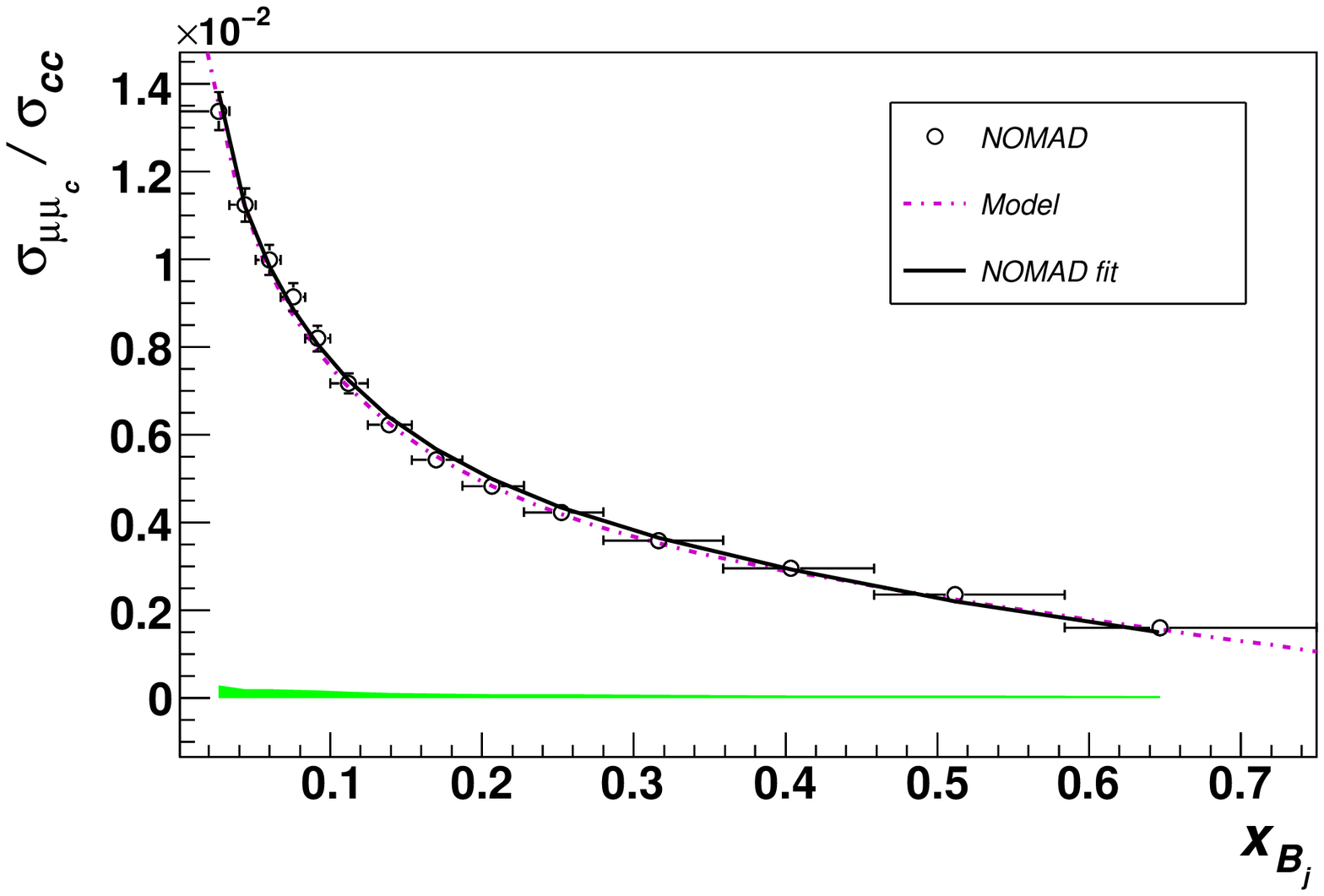,width=1.0\linewidth}} 
   \mbox{\epsfig{file=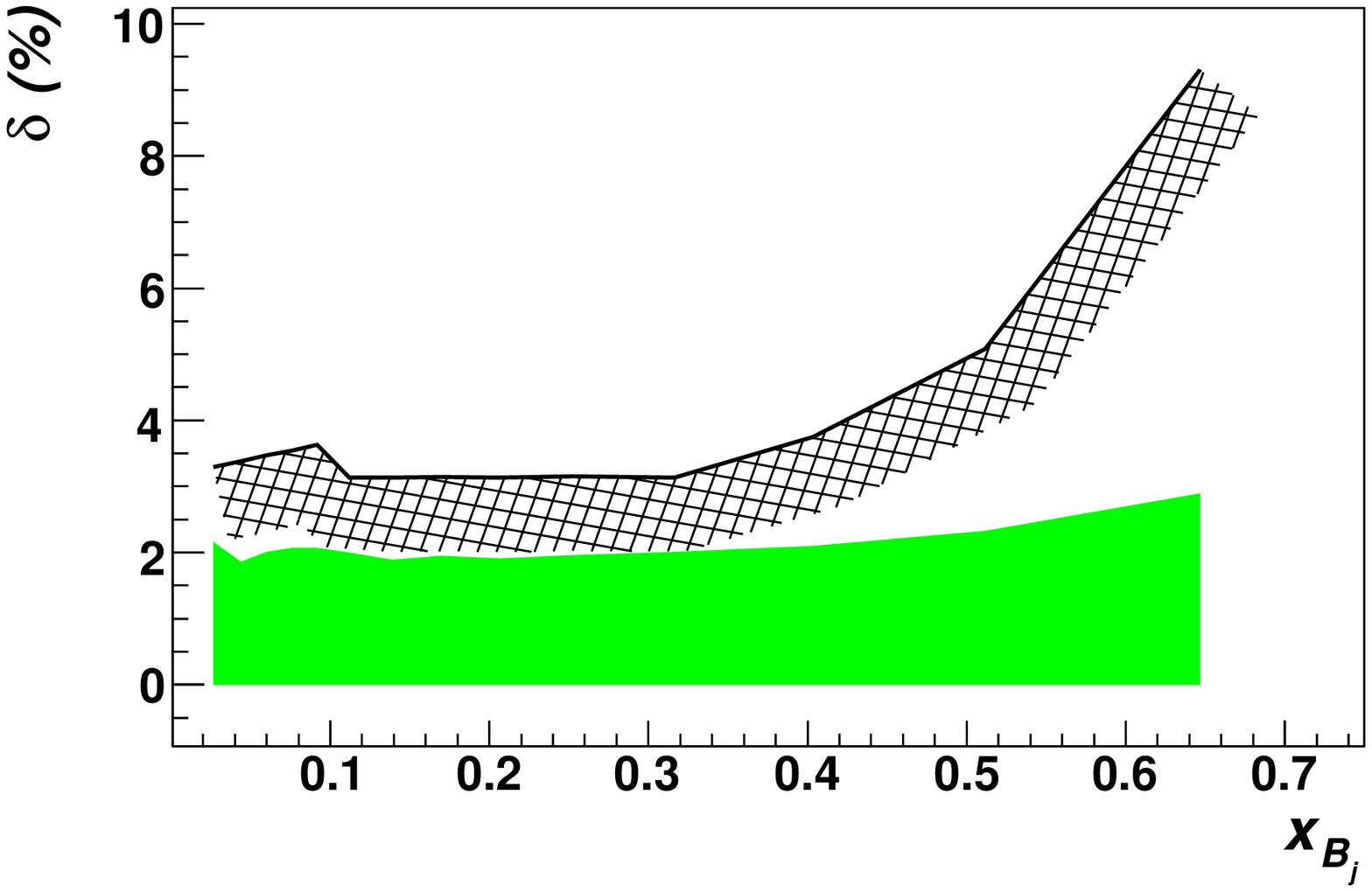,width=1.0\linewidth}}
  \caption {\it Final ratio ${\mathcal{R}}_{\mu\mu}$ between charm dimuon cross-section and inclusive
$\nu_\mu$ CC cross-section as a function of $x$-Bjorken. Both statistical
and systematic uncertainties are shown and a bin centering correction was applied.
The bottom plot gives the relative statistical (black curve) and systematic (green band)
uncertainties in percentage. 
The solid curve represents the result of our QCD fit to NOMAD and E531 data, while the dashed line 
describes an analytical calculation fully independent from NOMAD data and
based upon the cross-section model of Section~\ref{sec:MCwei}. 
}  
  \label{fig:prodcharmfull2}
 \end{center}
\end{figure}

\begin{figure}[p]
 \begin{center}
   \mbox{\epsfig{file=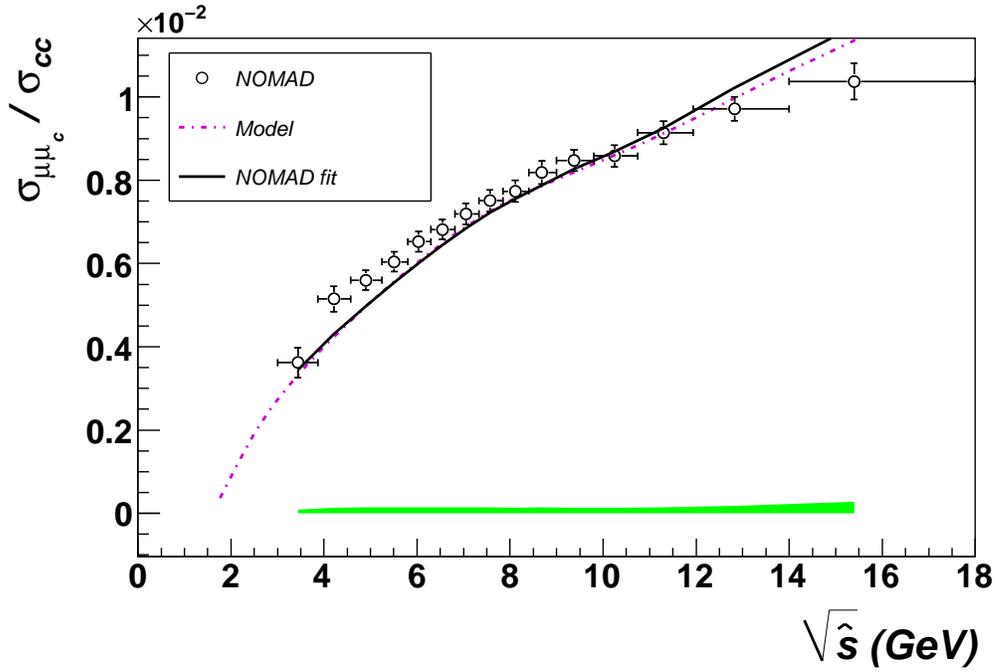,width=1.0\linewidth}} 
   \mbox{\epsfig{file=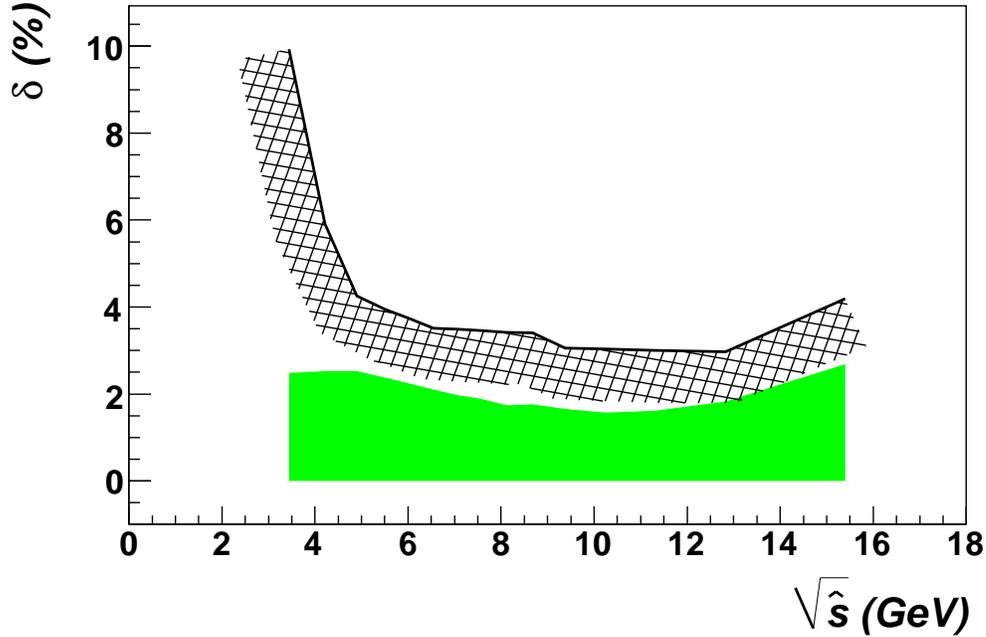,width=1.0\linewidth}}
  \caption {\it Final ratio ${\mathcal{R}}_{\mu\mu}$ between charm dimuon cross-section and inclusive
$\nu_\mu$ CC cross-section as a function of the center of mass energy $\sqrt{\hat{s}}$.
Both statistical
and systematic uncertainties are shown and a bin centering correction was applied.
The bottom plot gives the relative statistical (black curve) and systematic (green band)
uncertainties in percentage. 
The solid curve represents the result of our QCD fit to NOMAD and E531 data, while the dashed line 
describes an analytical calculation fully independent from NOMAD data and
based upon the cross-section model of Section~\ref{sec:MCwei}. 
}  
  \label{fig:prodcharmfull3}
 \end{center}
\end{figure}

The final results for the measured ratio ${\mathcal R}_{\mu \mu}=\sigma_{\mu\mu}/\sigma_{cc}$ as a function 
of the kinematic variables $E_\nu, x_{Bj}$ and $\sqrt{\hat{s}}$ are shown in 
Figures~\ref{fig:prodcharmfull1}-\ref{fig:prodcharmfull3}.
The curves represent our model calculation based upon the
global PDF fit including only NuTeV and CCFR data~\cite{Alekhin:2008mb}.
Our new NOMAD measurement is in agreement with the independent predictions
obtained without any input from NOMAD data.

We evaluate the overall average dimuon production rate in NOMAD by integrating
the measured cross-sections after the unfolding and obtain:
\begin{equation} 
\label{eqn:rmumu} 
\frac{\int \sigma_{\mu \mu} (E_\nu) \phi(E_\nu)  dE_\nu}
{\int \sigma_{cc}(E_\nu) \phi(E_\nu) dE_\nu} 
 =  (5.15 \pm 0.05 \pm 0.07) \times 10^{-3} 
\end{equation} 
where the first uncertainty is due to statistics and the second one to systematics (see Section~\ref{sec:syst}). 
After verifying the consistency of the normalization of each kinematic distribution,
we use this average value obtained from ${\mathcal R}_{\mu \mu}(E_\nu)$ to constrain the 
normalization of all the cross-section ratios ${\mathcal R}_{\mu \mu}$.

Tables~\ref{tab:prodcharm_enu_19a}-\ref{tab:prodcharm_shat_15a} summarize our final results for the ratio 
${\mathcal R}_{\mu \mu}$ for $Q^2>1$~GeV$^2$/c$^2$ as a function of the kinematic variables with the 
corresponding statistical and systematic uncertainties.  

\newpage 
\begin{table}[h]
\begin{center}
\small
\begin{tabular}{||c|c|c|c|c||}
\hline
\hline
       $E_{\nu}$ (GeV)     & Bin center &  $\sigma_{\mu\mu}/\sigma_{cc} \pm \delta^{stat} \pm \delta^{syst}$ ($10^{-3}$) & $\delta^{stat}$, \% & $\delta^{syst}$, \% \\
\hline
   6.000   -  22.00    &   15.91    &  2.807 $\pm$ 0.287 $\pm$ 0.083             &   10.22    &   3.05    \\
   22.00    -  27.00    &   24.38    &  4.118 $\pm$ 0.273 $\pm$ 0.098             &   6.63    &   2.39    \\
   27.00    -  31.00    &   28.85    &  4.489 $\pm$ 0.257 $\pm$ 0.098             &   5.73    &   2.19    \\
   31.00    -  35.34    &   32.88    &  4.815 $\pm$ 0.233 $\pm$ 0.098             &   4.85    &   2.05    \\
   35.34    -  40.00    &   37.31    &  5.113 $\pm$ 0.227 $\pm$ 0.107             &   4.44    &   2.08    \\
   40.00    -  44.27    &   41.78    &  5.453 $\pm$ 0.248 $\pm$ 0.102             &   4.55    &   1.88    \\
   44.27    -  48.97    &   46.23    &  5.807 $\pm$ 0.259 $\pm$ 0.115             &   4.46    &   1.99    \\
   48.97    -  54.17    &   51.17    &  6.056 $\pm$ 0.265 $\pm$ 0.111             &   4.37    &   1.83    \\
   54.17    -  59.98    &   56.73    &  6.227 $\pm$ 0.269 $\pm$ 0.114             &   4.32    &   1.84    \\
   59.98    -  66.40    &   62.87    &  6.348 $\pm$ 0.269 $\pm$ 0.113             &   4.23    &   1.79    \\
   66.40    -  73.61    &   69.70    &  6.425 $\pm$ 0.266 $\pm$ 0.109             &   4.14    &   1.70    \\
   73.61    -  81.47    &   77.29    &  6.816 $\pm$ 0.268 $\pm$ 0.115             &   3.93    &   1.69    \\
   81.47    -  90.37    &   85.78    &  7.121 $\pm$ 0.260 $\pm$ 0.116             &   3.66    &   1.64    \\
   90.37    - 100.0     &   95.01    &  7.337 $\pm$ 0.259 $\pm$ 0.113             &   3.53    &   1.55    \\
  100.0     - 111.4     &  105.4     &  7.660 $\pm$ 0.255 $\pm$ 0.123             &   3.33    &   1.60    \\
  111.4     - 124.7     &  117.6     &  7.800 $\pm$ 0.260 $\pm$ 0.120             &   3.33    &   1.53    \\
  124.7     - 142.9     &  133.0     &  7.989 $\pm$ 0.267 $\pm$ 0.135             &   3.34    &   1.69    \\
  142.9     - 171.4     &  155.4     &  8.368 $\pm$ 0.278 $\pm$ 0.153             &   3.32    &   1.83    \\
  171.4     - 300.0     &  205.5     &  8.859 $\pm$ 0.292 $\pm$ 0.192             &   3.29    &   2.17    \\
\hline
\hline
\end{tabular}
\normalsize
\caption {\it Measured ${\mathcal{R}_{\mu \mu}}$ as a function of visible neutrino energy $E_{\nu}$ including both statistical and total systematic uncertainties. The last two columns provide the corresponding relative uncertainties.
}   
\label{tab:prodcharm_enu_19a}
\end{center}
\end{table}

\newpage 
\begin{table}[h]
\begin{center}
\small
\begin{tabular}{||c|c|c|c|c||}
\hline
\hline
       $x_{\rm Bj}$      & Bin center &  $\sigma_{\mu\mu}/\sigma_{cc} \pm \delta^{stat} \pm \delta^{syst}$ ($10^{-3}$) & $\delta^{stat}$, \% & $\delta^{syst}$, \% \\
\hline
    0.0000 -   0.0336 &    0.0267 & 13.383 $\pm$ 0.441 $\pm$ 0.289             &   3.30    &   2.16    \\
    0.0336 -   0.0511 &    0.0440 & 11.245 $\pm$ 0.380 $\pm$ 0.210             &   3.38    &   1.87    \\
    0.0511 -   0.0672 &    0.0598 &  ~9.991 $\pm$ 0.347 $\pm$ 0.201             &   3.47    &   2.03    \\
    0.0672 -   0.0836 &    0.0756 &  ~9.141 $\pm$ 0.324 $\pm$ 0.189             &   3.55    &   2.08    \\
    0.0836 -   0.1000  &    0.0917 &  ~8.198 $\pm$ 0.297 $\pm$ 0.169             &   3.63    &   2.08    \\
    0.1000  -   0.1246  &    0.1122  &  ~7.176 $\pm$ 0.225 $\pm$ 0.144             &   3.13    &   2.02    \\
    0.1246  -   0.1535  &    0.1389  &  ~6.229 $\pm$ 0.195 $\pm$ 0.118             &   3.14    &   1.90    \\
    0.1535  -   0.1870  &    0.1699  &  ~5.427 $\pm$ 0.171 $\pm$ 0.106             &   3.15    &   1.96    \\
    0.1870  -   0.2277  &    0.2066  &  ~4.837 $\pm$ 0.151 $\pm$ 0.093             &   3.13    &   1.92    \\
    0.2277  -   0.2800  &    0.2524  &  ~4.235 $\pm$ 0.133 $\pm$ 0.083             &   3.15    &   1.97    \\
    0.2800  -   0.3590  &    0.3165  &  ~3.595 $\pm$ 0.113 $\pm$ 0.072             &   3.13    &   2.02    \\
    0.3590  -   0.4583  &    0.4036  &  ~2.955 $\pm$ 0.111 $\pm$ 0.062             &   3.75    &   2.11    \\
    0.4583  -   0.5838  &    0.5116  &  ~2.355 $\pm$ 0.120 $\pm$ 0.055             &   5.08    &   2.36    \\
    0.5838  -   0.7500  &    0.6465  &  ~1.607 $\pm$ 0.150 $\pm$ 0.047             &   9.31    &   2.96    \\
\hline
\hline
\end{tabular}
\normalsize
\caption {\it Measured ${\mathcal{R}_{\mu \mu}}$ as a function of $x_{\rm Bj}$ including both statistical and total systematic uncertainties. The last two columns provide the corresponding relative uncertainties.
}   
\label{tab:prodcharm_xbj_14a}
\end{center}
\end{table}

\newpage 
\begin{table}[h]
\begin{center}
\small
\begin{tabular}{||c|c|c|c|c||}
\hline
\hline
       $\sqrt{\hat{s}}$ (GeV)     & Bin center &  $\sigma_{\mu\mu}/\sigma_{cc} \pm \delta^{stat} \pm \delta^{syst}$ ($10^{-3}$) & $\delta^{stat}$, \% & $\delta^{syst}$, \% \\
\hline
    3.000   -   3.870   &    3.440   &  3.620 $\pm$ 0.360 $\pm$ 0.090             &   9.93    &   2.66    \\
    3.870   -   4.570   &    4.213   &  5.148 $\pm$ 0.304 $\pm$ 0.130             &   5.91    &   2.60    \\
    4.570   -   5.250   &    4.897   &  5.600 $\pm$ 0.238 $\pm$ 0.142             &   4.26    &   2.55    \\
    5.250   -   5.800   &    5.509   &  6.041 $\pm$ 0.239 $\pm$ 0.144             &   3.95    &   2.39    \\
    5.800   -   6.301   &    6.035   &  6.523 $\pm$ 0.244 $\pm$ 0.146             &   3.74    &   2.24    \\
    6.301   -   6.818   &    6.543   &  6.815 $\pm$ 0.239 $\pm$ 0.144             &   3.51    &   2.11    \\
    6.818   -   7.326   &    7.049   &  7.190 $\pm$ 0.251 $\pm$ 0.142             &   3.49    &   1.97    \\
    7.326   -   7.849   &    7.567   &  7.507 $\pm$ 0.260 $\pm$ 0.142             &   3.46    &   1.89    \\
    7.849   -   8.407   &    8.110   &  7.738 $\pm$ 0.264 $\pm$ 0.135             &   3.41    &   1.74    \\
    8.407   -   9.000   &    8.683   &  8.187 $\pm$ 0.278 $\pm$ 0.144             &   3.40    &   1.76    \\
    9.000   -   9.801   &    9.375   &  8.475 $\pm$ 0.259 $\pm$ 0.140             &   3.06    &   1.65    \\
    9.801   -  10.74    &   10.24    &  8.583 $\pm$ 0.261 $\pm$ 0.135             &   3.04    &   1.57    \\
   10.74    -  11.93    &   11.30    &  9.142 $\pm$ 0.274 $\pm$ 0.147             &   3.00    &   1.62    \\
   11.93    -  14.00    &   12.82    &  9.713 $\pm$ 0.289 $\pm$ 0.177             &   2.97    &   1.84    \\
   14.00    -  18.00    &   15.39    & 10.373 $\pm$ 0.435 $\pm$ 0.279             &   4.19    &   2.70    \\
\hline
\hline
\end{tabular}
\normalsize
\caption {\it Measured ${\mathcal{R}_{\mu \mu}}$ as a function of the center of mass energy $\sqrt{\hat{s}}$ including both statistical and total systematic uncertainties. The last two columns provide the corresponding relative uncertainties.
}   
\label{tab:prodcharm_shat_15a}
\end{center}
\end{table}
\newpage

\section{Systematic uncertainties}
\label{sec:syst}  

The use of the {\it ratio} ${\mathcal R}_{\mu \mu}$ allows a substantial reduction of 
systematic uncertainties since all the effects related to both the numerator (charm 
dimuons) and the denominator (inclusive $\nu_\mu$ CC) largely cancel out. 
This cancellation applies to the experimental systematics as well as to the model 
systematics. 

Figures~\ref{fig:prodcharmfull1}-\ref{fig:prodcharmfull3} show the final total systematic uncertainties, 
which are smaller than the statistical ones for all bins and kinematic variables.  
In most cases the overall systematic uncertainty can be kept below 2\%, as 
summarized in Tables~\ref{tab:prodcharm_enu_19a}-\ref{tab:prodcharm_shat_15a}. 
The total systematic uncertainty is 
dominated by the contributions directly related to the determination of the 
charm dimuon signal and therefore affecting only the numerator of the 
ratio ${\mathcal R}_{\mu \mu}$: background scale, charm fragmentation and mass 
of the charm quark $m_c$. The first one enters directly into the background 
subtraction procedure, while the last two enter only through the acceptance 
correction. We note all these three systematic uncertainties are actually determined by the 
limited statistics of our data samples and therefore they could be improved in future 
measurements using high resolution detectors similar to NOMAD~\cite{HiResMnu}. 

A detailed breakdown of the contributions from each source of systematic 
uncertainty is given in Tables~\ref{tab:prodcharmsyst_enu_19a}-\ref{tab:prodcharmsyst_shat_15a}.   
The sign in front of the numbers refers to a variation of $+1\sigma$ of the 
corresponding effect and shows the bin-to-bin correlation.  
The magnitude of each systematic uncertainty is estimated as the average between 
positive ($+1\sigma$) and negative ($-1\sigma$) variations of the relevant parameters. 
We also change the number of bins in each kinematic variable from the nominal 
value to 25 and 45 (three complete estimates) in order to check potential 
biases related to the bin size.  

In the following we will describe the procedure used to estimate 
the systematic uncertainties.

\subsection{Variation of the analysis cuts}

As discussed in Section~\ref{sec:cuts}, there is a good agreement between data and 
MC for the variables used in the selection procedure and for the final kinematic 
distributions in both charm dimuon events and inclusive $\nu_\mu$ CC events.  
Therefore, the systematic uncertainty associated to each analysis cut is 
essentially defined by the experimental resolution of the relevant variable 
for values close to the chosen cut. We evaluate the experimental resolutions 
from the difference between reconstructed and simulated variables in MC events 
close to the chosen cuts. We fit the corresponding distributions with 
Gaussian functions and vary each cut by the resulting standard deviation 
from the fit.  

The effect of a variation of the analysis cuts according to the experimental 
resolution is very small on the ratio ${\mathcal R}_{\mu \mu}$. The following 
effects have been taken into account:   
\begin{itemize} 
\item[$\delta_1$] $|x^{PV}_{ext}|<80$~cm. \\
                  We find $\delta (x^{PV}_{ext}) \simeq 0.6$ cm (0.75\%). 
\item[$\delta_2$] $|y^{PV}_{ext}|<90$~cm. \\
                  We find $\delta (y^{PV}_{ext}) \simeq 0.7$ cm (0.75\%). 
\item[$\delta_3$] Time correlation between two muons less than $5$~ns. \\ 
                  The timing of the muons is provided by the $t_0$ measurement at the 
                  first track hit in the drift chambers resulting 
                  in $\delta(t_0) \sim 1$ ns (20\%).  
\item[$\delta_4$] Energy of the current muon more than $3$~GeV. \\
                  We find $\delta (E_{\mu_{cc}}) = 162$ MeV (5.4\%). 
\item[$\delta_5$] Energy of the secondary muon from charm decay $E_{\mu_c}> 3$~GeV ($E_{had} > 3$~GeV). \\ 
                  We find $\delta (E_{\mu_c}) = 165$ MeV (5.5\%). 
\item[$\delta_6$] $Q^2 > 1$~GeV$^2/$c$^2$. \\  
                  We find $\delta (Q^2) = 0.30$ GeV$^2$ (30\%). 
\item[$\delta_7$] $E_{had} - E_{\mu^+} < 100$~GeV. \\ 
                  According to the FCAL energy resolution $\sigma(E)/E = 104\%/\sqrt{E}$ we find 
                  $\delta (E_{had} - E_{\mu^+}) = 10.4$ GeV at 100 GeV (10\%).  
\end{itemize}

\subsection{Energy scales and flux}   
\label{sec:flux} 

The impact of the energy scales and flux uncertainties on the ratio ${\mathcal R}_{\mu \mu}$ 
is very small due to the large cancellation between charm dimuon events and $\nu_\mu$ CC 
events. 

\begin{itemize} 
\item[$\delta_8$] Muon energy scale. \\ 
                  The measurement of the muon momentum is performed by fitting the 
                  curvature of the track in the low density tracking region equipped 
                  with drift chambers (DCH).
		  The $E_\mu$ scale was determined by a precise B-field mapping and 
                  DCH alignment accomplished by using several million beam muons traversing 
                  the detector throughout the neutrino runs. The momentum scale was
                  checked by using the invariant mass of over 30000 reconstructed
                  $K^0 \rightarrow \pi^+ \pi^-$ decays in the CC and NC data. 
                  The systematic uncertainty on the $E_\mu$ scale 
                  from DCH was determined to be 0.2\%.

                  \noindent 
                  The momentum of the muons at the first hit of the track in DCH is extrapolated 
                  back to the position of the primary vertex in FCAL by adding the 
                  corresponding energy loss in the FCAL material. Assuming a uniform 
                  distribution of the vertex position within each stack, we obtain a     
                  corresponding uncertainty of $\Delta E_{\rm loss} / \sqrt{12}$ due to 
                  the variable amount of material traversed by the muon. 
                  This contribution is dominant over the $E_\mu$ scale uncertainty from DCH 
                  at low energy. However, in our analysis we always assign to the event a 
                  {\em fixed z position} equal to the middle point of the stack in which the primary 
                  vertex is located. This fact implies that {\em on average} the energy loss in the 
                  FCAL material is correctly taken into account. Therefore, the overall 
                  uncertainty on the $E_\mu$ energy scale is basically defined by the DCH 
                  contribution.

\item[$\delta_9$] Hadronic energy scale.  \\ 
		  For the estimate of the systematic uncertainty on the global hadronic 
                  energy scale we start from the results obtained in Section~\ref{sec:ehadcal}. 
                  We repeat the calibration of the hadronic energy scale after 
                  restricting the fiducial volume to $\mid x,y \mid < 70$ cm, after removing the 
                  last stack (stack 4), and after 
                  changing the kinematic cuts used in the selection. We then compare 
                  the variations observed for the scale factor with the uncertainty 
                  band obtained in the original fit with $\Delta \chi^2 = 1$. 
                  We define our final uncertainty 
                  band by taking an outer envelope over the average of positive ($+1\sigma$) 
                  and negative ($-1\sigma$) variations. This band is consistent 
                  with the variations observed after changing the analysis cuts.  
                  The size of the $E_{\rm had}$ scale uncertainty goes from 0.3\% in the 
                  first energy bin to about 1\% in the last energy bin.      

\item[$\delta_{10}$] Beam flux prediction. \\ 
                     In our analysis we use a beam flux calculation based upon 
                     Ref.~\cite{Astier:2003rj}. The spectra for FCAL are 
                     slightly harder due to the restricted transverse size of the 
                     FCAL fiducial volume. We obtain the FCAL flux by applying the 
                     same fiducial cuts used in our analysis to the flux calculation 
                     of Ref.~\cite{Astier:2003rj} and we use the flux uncertainty  
                     as a function of energy from Ref.~\cite{Astier:2003rj} to estimate 
                     the corresponding uncertainty of the FCAL flux.  
                     Figure~\ref{fig:fcalflux} shows the expected $\nu_\mu$ flux 
                     within the fiducial volume of FCAL used in this analysis.

\begin{figure}[htb]
 \begin{center}
  \begin{tabular}{cc}
   \mbox{\epsfig{file=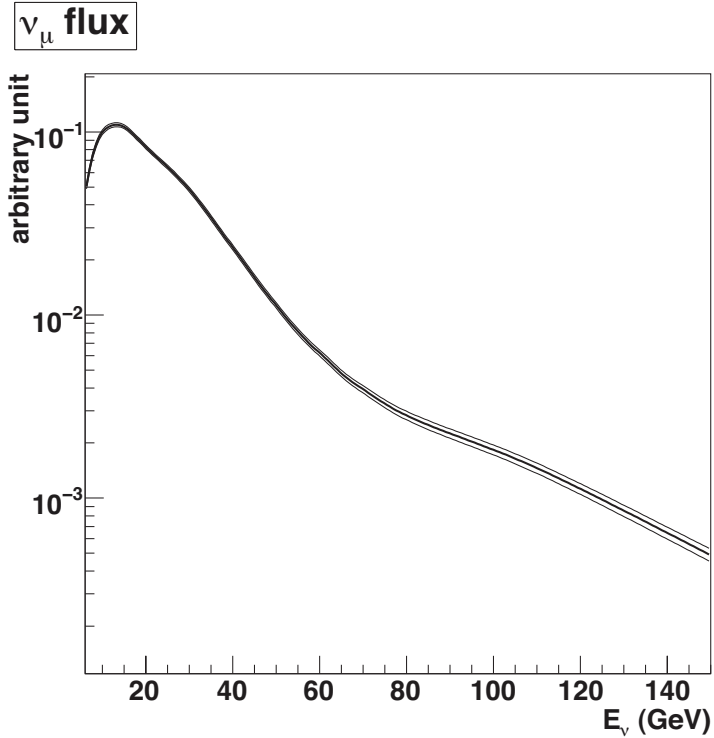,width=0.75\linewidth}} &
  \end{tabular}
  \caption {\it Flux predictions for $\nu_\mu$ in the FCAL~\cite{Astier:2003rj}.}
  \label{fig:fcalflux}
 \end{center}
\end{figure}

\end{itemize} 

\subsection{Model systematic uncertainties}

The modeling of the charm dimuon production and the background subtraction 
procedure in the dimuon sample are the dominant sources of systematic 
uncertainties. Other systematic effects related to the modeling of the structure functions 
affecting both the charm dimuon sample and the inclusive $\nu_\mu$ CC sample 
give very small contributions. For each contribution, we repeat the complete analysis 
after changing the relevant paramenters by $\pm 1\sigma$. In the following we 
describe the variations of the parameters used.    

\begin{itemize}
\item[$\delta_{11}$] Background scale.  \\ 
                     As discussed in Section~\ref{sec:analysis}, our background subtraction  
                     procedure in entirely based upon NOMAD data. The uncertainty on the 
                     background scale, i.e. on the ratio 
                     $N_{\mbox{\tiny $\mu\mu^+_{bg}$}} / N_{\mbox{\tiny $\mu\mu^-$}}$, 
                     is constrained by the measurement of the ratio $h^+/h^-$ of positive 
                     to negative hadrons in DCH. We use the uncertainty band from the fit 
                     to the measured ratio, including the full correlation matrix, 
                     as shown in Figure~\ref{fig:had_dch}. 
  
\item[$\delta_{12}$] Charm fragmentation. \\  
                     We vary the $\varepsilon_{\rm c}$ parameter in the Collins-Spiller fragmentation 
                     function within the uncertainty obtained from the fit to NOMAD+E531 data, 
                     $\pm \Delta \varepsilon_{\rm c} = 0.025$ (see Section~\ref{sec:charmfrag}). 

                     \noindent 
                     The acceptance correction is obtained from the MC simulation, 
                     which takes into account the decay of each charmed hadron according 
                     to recently measured branching ratios~\cite{PDG12}. The charmed fractions $f_h$ 
                     are the ones provided by the E531 data~\cite{Bolton:1997pq}.  
                     We varied the charmed fractions $f_h$ and the muon branching ratios 
                     of individual hadrons within their experimental uncertainties in 
                     our MC. The corresponding changes in the acceptance correction 
                     were found to be negligible.  

\item[$\delta_{13}$] Mass of charm quark $m_c$. \\ 
                     We vary the value of the mass of the charm quark by 
                     $\pm \Delta m_c = 75$ MeV/c$^2$, which is the uncertainty band 
                     obtained from the global PDF fit of Ref.~\cite{Alekhin:2008mb} to NOMAD dimuon 
                     data as discussed in Section~\ref{sec:charmfit}.   
\item[$\delta_{14}$] Structure functions (Leading Twist). \\    
                     We change {\em all} parton density functions obtained from 
                     a global fit to charged lepton DIS data, CHORUS (anti)neutrino 
                     DIS data, NuTeV and CCFR dimuon data and Drell-Yan data within 
                     their uncertainties~\cite{Alekhin:2008mb,Alekhin:2007fh}. 
                     The variations include strange quark sea distributions.  
\item[$\delta_{15}$] High twists. \\ 
                     We include twist-4 power corrections to the neutrino 
                     structure functions following the results of Ref.~\cite{Alekhin:2007fh}. 
                     For $F_2$ and $F_T$ the High Twist contributions are obtained 
                     from charged lepton scattering DIS data after rescaling for the 
                     quark charges (18/5). For $xF_3$ the twist-4 term is obtained from 
                     the (anti)neutrino differential cross-section measured by the CHORUS 
                     experiment. We use the uncertainties obtained from the global 
                     fits of Refs.~\cite{Alekhin:2007fh,Alekhin:2008mb} to estimate 
                     the systematic uncertainties related to High Twists.  
\item[$\delta_{16}$] Electroweak corrections. \\ 
                     Radiative corrections to neutrino DIS are calculated according 
                     to the code developed for the NOMAD analysis~\cite{Arbuzov-Bardin}. 
                     The measured ratio ${\mathcal R}_{\mu \mu}$ is not corrected for 
                     electroweak radiative effects in order to avoid model dependent 
                     corrections to the data. Therefore, the electroweak corrections 
                     only affect the measurement in an indirect way, through the 
                     detector acceptance. We evaluate the corresponding systematic uncertainties 
                     by varying the electroweak corrections within the uncertainty 
                     range from Ref.~\cite{Arbuzov-Bardin}. 
 
\item[$\delta_{17}$] Nuclear corrections. \\ 
                     We apply nuclear corrections using a detailed 
                     model~\cite{Kulagin:2007ju,Kulagin:2004ie,Kulagin:2010gd}  
                     taking into account a number of different effects including 
                     the nucleon Fermi motion and binding, neutron excess, nuclear shadowing, 
                     nuclear pion excess and the off-shell correction to bound 
                     nucleon structure functions. We use the uncertainties on the corresponding 
                     parameters provided by the analysis of charged lepton 
                     data in Ref.~\cite{Kulagin:2004ie}. The uncertainties include 
                     target mass corrections~\cite{Georgi:1976ve}, which are included 
                     into the nuclear convolution.   
\end{itemize}

\begin{sidetable}[htb]
\begin{center}
\small 
\begin{tabular}{||c|c|c|c|c|c|c|c|c|c|c|c|c|c|c|c|c|c||}
\hline
\hline
 Bin / $\delta^{syst}$, \% & $\delta_{1}$ & $\delta_{2}$ & $\delta_{3}$ & $\delta_{4}$ & $\delta_{5}$ & $\delta_{6}$ & $\delta_{7}$ & $\delta_{8}$ & $\delta_{9}$ & $\delta_{10}$ & $\delta_{11}$ & $\delta_{12}$ & $\delta_{13}$ & $\delta_{14}$ & $\delta_{15}$ & $\delta_{16}$ & $\delta_{17}$ \\
\hline
    6.000   -  22.00    &   0.24  &  -0.44  &   0.66  &   0.14  &  -0.53  &  -1.42  &   0.00  &   0.45  &   0.18  &  -0.18  &  -1.01  &   1.48  &  -1.67  &   0.02  &  -0.04  &  -0.01  &   0.11  \\
   22.00    -  27.00    &   0.22  &  -0.21  &   0.18  &   0.13  &  -0.40  &  -0.58  &   0.00  &   0.26  &   0.28  &  -0.16  &  -0.96  &   1.84  &  -0.78  &   0.00  &   0.02  &  -0.01  &   0.04  \\
   27.00    -  31.00    &   0.20  &  -0.16  &   0.26  &   0.13  &  -0.23  &  -0.28  &   0.00  &  -0.13  &   0.29  &  -0.10  &  -0.89  &   1.81  &  -0.60  &   0.00  &   0.01  &  -0.01  &   0.03  \\
   31.00    -  35.34    &  -0.11  &  -0.06  &   0.24  &   0.13  &  -0.20  &  -0.08  &   0.01  &  -0.17  &   0.25  &  -0.08  &  -0.88  &   1.74  &  -0.41  &   0.00  &   0.02  &  -0.01  &  -0.02  \\
   35.34    -  40.00    &  -0.07  &  -0.08  &   0.25  &   0.10  &  -0.25  &   0.05  &   0.01  &  -0.16  &   0.24  &  -0.09  &  -0.88  &   1.79  &  -0.36  &   0.00  &   0.02  &  -0.01  &   0.02  \\
   40.00    -  44.27    &  -0.09  &  -0.10  &   0.31  &   0.07  &  -0.26  &  -0.03  &   0.02  &  -0.09  &   0.38  &  -0.07  &  -0.90  &   1.49  &  -0.40  &   0.00  &   0.02  &  -0.01  &   0.03  \\
   44.27    -  48.97    &  -0.05  &  -0.03  &  -0.30  &   0.04  &  -0.27  &  -0.13  &   0.03  &  -0.06  &   0.56  &  -0.05  &  -0.93  &   1.57  &  -0.36  &   0.00  &   0.02  &  -0.01  &   0.02  \\
   48.97    -  54.17    &  -0.07  &  -0.06  &  -0.31  &   0.02  &  -0.26  &  -0.20  &   0.03  &   0.06  &   0.42  &  -0.07  &  -0.95  &   1.39  &  -0.34  &   0.00  &   0.01  &  -0.01  &  -0.00  \\
   54.17    -  59.98    &   0.09  &   0.10  &  -0.25  &   0.01  &  -0.27  &  -0.18  &   0.03  &   0.05  &   0.23  &  -0.08  &  -0.99  &   1.44  &  -0.28  &   0.00  &   0.01  &  -0.00  &  -0.00  \\
   59.98    -  66.40    &   0.02  &   0.13  &  -0.21  &   0.01  &  -0.30  &  -0.15  &   0.03  &   0.12  &   0.24  &  -0.07  &  -1.01  &   1.36  &  -0.26  &   0.01  &   0.01  &  -0.01  &   0.02  \\
   66.40    -  73.61    &   0.07  &  -0.15  &  -0.14  &   0.02  &  -0.29  &   0.11  &   0.04  &   0.10  &   0.34  &  -0.03  &  -1.05  &   1.21  &  -0.22  &   0.00  &   0.02  &  -0.01  &   0.03  \\
   73.61    -  81.47    &   0.08  &  -0.10  &  -0.09  &   0.02  &  -0.26  &   0.14  &   0.02  &  -0.04  &   0.28  &  -0.03  &  -1.08  &   1.20  &  -0.22  &   0.00  &   0.01  &  -0.01  &   0.02  \\
   81.47    -  90.37    &   0.23  &  -0.08  &  -0.15  &   0.01  &  -0.24  &   0.13  &   0.04  &   0.03  &   0.28  &   0.06  &  -1.10  &   1.10  &  -0.13  &   0.00  &   0.01  &  -0.00  &   0.02  \\
   90.37    - 100.0     &   0.15  &  -0.11  &  -0.16  &   0.01  &  -0.24  &   0.14  &   0.09  &   0.05  &   0.37  &   0.06  &  -1.13  &   0.89  &  -0.19  &   0.00  &   0.01  &  -0.01  &   0.02  \\
  100.0     - 111.4     &  -0.17  &  -0.08  &  -0.12  &  -0.00  &  -0.26  &   0.15  &   0.14  &   0.05  &   0.27  &   0.06  &  -1.17  &   0.97  &  -0.16  &   0.01  &   0.02  &  -0.00  &  -0.04  \\
  111.4     - 124.7     &  -0.11  &  -0.07  &  -0.13  &  -0.01  &  -0.24  &   0.13  &   0.19  &   0.06  &   0.28  &   0.10  &  -1.20  &   0.81  &  -0.10  &   0.00  &   0.01  &  -0.00  &   0.01  \\
  124.7     - 142.9     &  -0.19  &  -0.12  &  -0.21  &  -0.00  &  -0.23  &   0.15  &   0.59  &   0.04  &   0.34  &   0.05  &  -1.23  &   0.84  &  -0.02  &   0.00  &   0.01  &  -0.00  &   0.01  \\
  142.9     - 171.4     &  -0.15  &  -0.20  &  -0.25  &  -0.00  &  -0.12  &   0.17  &   0.73  &  -0.05  &   0.67  &  -0.14  &  -1.24  &   0.80  &   0.01  &   0.00  &   0.01  &   0.00  &   0.00  \\
  171.4     - 300.0     &  -0.22  &   0.18  &  -0.21  &  -0.00  &   0.11  &   0.20  &  -0.40  &   0.10  &   1.13  &  -0.81  &  -1.27  &   0.88  &   0.18  &   0.00  &   0.03  &   0.01  &   0.01  \\
\hline
\hline
\end{tabular}
\normalsize
\caption {\it Summary of the relative systematic uncertainties (in percentage) on the measurement of ${\mathcal{R}}_{}\mu \mu$ as a function of $E_{\nu}$. Each column gives the relative change $\delta_i$ resulting from a $+1\sigma$ variation of the corresponding parameter $i$.}
\label{tab:prodcharmsyst_enu_19a}
\end{center}
\end{sidetable}

\begin{sidetable}[htb]
\begin{center}
\small 
\begin{tabular}{||c|c|c|c|c|c|c|c|c|c|c|c|c|c|c|c|c|c||}
\hline
\hline
 Bin / $\delta^{syst}$, \% & $\delta_{1}$ & $\delta_{2}$ & $\delta_{3}$ & $\delta_{4}$ & $\delta_{5}$ & $\delta_{6}$ & $\delta_{7}$ & $\delta_{8}$ & $\delta_{9}$ & $\delta_{10}$ & $\delta_{11}$ & $\delta_{12}$ & $\delta_{13}$ & $\delta_{14}$ & $\delta_{15}$ & $\delta_{16}$ & $\delta_{17}$ \\
\hline
    0.0000 -   0.0336 &   0.07  &  -0.40  &   0.18  &  -0.06  &   0.07  &   1.38  &   0.18  &   0.20  &   0.76  &  -0.03  &  -0.62  &   1.18  &   0.24  &  -0.10  &   0.05  &  -0.23  &   0.10  \\
    0.0336 -   0.0511 &   0.18  &   0.05  &  -0.08  &  -0.04  &  -0.21  &  -0.57  &  -0.31  &  -0.21  &   0.56  &  -0.05  &  -0.65  &   1.46  &  -0.26  &  -0.08  &   0.02  &  -0.03  &   0.06  \\
    0.0511 -   0.0672 &  -0.13  &   0.14  &  -0.20  &   0.02  &  -0.29  &  -0.75  &   0.21  &  -0.21  &   0.49  &  -0.13  &  -0.72  &   1.50  &  -0.51  &  -0.07  &   0.02  &  -0.03  &   0.07  \\
    0.0672 -   0.0836 &  -0.03  &   0.13  &  -0.31  &   0.04  &  -0.35  &  -0.71  &   0.15  &   0.05  &   0.51  &  -0.19  &  -0.82  &   1.51  &  -0.56  &  -0.06  &   0.02  &  -0.03  &   0.04  \\
    0.0836 -   0.1000  &  -0.08  &  -0.17  &  -0.34  &   0.07  &  -0.34  &  -0.57  &  -0.35  &   0.15  &   0.50  &  -0.16  &  -0.90  &   1.45  &  -0.64  &  -0.04  &   0.02  &  -0.03  &   0.06  \\
    0.1000  -   0.1246  &  -0.09  &  -0.16  &  -0.25  &   0.09  &  -0.25  &  -0.43  &  -0.47  &   0.21  &   0.45  &  -0.21  &  -0.97  &   1.39  &  -0.59  &  -0.03  &   0.02  &  -0.03  &   0.02  \\
    0.1246  -   0.1535  &   0.04  &  -0.11  &  -0.13  &   0.09  &  -0.19  &  -0.34  &  -0.38  &   0.21  &   0.38  &  -0.16  &  -1.04  &   1.31  &  -0.53  &  -0.01  &   0.02  &  -0.03  &  -0.01  \\
    0.1535  -   0.1870  &  -0.03  &  -0.24  &  -0.13  &   0.07  &  -0.23  &  -0.27  &   0.34  &  -0.14  &   0.34  &  -0.20  &  -1.08  &   1.41  &  -0.43  &   0.00  &   0.02  &  -0.03  &  -0.07  \\
    0.1870  -   0.2277  &  -0.03  &  -0.15  &  -0.14  &   0.05  &  -0.34  &  -0.23  &   0.26  &   0.08  &   0.33  &  -0.15  &  -1.13  &   1.36  &  -0.38  &   0.02  &   0.02  &  -0.03  &  -0.07  \\
    0.2277  -   0.2800  &  -0.08  &  -0.24  &  -0.17  &   0.04  &  -0.45  &  -0.22  &   0.17  &   0.14  &   0.29  &  -0.13  &  -1.20  &   1.33  &  -0.41  &   0.03  &   0.03  &  -0.02  &  -0.08  \\
    0.2800  -   0.3590  &  -0.09  &  -0.07  &   0.33  &   0.04  &  -0.46  &  -0.22  &   0.30  &   0.11  &   0.25  &  -0.08  &  -1.26  &   1.31  &  -0.46  &   0.04  &   0.04  &   0.00  &  -0.07  \\
    0.3590  -   0.4583  &   0.05  &  -0.20  &   0.71  &   0.04  &  -0.38  &  -0.22  &  -0.28  &   0.06  &   0.22  &  -0.02  &  -1.30  &   1.26  &  -0.58  &   0.04  &   0.02  &   0.03  &  -0.03  \\
    0.4583  -   0.5838  &   0.07  &  -0.29  &   0.90  &   0.03  &  -0.27  &  -0.22  &  -0.19  &   0.10  &   0.15  &   0.15  &  -1.32  &   1.34  &  -0.95  &   0.03  &  -0.11  &   0.07  &  -0.08  \\
    0.5838  -   0.7500  &   0.08  &  -0.43  &   0.99  &   0.03  &  -0.07  &  -0.22  &  -0.24  &   0.12  &   0.11  &   0.90  &  -1.42  &   1.60  &  -1.34  &   0.03  &  -0.31  &   0.08  &  -0.39  \\
\hline
\hline
\end{tabular}
\normalsize
\caption {\it Summary of the relative systematic uncertainties (in percentage) on the measurement of ${\mathcal{R}}_{}\mu \mu$ as a function of $x_{\rm Bj}$. Each column gives the relative change $\delta_i$ resulting from a $+1\sigma$ variation of the corresponding parameter $i$.}
\label{tab:prodcharmsyst_xbj_14a}
\end{center}
\end{sidetable}

\begin{sidetable}[htb]
\begin{center}
\small 
\begin{tabular}{||c|c|c|c|c|c|c|c|c|c|c|c|c|c|c|c|c|c||}
\hline
\hline
 Bin / $\delta^{syst}$, \% & $\delta_{1}$ & $\delta_{2}$ & $\delta_{3}$ & $\delta_{4}$ & $\delta_{5}$ & $\delta_{6}$ & $\delta_{7}$ & $\delta_{8}$ & $\delta_{9}$ & $\delta_{10}$ & $\delta_{11}$ & $\delta_{12}$ & $\delta_{13}$ & $\delta_{14}$ & $\delta_{15}$ & $\delta_{16}$ & $\delta_{17}$ \\
\hline
    3.000   -   3.870   &   0.19  &  -0.11  &   0.58  &   0.05  &   0.18  &  -0.56  &  -0.07  &   0.17  &   0.30  &  -0.51  &  -1.08  &   0.41  &  -2.15  &   0.03  &  -0.06  &  -0.00  &   0.03  \\
    3.870   -   4.570   &   0.16  &  -0.06  &   0.38  &   0.03  &  -0.29  &  -0.54  &  -0.08  &   0.14  &   0.34  &  -0.29  &  -1.02  &   1.68  &  -1.46  &   0.02  &  -0.02  &  -0.02  &   0.07  \\
    4.570   -   5.250   &   0.15  &  -0.14  &   0.23  &   0.05  &  -0.41  &  -0.46  &  -0.09  &   0.12  &   0.39  &  -0.19  &  -0.96  &   2.03  &  -0.89  &   0.01  &   0.01  &  -0.02  &   0.05  \\
    5.250   -   5.800   &  -0.08  &  -0.16  &   0.09  &   0.06  &  -0.43  &  -0.33  &  -0.09  &   0.07  &   0.45  &  -0.12  &  -0.94  &   2.00  &  -0.53  &  -0.00  &   0.02  &  -0.03  &   0.01  \\
    5.800   -   6.301   &   0.07  &  -0.15  &  -0.07  &   0.07  &  -0.39  &  -0.22  &  -0.08  &   0.04  &   0.44  &  -0.11  &  -0.93  &   1.89  &  -0.36  &  -0.01  &   0.03  &  -0.03  &  -0.01  \\
    6.301   -   6.818   &  -0.07  &  -0.25  &  -0.09  &   0.08  &  -0.34  &  -0.17  &  -0.07  &   0.04  &   0.45  &  -0.09  &  -0.93  &   1.76  &  -0.20  &  -0.01  &   0.04  &  -0.03  &  -0.03  \\
    6.818   -   7.326   &  -0.05  &  -0.19  &  -0.11  &   0.08  &  -0.30  &  -0.14  &  -0.06  &   0.06  &   0.45  &  -0.08  &  -0.95  &   1.61  &  -0.12  &  -0.02  &   0.05  &  -0.03  &  -0.03  \\
    7.326   -   7.849   &  -0.13  &  -0.16  &  -0.13  &   0.07  &  -0.28  &  -0.10  &  -0.04  &   0.09  &   0.44  &  -0.03  &  -0.98  &   1.50  &  -0.01  &  -0.02  &   0.05  &  -0.03  &  -0.03  \\
    7.849   -   8.407   &   0.03  &  -0.20  &  -0.19  &   0.07  &  -0.27  &  -0.11  &  -0.02  &   0.11  &   0.44  &  -0.08  &  -1.02  &   1.27  &   0.06  &  -0.02  &   0.06  &  -0.02  &  -0.04  \\
    8.407   -   9.000   &  -0.10  &   0.16  &  -0.23  &   0.06  &  -0.26  &  -0.10  &   0.01  &   0.12  &   0.42  &  -0.09  &  -1.06  &   1.26  &   0.14  &  -0.02  &   0.06  &  -0.02  &  -0.07  \\
    9.000   -   9.801   &   0.08  &   0.14  &  -0.25  &   0.06  &  -0.27  &  -0.10  &   0.05  &   0.12  &   0.42  &  -0.10  &  -1.11  &   1.04  &   0.22  &  -0.02  &   0.07  &  -0.02  &  -0.06  \\
    9.801   -  10.74    &   0.07  &   0.22  &  -0.24  &   0.05  &  -0.30  &  -0.10  &   0.14  &   0.11  &   0.41  &  -0.05  &  -1.15  &   0.80  &   0.30  &  -0.02  &   0.07  &  -0.02  &  -0.07  \\
   10.74    -  11.93    &  -0.22  &  -0.22  &  -0.24  &   0.05  &  -0.33  &  -0.10  &   0.32  &   0.09  &   0.49  &  -0.07  &  -1.19  &   0.65  &   0.37  &  -0.03  &   0.07  &  -0.03  &  -0.09  \\
   11.93    -  14.00    &  -0.22  &  -0.13  &  -0.33  &   0.04  &  -0.36  &  -0.08  &   0.79  &   0.07  &   0.68  &  -0.11  &  -1.20  &   0.54  &   0.46  &  -0.05  &   0.07  &  -0.05  &  -0.09  \\
   14.00    -  18.00    &   0.13  &  -0.17  &  -0.48  &  -0.02  &  -0.39  &  -0.06  &   1.88  &   0.09  &   1.06  &  -0.09  &  -1.21  &   0.58  &   0.61  &  -0.08  &   0.08  &  -0.12  &  -0.08  \\
\hline
\hline
\end{tabular}
\normalsize
\caption {\it Summary of the relative systematic uncertainties (in percentage) on the measurement of ${\mathcal{R}}_{}\mu \mu$ as a function of the center of mass energy $\sqrt{\hat{s}}$. Each column gives the relative change $\delta_i$ resulting from a $+1\sigma$ variation of the corresponding parameter $i$.}
\label{tab:prodcharmsyst_shat_15a}
\end{center}
\end{sidetable}

\section{Determination of Charm Production Parameters}
\label{sec:charmfit}  

The unfolding correction factorizes out the detector acceptance from the measurement. 
Therefore, the resulting cross-sections can be directly compared with the analytical 
model to extract the charm production parameters, which include the mass of the 
charm quark, $m_c$, the effective semileptonic branching ratio, $B_\mu$, and the 
strange sea parton distribution function, $s(x)$. 

\begin{table}[t]
\begin{centering}
\begin{tabular}{l|ccc} \hline
Parameter & $A_s$ & $a_s$ & $b_s$ \\ \hline\hline
NOMAD+E531 & $0.0899 \pm 0.0029$  & $-0.240$  &  $8.75 \pm 0.43$ \\ \hline
\end{tabular}
\caption{\it Parameters describing the strange quark distribution according to 
Eq.~(\ref{eqn:ssea}) obtained from our QCD fit to NOMAD dimuon data and E531 data.}  
\label{tab:charmfit}
\end{centering}
\end{table}

We fit the NOMAD ${\mathcal R}_{\mu \mu}$ data within the framework described in 
Ref.~\cite{Alekhin:2008mb} to satisfy QCD sum rules. 
We parameterize the $x$-dependence of the strange quark distribution as: 
\begin{equation} 
\label{eqn:ssea} 
xs(x,Q^2_0) = A_s x^{a_s} \left( 1 - x \right)^{b_s} 
\end{equation} 
at the starting value of the QCD evolution $Q^2_0=9$ GeV$^2$. The low-$x$ exponent $a_s$ is assumed 
to be the same as the one for non-strange sea quark distributions, since the existing dimuon data are 
not sensitive to this parameter~\cite{Alekhin:2008mb}. The remaining parameters $A_s$ and $b_s$ are 
extracted from a fit to the NOMAD ${\mathcal R}_{\mu \mu}$ distributions simultaneously with the mass of the 
charm quark $m_c$, and the parameters describing the energy dependence of the charm semileptonic 
branching ratio (see Section~\ref{sec:bmu}). 
In order to determine the charm production parameters from NOMAD dimuon data alone, 
our fit does not include the dimuon samples from NuTeV and CCFR. We add the E531 data to constrain 
$B_\mu$, as discussed in Section~\ref{sec:bmu}. The non-strange parton distribution functions are 
initially fixed to the ones extracted from a global fit to charged lepton DIS data and 
Drell-Yan data~\cite{Alekhin:2012ig}. In the fit of NOMAD data we take into account both 
statistical and systematic uncertainties, including the detailed bin to bin correlations from 
Tables~\ref{tab:prodcharmsyst_enu_19a}-\ref{tab:prodcharmsyst_shat_15a}.  
For the calculation of the heavy quark contributions to the structure functions we use the framework 
introduced in Ref.~\cite{Alekhin:2010sv}, which uses the running mass in the $\overline{\rm MS}$ scheme 
for DIS charm production. 
The calculation is performed within the fixed 3-flavor scheme at the NLO approximation for the 
coefficients of the heavy quark DIS structure functions~\footnote{\it A NNLO calculation of the 
heavy quark coefficient functions for the neutrino DIS charged current structure functions is not available yet.} 
and at the NNLO approximation for the light quark DIS contributions. 

Table~\ref{tab:charmfit} summarizes the parameters of the strange quark distribution obtained from 
our fit to the NOMAD dimuon data and the E531 data. 
We obtain the following results for the $\overline{\rm MS}$ running mass, 
$m_c(m_c)$, and the strange quark sea suppression factor, $\kappa_s$: 
\begin{eqnarray} 
m_c(m_c) & \;\;\; = \;\;\; & 1.159 \pm 0.075~{\rm GeV/c^2} \label{eqn:mc} \\ 
\kappa_s(Q^2=20~{\rm GeV}^2/{\rm c^2}) & \;\;\; = \;\;\; & 0.591 \pm 0.019  \label{eqn:kappa} 
\end{eqnarray} 
where the renormalization scale has been chosen $\mu_r = m_c$ and both statistical and systematic 
uncertainties have been included. The best fit corresponds to a value 
of $\chi^2/DOF=53/48$. In this fit the strange quark sea distribution and the value of 
$m_c$ are entirely defined by the new NOMAD dimuon data. We also performed 
a fit by using the NLO approximation for both the light and heavy quark coefficient functions 
and for the PDFs, obtaining results very similar to our standard fit.  

Our results show that the NOMAD dimuon measurement described in this paper allows a reduction by 
more than a factor of two of the uncertainty on the strage sea distribution and on $\kappa_s$,  
with respect to the ones obtained in Ref.~\cite{Alekhin:2008mb} from NuTeV and CCFR.  
It is worth noting that we can obtain a further reduction on the uncertainty of the strange quark 
sea distributions by including both NOMAD and NuTeV/CCFR in a global fit~\cite{DIS11}, 
since the kinematic coverage of NOMAD dimuon data is complementary to the one in NuTeV and CCFR. 
Details of this global fit are 
outside the scope of this paper and will be described in a separate publication.   

The uncertainty on $m_c$ is also substantially reduced with respect to the one obtained in 
Ref.~\cite{Alekhin:2008mb} from NuTeV and CCFR. 
We note that our value of $m_c(m_c)$ is somewhat lower than the average one reported in Ref.~\cite{PDG12}.  
This fact can be explained with the different QCD approximations used to define the running mass - one loop NLO for 
our fit and 2 loops NNLO for Ref.~\cite{PDG12} - since $m_c(m_c)$ is expected to increase with the 
number of loops considered. 

From the same fit to NOMAD dimuon data including the E531 data from Figure~\ref{fig:bmu} 
we also extract the semileptonic branching ratio 
$B_\mu$, parameterized according to Eq.~\ref{eqn:bmu}. The corresponding parameters   
$a$ and $b$ describing the energy dependence of $B_\mu$ are summarized in the last 
row of Table~\ref{tab:bmufit}. The NOMAD data allow a substantial reduction of uncertainties 
in the determination of $B_\mu$ with respect to previous experiments, due to the much lower 
energy threshold.

\section{Summary} 
\label{sec:summary} 

We have presented a new measurement of charm dimuon production using events originated in the 
Front Calorimeter of the NOMAD experiment (Fe target).    
Our final data sample has the largest statistics - 15,344 charm dimuon events -  
as well as the lowest energy threshold - $E_\nu \sim 6~GeV$ - among the existing measurements. 
We find the charm dimuon production rate, averaged over the NOMAD flux for $Q^2>1$~GeV$^2$/c$^2$,  
to be $(5.15\pm0.05 \pm 0.07) \times 10^{-3}$ compared to the rate of the inclusive $\nu_\mu$ CC interactions. 

A key feature of our analysis is the extraction of the whole background predictions from the  
data themselves, together with a substantial reduction of the minimal energy thresholds on the 
observed muons with respect to previous measurements. Both achievements have been possible only 
through the use of a low density and high resolution magnetic spectrometer, allowing a detailed 
reconstruction of the energy and momentum of individual secondary particles produced in 
neutrino interactions. With such a high resolution detector and a control sample of more than 
$9\times 10^6$ reconstructed $\nu_\mu$ CC events, we could keep the total systematic uncertainties 
- including 17 different sources - on the ratio of charm dimuon to inclusive CC 
cross-sections at the level of about $2\%$. 
This value  makes our analysis the most precise measurement of charm 
dimuon production in neutrino interactions.  

Finally, we used the new NOMAD data on charm dimuon production to determine the charm 
production parameters and the strange quark sea content of the nucleon within the context of a QCD fit.
For the calculation of the charm contributions to the structure functions we use the 
running mass in the $\overline{\rm MS}$ scheme and include NLO (NNLO) corrections for 
the heavy (light) quark QCD coefficients functions. 
We obtain a mass of the charm quark $m_c(m_c) = 1.159 \pm 0.075$~GeV/c$^2$ 
and a strange quark sea suppression factor $\kappa_s = 0.591 \pm 0.019$ at $Q^2=20$~GeV$^2$/c$^2$. 
We also obtain a value $\varepsilon_{\rm c}=0.165\pm0.025$ for the free parameter in the Collins-Spiller 
charm fragmentation function, and a parameterization of the semileptonic branching ratio 
as a function of the neutrino energy $B_\mu(E_\nu) = (0.097\pm0.003) / [1+(6.7\pm1.8~{\rm GeV})/E_\nu]$. 
All our results on $m_c, k_s, \varepsilon_{\rm c}$ and $B_\mu$ are the most precise measurements  
from neutrino data.

\section*{Acknowledgments}

We extend our grateful appreciations to the CERN SPS staff for the magnificent 
performance of the neutrino beam. 
The experiment was supported 
by the following agencies: 
ARC and DIISR of Australia; IN2P3 and CEA of France, BMBF of 
Germany, INFN of Italy, JINR and INR of Russia, FNSRS of 
Switzerland, DOE, NSF, Sloan, and Cottrell Foundations of 
USA, and VP Research Office of the University of South Carolina. 
This work was partially supported by the University of South Carolina, 
by the DOE grant DE-FG02-95ER40910, by the Russian Federal grant 02.740.11.5220 and 
MK-432.2013.2 and JINR grant 13-201-01.

\end{document}